\documentclass[english]{article}
\usepackage[T1]{fontenc}
\usepackage{geometry}
\usepackage{amsmath}
\usepackage{amssymb}
\usepackage{dsfont}
\usepackage{mathtools}
\usepackage{url}
\usepackage{xcolor}
\usepackage{cite}
\usepackage{comment}
\usepackage{caption}
\usepackage{subcaption}
\usepackage{svg}

\definecolor{EdwardsLinkColor}{rgb}{0.0,0.0,0.3}

\usepackage[unicode=true,
 bookmarks=true,bookmarksnumbered=false,bookmarksopen=false,
 breaklinks=false,pdfborder={0 0 0},pdfborderstyle={},backref=false,colorlinks=true]
 {hyperref}
\hypersetup{pdftitle={Open-Closed-Open Triality},colorlinks=true,urlcolor=EdwardsLinkColor,anchorcolor=EdwardsLinkColor,citecolor=EdwardsLinkColor,filecolor=EdwardsLinkColor,linkcolor=EdwardsLinkColor,menucolor=EdwardsLinkColor,linktocpage=true}

\DeclareMathOperator{\Tr}{Tr}

\makeatletter
\usepackage{braket}
\let\mc=\mathcal
\let\mb=\mathbb

\newcommand\be{\begin{equation}}
\newcommand\ee{\end{equation}}

\makeatother

\usepackage{babel}

\begin{document}

\title{\bf{Deriving the Simplest Gauge-String Duality -- I: Open-Closed-Open Triality}}

\author{Rajesh Gopakumar${^*}$ \& Edward A. Mazenc${^{**}}$}
\date{}

\maketitle

\begin{center}
    \small{*\it{International Centre for Theoretical Sciences-TIFR,\\ Shivakote, Hesaraghatta Hobli, Bengaluru North 560089, India}}\\
\vspace{0.1cm}
  \small{**\it{Kadanoff Center for Theoretical Physics, University of Chicago,\\ Chicago, IL 60637, USA }}
\end{center}

\vspace{2cm}

\begin{abstract}
We lay out an approach to derive the closed string dual to the simplest possible gauge theory, a single hermitian matrix integral, in the conventional 't Hooft large $N$ limit. In this first installment of three papers, we propose and verify an explicit correspondence with a (mirror) pair of closed topological string theories. On the A-model side, this is a supersymmetric $SL(2, \mathbb{R})_1/U(1)$
Kazama-Suzuki coset (with background momentum modes turned on). The mirror B-model is a topological Landau-Ginzburg theory with superpotential $W(Z)=\frac{1}{Z}$ and its deformations. We arrive at these duals through an "open-closed-open triality". This is the notion that \textit{two} open string descriptions ought to exist for the same closed string theory depending on how closed strings manifest themselves from open string modes. 
Applying this idea to the hermitian matrix model gives an exact mapping to the Imbimbo-Mukhi matrix model. The latter model is known to capture the physical correlators of the $c=1$ string theory at self-dual radius, which, in turn, has the equivalent topological string descriptions given above. This enables us to establish the equality of correlators, to all genus, between single trace operators in our original matrix model and those of the dual closed strings. Finally, we comment on how this simplest of dualities might be fruitfully viewed in terms of an embedding into the full AdS/CFT correspondence.


\end{abstract}

\pagebreak

\tableofcontents

\pagebreak

\section{The Big Picture and A Roadmap} \label{sec:bigpicture}

\subsection{Introduction}
Well before the advent of the AdS/CFT correspondence, 't Hooft proposed that the large $N$ expansion of gauge theories is the genus expansion of a dual string theory \cite{tHooftPlanar}. This was suggested by his insight that large $N$  Feynman diagrams of a gauge theory can be viewed as triangulations of a Riemann surface of the appropriate genus. While pictorially suggestive, this connection between Feynman diagrams and worldsheets is not easy to make precise into a calculational framework that enables one to derive the dual string theory or even qualitatively describe it. As a result, this picture is often relegated to mere motivation in our modern view of gauge/string duality, which has shifted its focus onto the transformative concept of holography.  

The integral over a single hermitian $N \times N$ matrix is arguably the simplest large $N$ gauge theory. As was advocated, more than ten years ago, by one of the authors, it provides an ideal testing ground to realize 't Hooft's original picture in full detail, and uncover the mechanics of gauge-string duality \cite{gopakumar2011simplest, gopakumar2013correlators}. In particular, correlators of single trace operators in the famous Gaussian or Wigner matrix model capture the combinatorics of the large $N$ Wick contractions. From that point of view, the Gaussian matrix model is the simplest instance of the free gauge theories which are dual to tensionless string theories on $AdS$. In recent years, there has been some success in understanding and even deriving the AdS/CFT correspondence for the symmetric product orbifold $CFT_2$ \cite{Gaberdiel:2018rqv, eberhardt2020deriving,eberhardt2019worldsheet,dei2021free}, for free 4d ${\cal N}=4$ Super Yang-Mills \cite{gaberdiel2021string,gaberdiel2021worldsheet} and large $N$ vector models \cite{Aharony:2020omh}. It is thus an opportune moment to revisit the simplest gauge string duality.

\begin{figure}[h!!]
     \centering
         \includegraphics[width=\textwidth]{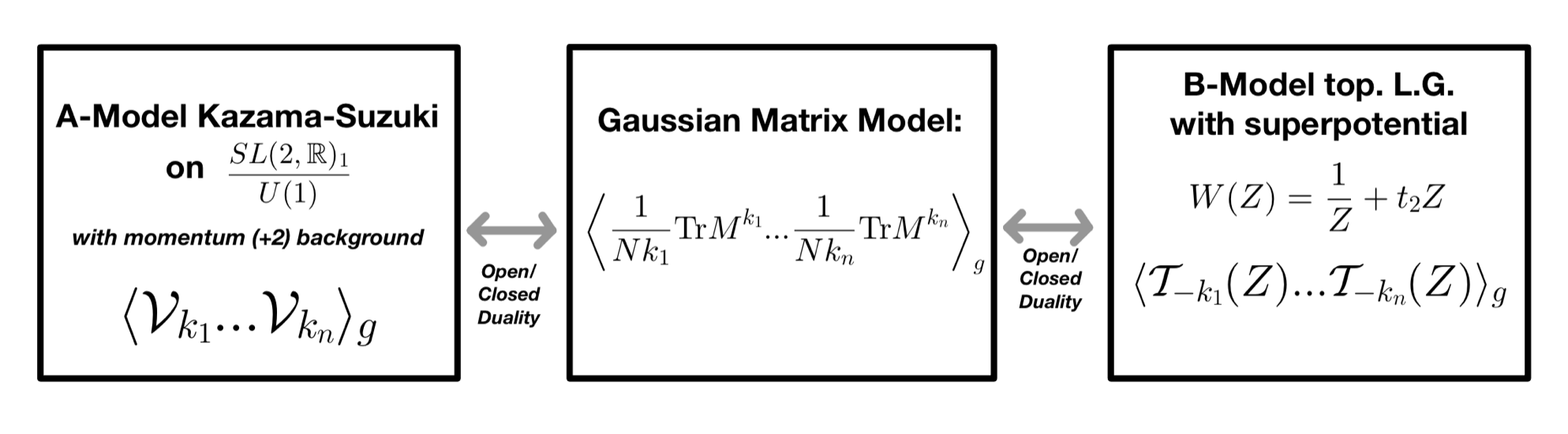}
        \caption{\small{\textbf{The Closed String Dual(s) to the Gaussian Matrix Model:} For concreteness, we often study the Gaussian matrix model (though we stress our methods extend beyond the free case). We will see in this paper that open-closed-open triality guarantees that the correlation functions of single trace operators agree to all orders in the genus expansion with 1) correlators of physical vertex operators in the A-twisted Kazama-Suzuki coset, with a particular $(+2)$ momentum mode background and 2) with physical correlators in a B-model topological Landau-Ginzburg theory with the shown deformed superpotential. More complicated matrix potentials simply turn on backgrounds of other momentum modes in the A-model, and give rise to further powers of $Z$ in the B-model superpotential.}}
        \label{fig:Gaussianproposal}
\end{figure}

In this and accompanying two papers \cite{DSDII, DSDIII}, we propose, verify and go quite some way towards deriving two (mirror) worldsheet closed string theories dual to the Gaussian matrix model\footnote{As will become clear, much of what we will say will also go through for a one matrix model $M$ with an arbitrary potential ${\rm Tr}V(M)$, under the `one-cut' assumption (i.e. the semi-classical eigenvalue distribution is supported on a single interval).}. More precisely, we will argue here, somewhat indirectly, that all $n$-point functions of single trace operators ${\rm Tr}M^k$ agree with a corresponding set of physical correlators in an A- and (mirror) B-model topological string. See Fig \ref{fig:Gaussianproposal}. The agreement holds to all orders in the $\frac{1}{N}$ expansion. Later papers will describe how to directly derive these two string duals starting from the matrix model.

Of course, matrix models and string theory have enjoyed a fruitful marriage for over thirty years, primarily in the context of non-critical or minimal model backgrounds - see \cite{ginsparg1993lectures, klebanov1991string} for reviews and 
\cite{Anninos:2020ccj} for a recent overview. However, we should emphasize that, in these developments, the mechanism through which the matrix Feynman diagrams connect to the closed string worldsheet is not quite within the standard 't Hooft paradigm. Rather one views the matrix ribbon graphs as a discretization of a worldsheet and therefore not universal (in the sense of critical phenomena). The philosophy then is that one needs to take a double-scaling limit to a critical potential, to recover the continuum Riemann surface with its string correlators \cite{gross-migdal, brezin1990exactly,KAZAKOV1990212,douglas1990strings}. This is in contrast to our most well-understood examples of the AdS/CFT correspondence, where we do not resort to such a limit. 

In the program \cite{freefieldsadsI,freefieldsadsII,freefieldsIII} to derive string theories from free field theories one instead associates a string worldsheet to {\it each} large $N$ Feynman diagram.  This exploits the Strebel parametrization of the moduli space of Riemann surfaces which allows us to map each Feynman diagram to a point on $\mc{M}_{g,n}$. In that sense it is a refinement of the 't Hooft assignment of a genus to a diagram. The Strebel map translates the sum over Wick contractions in the gauge theory into a (sum or) integral over the dual string's moduli space. We can also view this construction as spelling out precisely how open string strips - which are the origin of the double-lines of the gauge theory diagrams - assemble into a (punctured) closed string worldsheet. This, therefore, is also a concrete realisation of open-closed string duality. In the present works, we will employ this paradigm to study matrix integrals in the standard 't Hooft limit\footnote{Dijkgraaf and Vafa \cite{DijkgraafVafa02, Aganagic_2005} were probably the first to emphasise that one can consider the string duals of these matrix models without double-scaling. Our duals bear a close relation to their proposal even though they are somewhat differently framed. We will comment further on this in Sec. 8.}. The closed string duals exhibited in Fig.~\ref{fig:Gaussianproposal} are as well defined as any noncritical string theory. We will see, in later papers,  how we can recover the more familiar noncritical closed string amplitudes by taking an appropriate BMN-like limit of large operators\footnote{Stringy descriptions of the Gaussian matrix model have been proposed before. Some examples are \cite{razamatGauss, razamat2010matrices, itzhaki2005large, berenstein2004toy, gopakumar2011simplest, gopakumar2013correlators}. The proposal in \cite{gopakumar2011simplest,gopakumar2013correlators} (see also \cite{deMelloKoch:2014khl} for a larger class of correlators) for an A-model topological string on $\mathbb{P}^1$ is quite close to the A-model dual proposed here. It would be good to understand the precise relation better.}.   

\subsection{The Roadmap}

\begin{figure}[h!]
     \centering
         \includegraphics[width=0.9\textwidth]{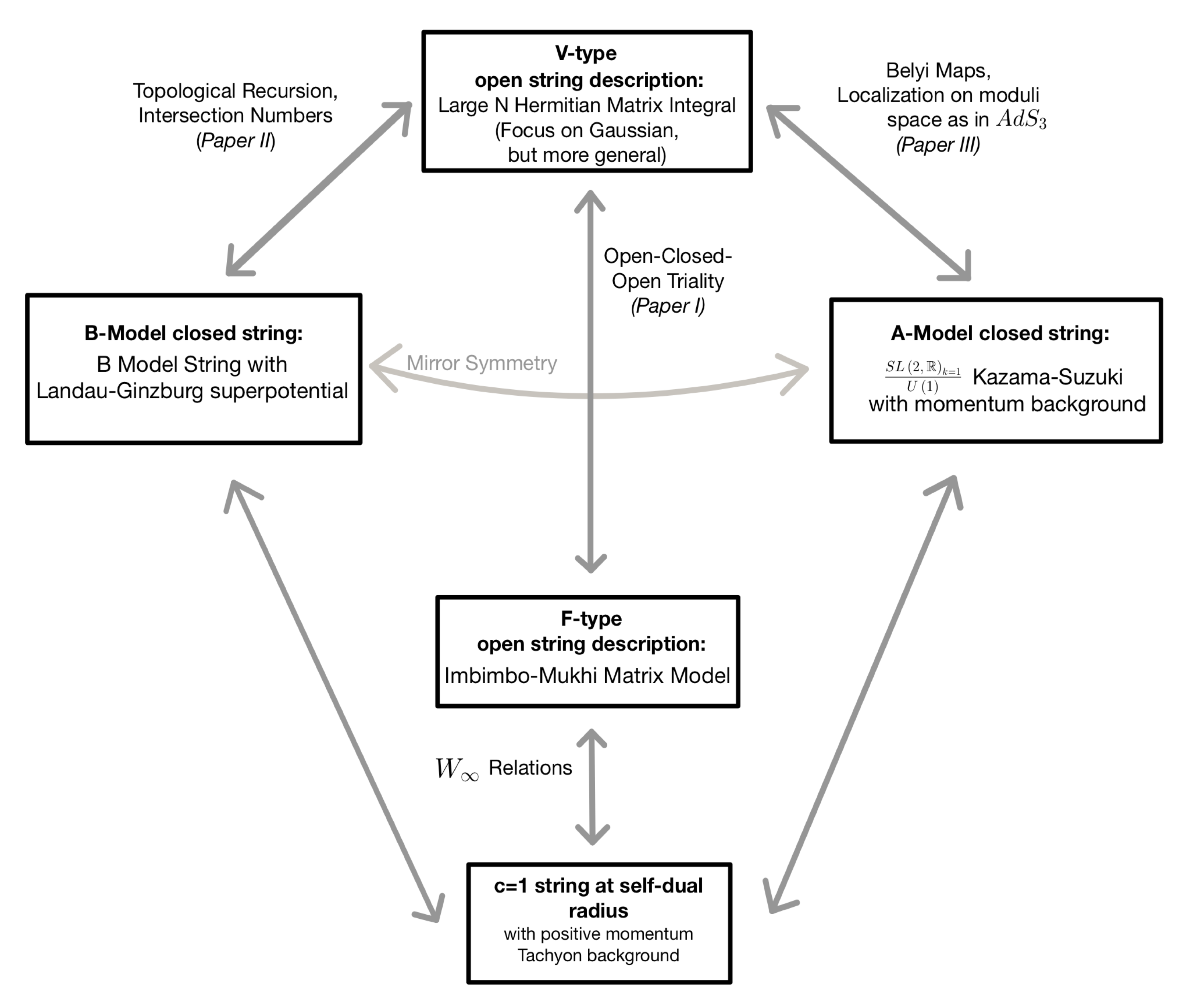}
        \caption{\small{\textbf{A web of dualities \& papers:} This current paper uses open-closed-open triality to establish an equality of the one matrix model with the Imbimbo-Mukhi matrix model, which encodes all correlation functions of tachyon vertex operators in the $c=1$ string at self-dual radius. Known topological string descriptions of this ancillary $c=1$ theory verify our proposed A- and B-model worldsheet theories and guarantee agreement of correlators to all orders in the genus expansion. Our subsequent papers aim to directly derive the A- and B-model duals without reference to the $c=1$ string.}}
        \label{fig:flowchart}
\end{figure}

Figure \ref{fig:flowchart} outlines the three pillars of our derivation of this `bare-bones' version of the gauge-string correspondence. 
We offer this figure to the reader as a roadmap to the work presented in this paper as well as a preview of the following two papers. There is a somewhat involved web of connections and existing dualities that we build upon and we trust this section will serve as some kind of a compass.

This first paper will be largely devoted to proposing the two topological closed string duals to the Gaussian model (and its interacting generalizations). We will also provide compelling evidence for the duality by arguing for the match of physical correlation functions of the string dual(s) to the correlators of matrix single traces. The argument  relies on a somewhat surprising connection of the matrix model to the $c=1$ string theory as indicated in the bottom box of Fig.~\ref{fig:flowchart}. The connection rests on the mathematical equivalence of two different matrix models as shown. However, we believe it is fruitful to see this as more than just a convenient trick but rather as a manifestation of the more general underlying phenomenon of what can be dubbed \textit{open-closed-open triality}. This paper will therefore have two distinct parts which can more or less be read independently. The first part will be on the broad philosophy of open-closed-open triality, its ramifications and how it fits snugly with the Strebel program of \cite{freefieldsadsII, freefieldsIII} for reconstructing the closed string. We will also illustrate the idea through the relation between a pair of two matrix integrals. We will use a special case of this in part II to make the connection to the $c=1$ string. We will then use known equivalences of the latter to the A-model and B-model topological strings to argue for the matching of correlators.

The second paper in this series \cite{DSDII} will focus on the equivalence to the dual B-model closed string, which is a topological Landau-Ginzburg theory coupled to two-dimensional topological gravity. Its superpotential can be read off from the spectral curve of the matrix model, as also suggested in \cite{Aganagic_2005}. Variations of the matrix model action correspond to deformations of this spectral curve. The dual proposed here is a generalisation of the theory proposed in \cite{ghoshalMukhiLG, hananyOzPlLG} for the $c=1$ string theory. We will show how to arrive at this B-model description, primarily through the use of the machinery of topological recursion \cite{Eynard:2007kz,eynard2008algebraic,eynard2bp}. This will naturally land us on an algebro-geometric description of the correlation functions of this theory in terms of intersection numbers of characteristic classes on two copies of moduli space. We will be able to view these integrals over moduli space as the localized B-model path integral, after integrating out the matter sector. The two critical points $W'(z)=0$ of the deformed superpotential/spectral curve, $W(z)=\frac{1}{z}+t_{2}z$ (and its generalization for interacting theories) give rise to the two copies (the two edges of a single cut eigenvalue saddle). We will find very explicit expressions for the dual of the single trace operators in terms of natural cohomology classes on $\mc{M}_{g,n}$.

The third instalment \cite{DSDIII} will be on the second string dual which is an A-twisted, supersymmetric Kazama-Suzuki coset $\frac{SL(2,\mb{R})}{U(1)}$, at (susy) level $k=1$. It is, of course, also coupled to topological gravity on the worldsheet. This is again a deformation of the background dual to the $c=1$ string at self-dual radius \cite{MukhiVafa}.  In this case, the matrix model potential determines a particular background momentum being turned on. The physical Hilbert space is built on the spectrally flowed $j=1/2$ representations of $\mathfrak{sl}(2, \mathbb{R})$ \cite{MOOgI}. We show how this A-model picture arises, in our third paper. The basic argument is that the matrix model Feynman diagrams can be viewed as giving a combinatorial description of certain holomorphic covering maps, branched over three points of the target space - the so-called Belyi maps \cite{bauer1993triangulations, koch2010matrix}. This connects very nicely with recent results on the tensionless limit of the $AdS_3/CFT_2$ correspondence, where  spectrally flowed $j=\frac{1}{2}$ representations of $\mathfrak{sl}(2,\mb{R})$ at level $k=1$ played a crucial role in the worldsheet theory \cite{eberhardt2020deriving, gaberdiel2021symmetric, eberhardt2019worldsheet}. By a similar argument as in that case \cite{eberhardt2020deriving, dei2021free}, the string path integral localizes to discrete points on the moduli space, as anticipated in \cite{aharony2007remarks, razamatGauss, gopakumar2011simplest}. These represent the `discretised worldsheets', which are now seen as arising from a conventional topological sigma model.   

\subsection{Plan of this paper}

Let us now give a more detailed breakup of the present paper.
As mentioned above, there are two parts to the paper, which can be read more or less independently of each other. Part I describes a certain broad, more generally applicable, approach to gauge-string duality and is much more discursive in nature. The reader who is impatient to get to the proposed closed string duals to the matrix model (and how the ancillary $c=1$ description helps match correlators) is welcome to skip directly to Part II, perhaps after looking over the matrix equivalence derived in Sec. 3.1. 

Part I fleshes out the notion of open-closed-open triality which was outlined by one of the authors in \cite{Joburg}. 
What is Open-Closed-Open triality? Succinctly, it states that there exist \textit{two} open string descriptions which encode the same closed string geometry, but in distinct ways. 
At its core, the idea is simple: by considering two separate stacks of branes and integrating out degrees of freedom of either one, we get two distinct, though equivalent open string descriptions. Moreover, these two descriptions differ in whether closed string insertions correspond to vertices or faces of the dual open string Feynman graphs. Accordingly they are dubbed as the V- or F-type dual \cite{Joburg}. This explains the terminology in Fig. \ref{fig:flowchart}. These ideas are broadly introduced in Section 2. 

Section 3 illustrates these ideas concretely through an illuminating case study. This arises in the context of open topological strings that reduce to matrix integrals \textit{\`a la} Dijkgraaf-Vafa \cite{DijkgraafVafa02, Aganagic_2005}. The  equivalence of the two open string descriptions is essentially the equality Eq.~(\ref{eq:source/det duality}) between a pair of two matrix integrals with sources and determinant insertions. The two integrals correspond to those of the open string degrees of freedom of a set of compact branes and non-compact branes. The steps in the equivalence have a meaning in terms of integrating in and out, degrees of freedom stretched between the two sets of branes \cite{Maldacena:2005hi}. We also give an alternate demonstration of the equality of the two integrals using their interpretation as brane wavefunctions. 

Section 4 shows that the V- and F-type matrix models of Sec. 3 are related by graph duality. This can be seen through a graphical interpretation of the various steps of integrating in and out. This naturally leads to the discussion in Sec. 5 which begins with a short pedagogical introduction to the mathematics of Strebel differentials. We then review the construction of the Strebel map between Feynman graphs and closed worldsheets in both V- and F-type duals. In the process we see how the graph duality of Sec. 4 sits well with an exchange of the so-called (critical) horizontal and (families of) vertical trajectories of the Strebel differential.  

Part II can gainfully be thought of as an application of the general ideas of Part I to a particular case, though it can also be viewed in a stand alone capacity as simply proposing and giving evidence for the closed string duals to the one Hermitian matrix model. 
Section 6 thus specialises the matrix integral equivalence of Sec. 3.1 such that one side reduces to a Gaussian matrix model perturbed by an arbitrary single trace potential. This generating function of single trace correlators is then mapped to another matrix integral, the Imbimbo-Mukhi matrix model \cite{ImbimboMukhi}. The latter exactly encodes all $n$-point functions of tachyon vertex operators in the two-dimensional $c=1$ string at self-dual radius. Single trace operators ${\rm Tr}M^k$ in the original matrix model correspond to negative momentum $(-k)$ tachyons in this description. The Gaussian potential corresponds to turning on a tachyon momentum $(+2)$ background.  As a sanity check, we explicitly match the all-genus expression for the one-point function of traces in the Gaussian matrix model to the expectation value of a negative momentum tachyon vertex operator in a background of positive ($+2$) momentum modes.

This unexpected relation to the $c=1$ string theory at self-dual radius is what allows us to go ahead in Sec. 7 and claim agreement of arbitrary correlators in the Gaussian model with correlators in the proposed topological closed string duals. Indeed the $c=1$-theory is known to admit two alternate, topological string formulations: a B-model topological Landau-Ginzburg theory with superpotential $W(Z)=1/Z$ \cite{ghoshalMukhiLG,ghoshalMukhiIm,hananyOzPlLG} and an A-twisted, supersymmetric Kazama-Suzuki coset $\frac{SL(2,\mb{R})}{U(1)}$ at level $k=1$ \cite{MukhiVafa, ashok2006topological}! Tachyon vertex operators are mapped onto corresponding observables, which therefore carries over into an operator dictionary for our original Gaussian single traces. The deformation by a tachyon momentum $(+2)$ background for the Gaussian potential can be mapped 
onto the necessary deformations of the A- and B-model duals. In fact, in the B-model this deforms the unperturbed $1/Z$ superpotential to $W(Z)=1/Z +t_{2}Z$. On the A-model side this corresponds to the $(+2)$ momentum mode background, mentioned above. We should stress here that the derivation in our subsequent papers does not reference the $c=1$ string which is an auxiliary crutch as far as we are concerned. However, open-closed-open triality embeds our results in this wider web of dualities.

\begin{figure}[h!]
     \centering
         \includegraphics[width=\textwidth]{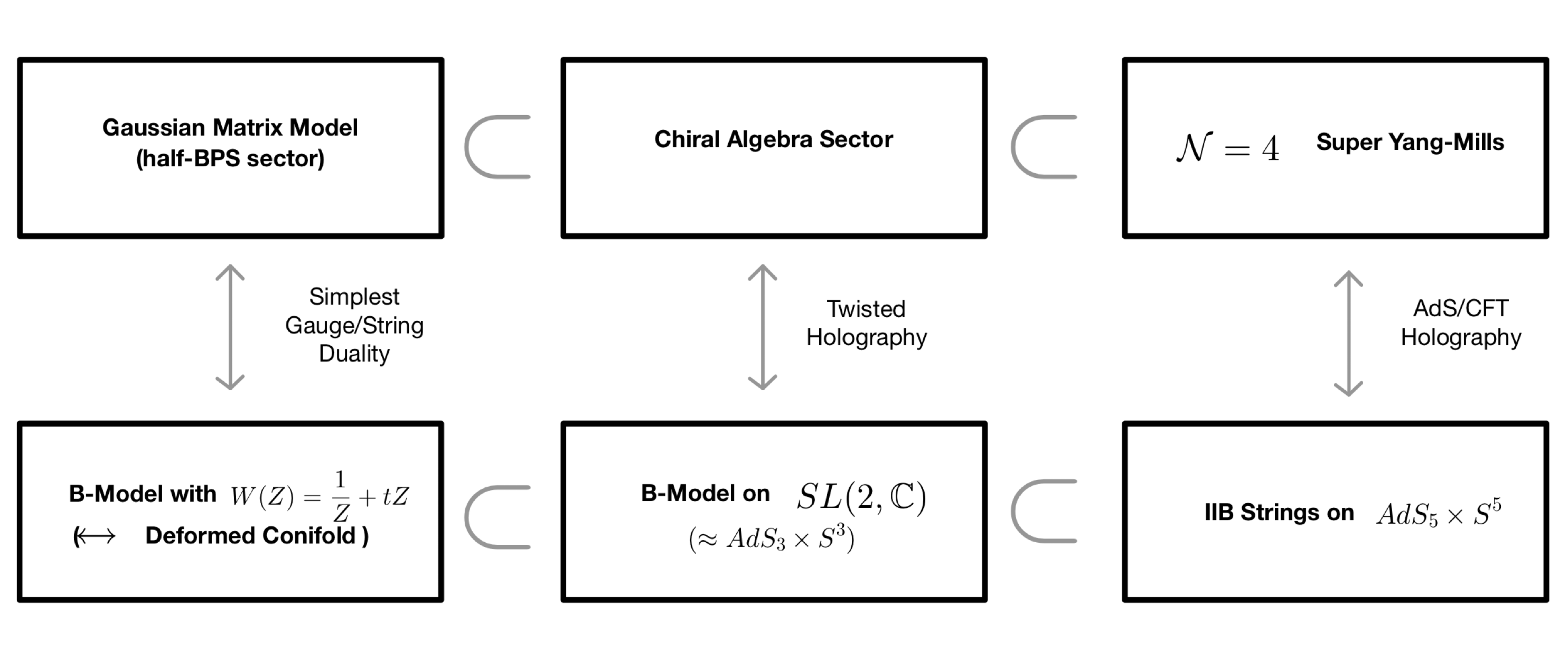}
        \caption{\small{\textbf{The even bigger picture:} The Gaussian matrix model embeds into the full-fledged AdS/CFT correspondence. It captures a half-BPS sector of $\mc{N}=4$ Super-Yang-Mills. Furthermore, its correlators encode a subset of topological observables in the chiral algebra sector. Twisted holography gives the holographic dual to this chiral sector in terms of a B-model string on the deformed conifold $SL(2,\mathbb{C})$}.}
        \label{fig:biggerpicture}
\end{figure}

Finally, section \ref{sec:discussion} contains some additional remarks by way of discussion of our results. One point that is perhaps worth mentioning already here is that this simple instance of gauge-string duality fits into the broader context of holography. Beyond being the simplest gauge theory, the Gaussian matrix model captures a particular half-BPS sector of $\mc{N}=4$ Super-Yang-Mills (SYM). In fact, the recent twisted holography program \cite{CostelloGaiottoTH,Costello:2020jbh} provides a further nesting within the AdS/CFT correspondence. Costello and Gaiotto considered the chiral subalgebra of $\mc{N}=4$ Super-Yang-Mills \cite{beemRastelli2015,bonettiRastelli2018}. They showed that it was holographically dual to a (closed) B-model string theory on $SL(2,\mathbb{C})$. The $n$-point functions of (normal ordered) traces in the Gaussian model turn out to calculate certain topological correlators in this chiral subalgebra, see section 6 of \cite{Budzik:2021fyh}. We summarize these relations in Fig. \ref{fig:biggerpicture}.

There are a few appendices which contain supplementary material as well as details of some calculations.

\pagebreak

\part{Open-Closed-Open Triality }

\section{Open-Closed-Open Triality: The General Framework}

Open-closed string duality underlies our best studied examples of the AdS/CFT correspondence. But what is the underlying mechanics here? The notion of Open-Closed-Open triality  will force us to sharpen our notion of gauge-string duality in that we will see two different open string descriptions of the same closed string theory \cite{Joburg}. As usual, from a target space perspective, the back reaction of D-branes modifies the geometry. From a worldsheet point of view, the holes `close up'. However, we will see that there are, in fact, {\it two distinct} ways the holes in open string worldsheets can close up, which we call V- and F-type. We thus distinguish V- and F-type open-closed duality. In fact, following 't Hooft, and more precisely the program articulated in \cite{freefieldsadsI,freefieldsadsII,freefieldsIII}, we can reconstruct the closed string worldsheet from the Feynman diagrams of either of the two different open string descriptions. Let us describe how this works. 

\begin{figure}[h!]
     \centering
      \begin{subfigure}[b]{0.4\textwidth}
         \centering
         \includegraphics[width=\textwidth]{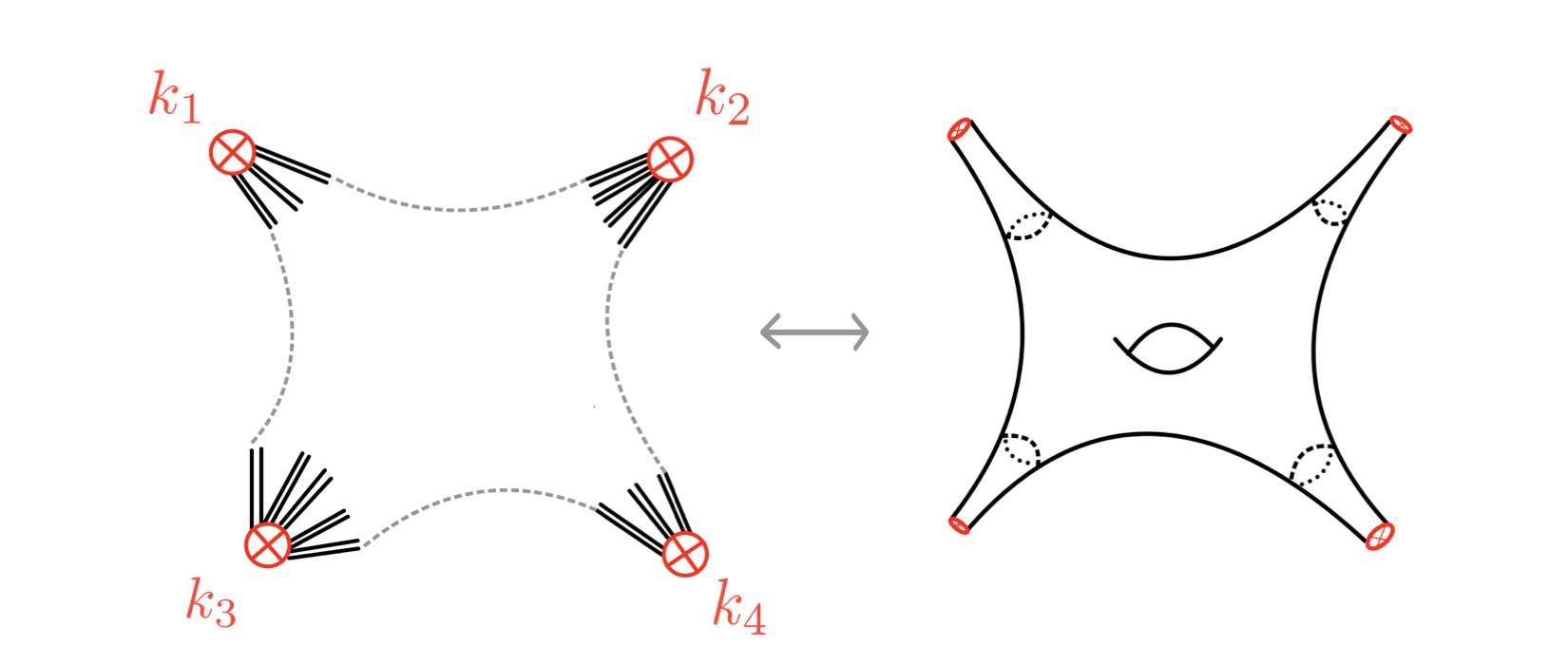}
         \caption{V-Type Duality}
         \label{fig:three sin x}
     \end{subfigure}
     \begin{subfigure}[b]{0.4\textwidth}
         \centering
         \includegraphics[width=\textwidth]{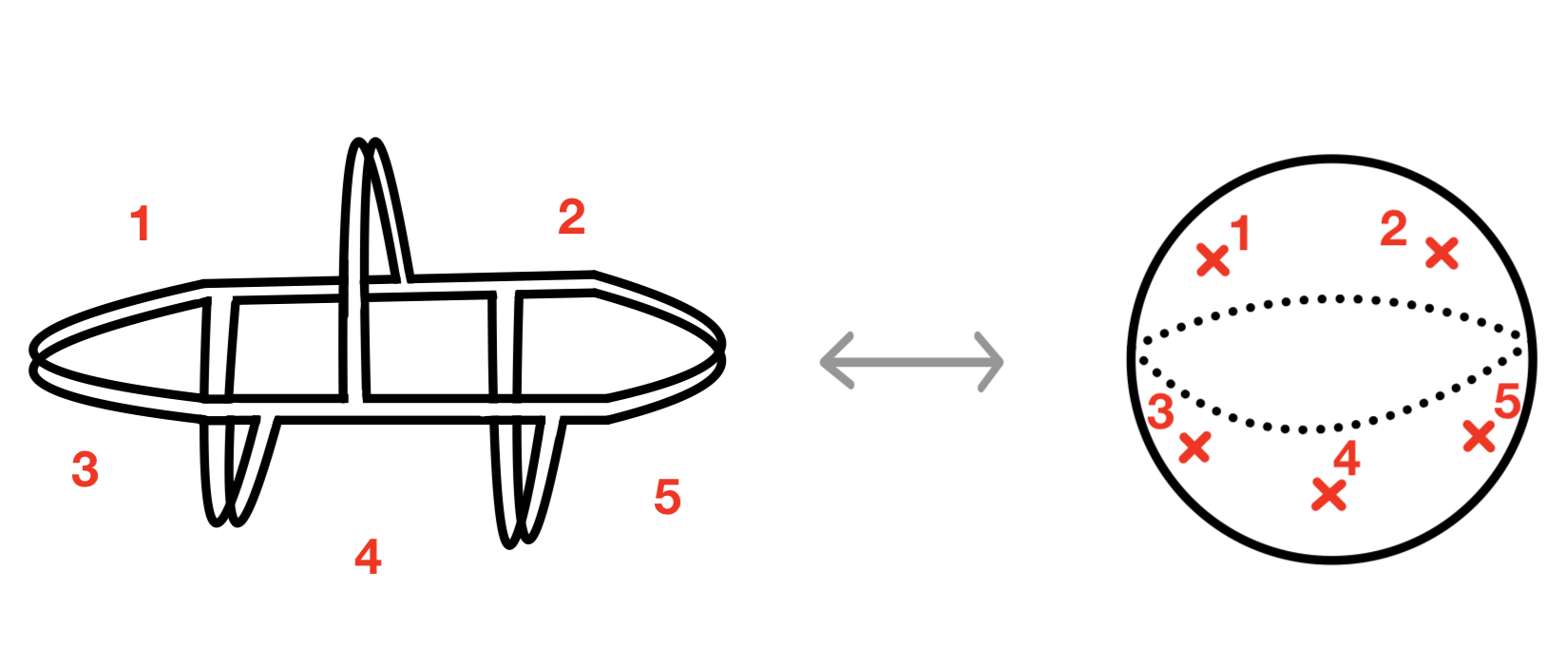}
         \caption{F-Type Duality}
         \label{fig:y equals x}
     \end{subfigure}
        \caption{ (a) In V-type open-closed duality, single trace operators create asymptotic closed string states. Put differently, the external vertices of the gauge theory Feynman diagram map onto the vertex operator insertions of the closed string worldsheet. Depicted (left) is a schematic Feynman diagram contributing to the four-point function of single trace operators. (b) In F-Type duality, the faces - which represent D-brane boundary states - shrink to closed string punctures. Depicted (right) is a genus zero Feynman diagram in, say, $SU(N)$ Chern-Simons theory and the corresponding A-model closed string worldsheet.}
        \label{fig:VversusF}
\end{figure}

In V-type duality, (external) vertices of the large $N$ gauge theory Feynman diagrams correspond to closed string insertions. The canonical duality between $\mc{N}=4$ SYM and IIB string theory on $AdS_5 \times S^5$ is of V-type. Thus we write the genus $g$ contribution to an $n$-point correlator of single trace operators as an $n$-point string correlator. Schematically,   $$\Braket{\mc{O}_{k_1}...\mc{O}_{k_n}}^{g}_{CFT} =\int_{\mc{M}_{g,n}} \Braket{\mc{V}_{k_1}(\xi_1)...\mc{V}_{k_n}(\xi_n)}_{ws}.$$ 
Here, we have suppressed most labels and indices except the $k_i$ which indicate the size of the composite gauge invariant operator or, equivalently, the coordination number of the external vertices, of the Feynman graphs, at which they are inserted. See the left hand side of Fig.~\ref{fig:VversusF}. We will also view internal vertices as generated by a perturbation expansion around the free theory correlator when we are in a weak coupling regime. This is as in textbook perturbative quantum field theory where all computations reduce to free field correlators (Wick contractions). We should view the closed string correlators in this case as computing small fluctuations around a fixed background which has already been generated by the backreaction of a set of branes. In other words, these correspond to normalisable states
in the single particle Hilbert space of the background geometry (at large $N$). 

In F-type dualities, on the other hand, it is the faces which get replaced by local vertex operators on the closed string worldsheet. This is also a natural mode of realisation of open-closed string duality, often directly arising from an Open String Field Theory (OSFT) description of D-branes. Here D-brane boundary states get replaced by (a sum of) closed string vertex operator insertions, which can be viewed as backreacting on the background. A well-known example of F-type duality is that between pure $SU(N)$ Chern-Simons theory on $S^3$ and the A-model topological (closed) string on the resolved conifold \cite{GV98, GV99}. In this case, the gauge theory is exactly the cubic open string field theory on $N$ topological A-branes wrapping the $S^3$ in the total space $T^{*}(S^3)$ \cite{WittenCSasOSFT}. In fact, the mechanism by which these D-brane boundaries or faces `open up', in a gauge linear sigma model description of the closed string background, at weak 't Hooft coupling, led to a derivation of this duality\cite{GV99, OV2002worldsheet}. Equivalently, the holes closing up is the statement of the D-branes backreacting on the background and deforming it \cite{OV2002worldsheet}. Note that D-brane boundary states are typically not normalisable. 
Figure \ref{fig:VversusF} shows examples of open string diagrams and how they map onto closed string worldsheets.

\begin{figure}[h!]
\centering
\includegraphics[scale=0.12]{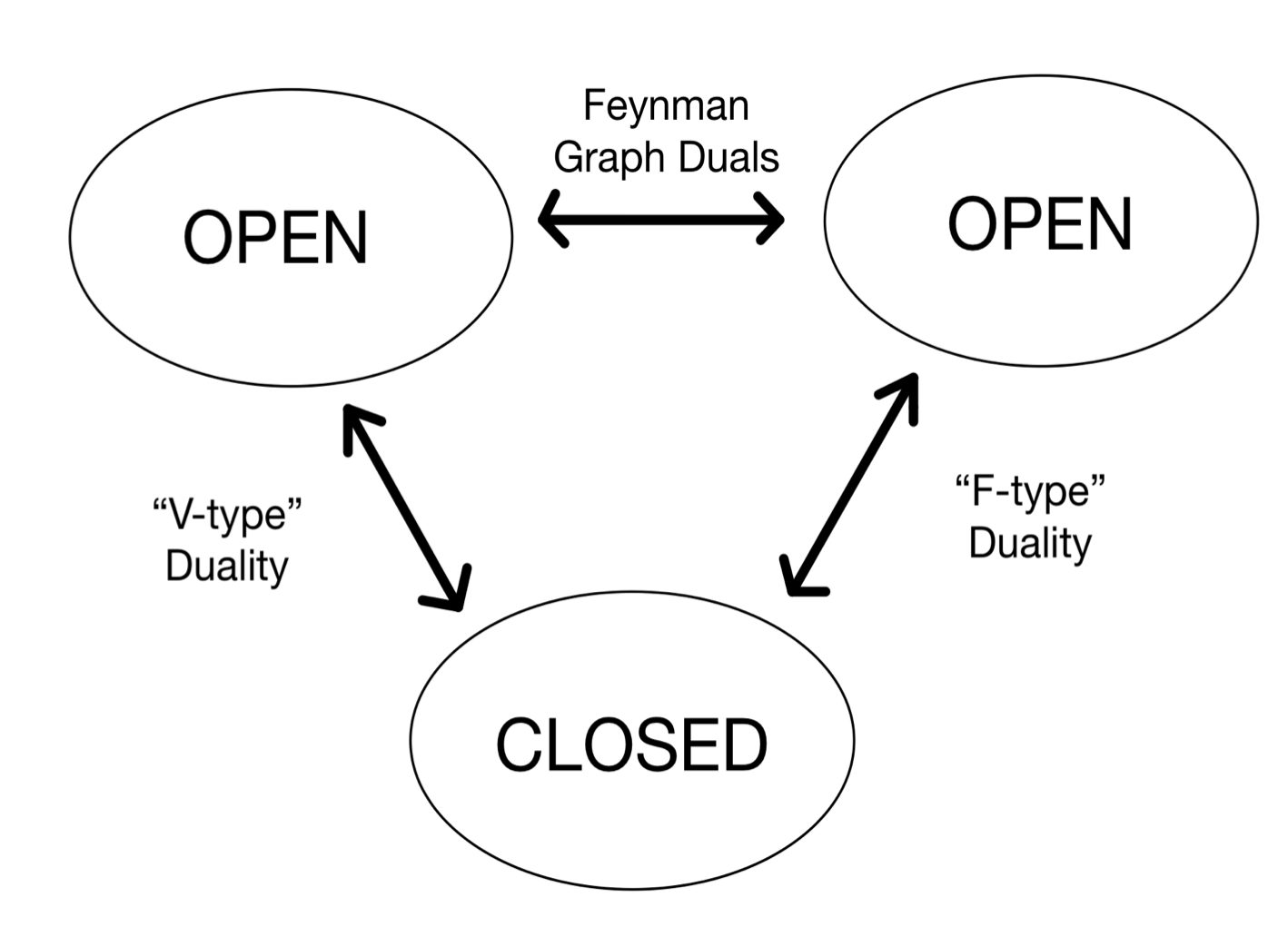}
\caption{\small{\textbf{Open-Closed-Open Triality}: Two distinct open string theories capture the same closed string geometry. The Feynman diagrams of the two open string theories are related by a dynamical graph duality, interchanging vertices and faces. In the topological string setup that we focus on in this paper - where we can make many of these ideas precise - V-type duals arise as the open string field theory on {\it compact branes}, whereas {\it non-compact branes} give rise to F-type duals.} }
\label{fig:OCOtriality}
\end{figure}

At first, one might view the above as merely a two-fold classification of open-closed \textit{dualities}, as V- or F-type depending on circumstances. Open-closed-open \textit{triality} makes the stronger claim  that \textit{both} open string duals exist for the same closed string theory \cite{Joburg}. For example, it was conjectured in \cite{Joburg} that open strings on giant gravitons might provide the F-type dual to IIB on $AdS_5 \times S^5$. The work of \cite{komatsuOCO} has provided evidence along these lines in the context of free (or weakly coupled) $\mc{N}=4$ Super Yang-Mills theory. 

The matrix models of two-dimensional gravity, studied intensely in the 90's, might provide to some readers a familiar setting where these two open descriptions are known to co-exist and describe the same closed string theory. In this case, the double-scaled Hermitian matrix models \cite{ginsparg1993lectures, gross-migdal, brezin1990exactly, douglas1990strings} are of V-type, while the Kontsevich type matrix integrals \cite{KontsevichAiry,DijkgraafIntersectionNotes, mukhireview} are of F-type. The relation between these two descriptions had remained rather mysterious until the work of Maldacena, Moore, Seiberg and Shih in \cite{exact}. They showed how the double-scaled one matrix model with $Q$ FZZT brane insertions could be precisely rewritten as the $Q \times Q$ cubic Kontsevich matrix integral. In \cite{Gaiotto:2003yb}, Gaiotto and Rastelli rederived the Kontsevich integral directly from cubic open string field theory on $Q$ FZZT branes in the $(2,1)$ minimal string, cementing this beautiful picture. Open-closed-open triality provides the general framework unifying these two different open string descriptions. 

\begin{figure}[h!]
\centering
\includegraphics[scale=0.09]{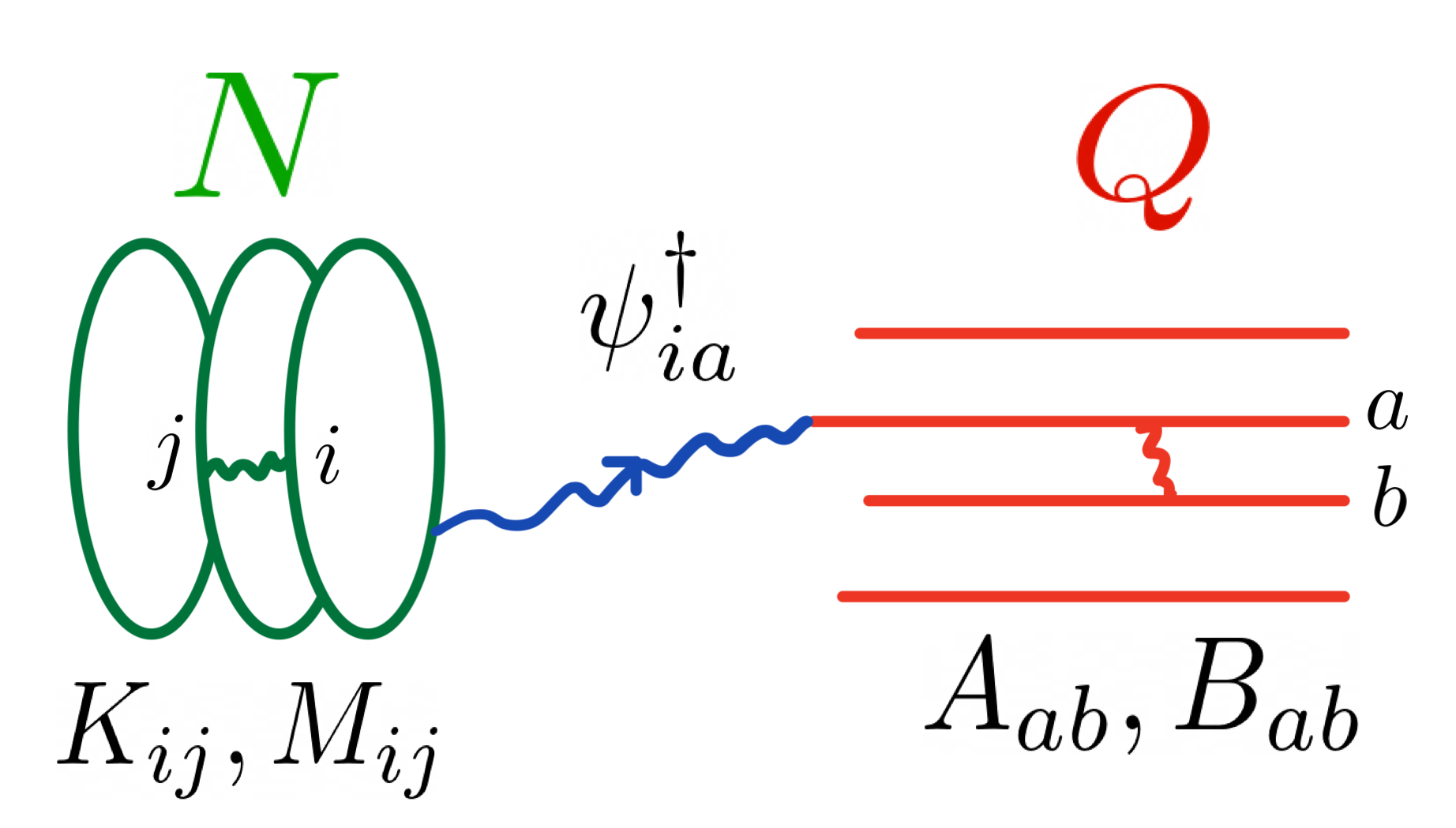}
\caption{\small{\textbf{Two Stacks of Branes and their Open Strings}: We consider two stacks of branes in topological string theory,  with $N$ compact and $Q$ non-compact branes. There are three types of open strings. The bosonic $N \times N$ matrices $K_{ij},M_{ij}$ describe the two transverse directions of the open strings stretched between the $i$- and $j$-th compact branes. The bosonic $Q \times Q$ matrices $A_{ab},B_{ab}$ play the analogous role for the strings between $Q$ non-compact branes. The Grassmann-valued $N \times Q$ $\psi^{\dagger}_{ia}$ transform in the bi-fundamental of $U(N) \times U(Q)$ and capture the open strings connecting the $i$-th compact and $a$-th non-compact branes. The two equivalent open string pictures arise from integrating out all open string strings ending on either stack.} }
\label{fig:2stacks}
\end{figure}

In Section 3, we will see such a triality in the more general setting of a two matrix model,  without taking any double scaling limit. These matrix integrals can be viewed as arising from the reduction of a topological open string theory. Thus these integrals directly describe systems of branes on non-compact Calabi-Yau threefolds in the B-model topological string\footnote{We will also discuss the interpretation of the matrix integrals as a system of D-branes from the A-model perspective in Section \ref{sec:discussion}.}.
In fact, building on the results of \cite{DijkgraafVafa02}, Aganagic et al. had also considered dual open string theories in \cite{Aganagic_2005}, tantamount to the V- and F-type duals for this special case. In a way, their approach went through the closed string picture, using the geometry defined by the V-type matrix model's spectral curve. Here, we will directly derive the F-type dual via the  integrating in-out-in-out procedure of \cite{exact}. These manipulations correspond to keeping track of various open strings between two stacks of branes, as summarized in Fig. \ref{fig:2stacks}.
Following the terminology of \cite{Aganagic_2005}, it will be the case for us that the V-type duals are associated with `compact branes' i.e. wrapping compact cycles in some geometry while the F-type duals are associated with `non-compact branes'. 

The simple classification of holographic dualities as V- and F-type has, in many cases, a striking corollary for the two open string theories. Their Feynman diagrams are related by graph duality i.e. vertices of one map onto faces of the other and vice-versa. We will verify this prediction in the setting of this paper in Sec. 4, where we can directly derive one open string description from the other. Furthermore, we will see in Sec. 5, how this graph duality fits in nicely with the program \cite{freefieldsadsI,freefieldsadsII,freefieldsIII} of reconstructing the closed string worldsheet using the Strebel parametrisation of the closed string moduli space of punctured Riemann surfaces $\mc{M}_{g,n}$. 
One remarkable facet of open-closed-open triality is that it holds at the level of individual string configurations. Each Feynman diagram of one open string description can be mapped onto a diagram of the other gauge theory via graph duality; both reconstruct the same closed string worldsheet. This argument does not rely on the string theory being topological and thus strongly suggests that open-closed-open triality should be rather generic in holography.

\section{Open-Closed-Open Triality and Matrix Models} \label{sec:twostacks}

In this section, we demonstrate the equality of the two matrix integrals shown below in Eq.~(\ref{eq:source/det duality}). The first is an $N \times N$ matrix model in the presence of an external source and $Q$ determinant insertions. The second is a $Q \times Q$ matrix integral, also with an external source, but now with $N$ determinant insertions. The source and determinant insertions get interchanged, as we will see, between the two matrix integrals.     
\begin{eqnarray} \label{eq:source/det duality}
   & &\frac{1}{Z_{N}}  \int dK dM_{N \times N} e^{+ \frac{1}{g} \Tr \left( V(K) - K(M-Y) \right)} \prod_{a=1}^{Q} \det(x_{a}-M)  \nonumber \\
     & = & \frac{(-1)^{NQ}}{Z_{Q}}\int dA dB_{Q \times Q} e^{- \frac{1}{g} \Tr \left( V(A) + A(B-X) \right)} \prod_{i=1}^{N} \det(y_{i}-B). 
\end{eqnarray}
We should view $X$ and $Y$ as diagonal source matrices
\begin{equation}
X={\rm diag}(x_a) ; \,\,\,\,\, Y={\rm diag} (y_i) .    
\end{equation}

\begin{figure}[h!]
\centering
\includegraphics[width=0.6\textwidth]{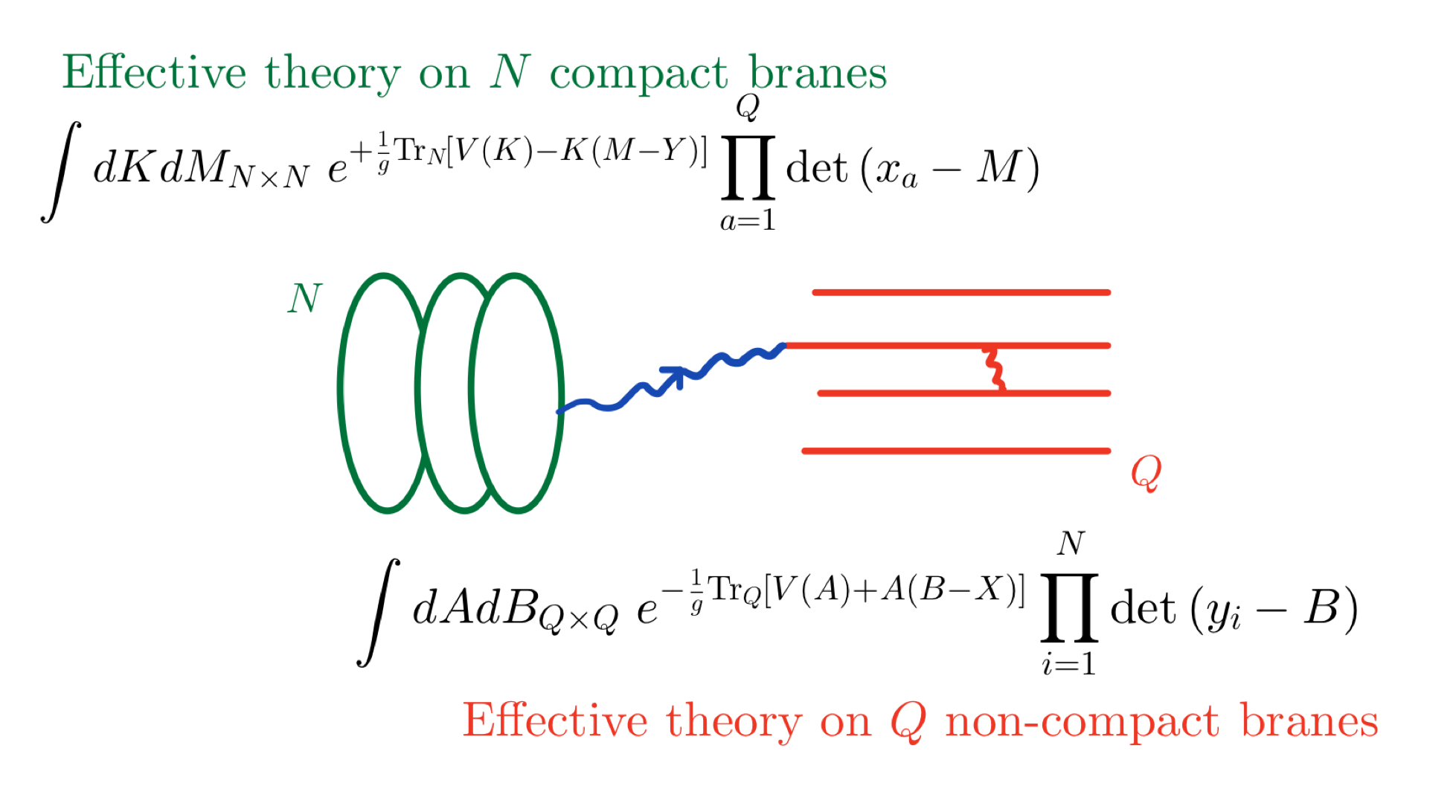}
\caption{\small{\textbf{V- and F-Type Open String Duals}: Starting with two stacks of branes, we can derive the effective theory on either one by integrating out all the open strings ending on the other set of branes. The $N \times N$ matrix integral corresponds to the open string description on the $N$ compact branes, and similarly for the $Q \times Q$ one. Since the integrating out procedure is exact, the two open string descriptions are physically identical, i.e. dual to each other. As shown in Section \ref{sec:graphduality}}, the manipulations leading from one matrix model to the other implements graph duality at the level of their Feynman diagrams. This ensures that the dual closed string insertions are indeed mapped to either the vertices or faces in the V-type or F-type open string description, respectively.  }
\label{fig:2matrix}
\end{figure}

Physically, these matrix integrals arise from two stacks of branes in a B-model topological string theory on a non-compact Calabi-Yau threefold. The first stack consists of $N$ branes which are compact, while the other $Q$, making up the second stack, are non-compact. 
See Fig. \ref{fig:2matrix}. 
The matrix integrals represent the effective open string theory of (Euclidean) D1 branes (or what are called two-dimensional B-branes in the topological string context). Roughly speaking, the two matrices can be viewed as the two transverse complex degrees of freedom in a dimensional reduction of the holomorphic 
Chern-Simons field theory. This further reduces to a zero dimensional integral. Following the discussion in \cite{DijkgraafVafa02, Aganagic_2005, marino2004houches}, we outline in Appendix \ref{sec:holCSarg} how these matrix integrals arise. We will also comment on the A-model interpretation of these two sets of branes in Sec. \ref{sec:discussion}. 

\subsection{Deriving the F-type dual from the V-type}

With this motivation of the brane picture underlying the matrix integrals in Eq.~(\ref{eq:source/det duality}), we now turn to deriving their equivalence. 
Our starting point will be the LHS of Eq.~(\ref{eq:source/det duality}) i.e. the addition of $Q$ determinants into our V-type matrix integral, normalized by the bare partition function $Z_N=(2\pi g)^{N^{2}}$  (i.e. without the determinant operators or the source $Y$). The determinants correspond to the insertion of $Q$ non-compact branes, extended along the normal direction denoted by $M$. 
\begin{equation}
    Z(X,Y) \equiv \frac{1}{Z_{N}}\int dK dM e^{\frac{1}{g} \Tr_{N} \left( V(K)-K(M-Y) \right)} \prod_{a=1}^{Q} \det\left(x_a -M \right).\label{eq:ZXY}
\end{equation}
Before delving into the details, let us outline the broad strategy, which is that of the Hubbard-Stratonovich trick. The derivation proceeds in four steps: 

\begin{enumerate}
    \item Integrate in fermions $\psi_{ia}$ as a rewrite of the determinant insertions
    \item Integrate out the original $N\times N $ matrices $K_{ij},M_{ij}$
    \item Integrate in $Q \times Q$ matrices $A_{ab},B_{ab}$
    \item Integrate out the fermions
\end{enumerate} 
This was done for the case of the Gaussian one matrix model in \cite{exact,Brezin:2016eax} and relied on the potential being quadratic. We extend these results using exact collective fields, inspired by similar rewrites in \cite{hartnoll2019topological}. The case of $Q=1$ appeared already in \cite{Hashimoto_2005}, in the context of the $(p,1)$ minimal string. We stress that here we are especially interested in the theory away from any double scaling limit. 

\begin{figure}[h!]
\centering
\includegraphics[width=0.9\textwidth]{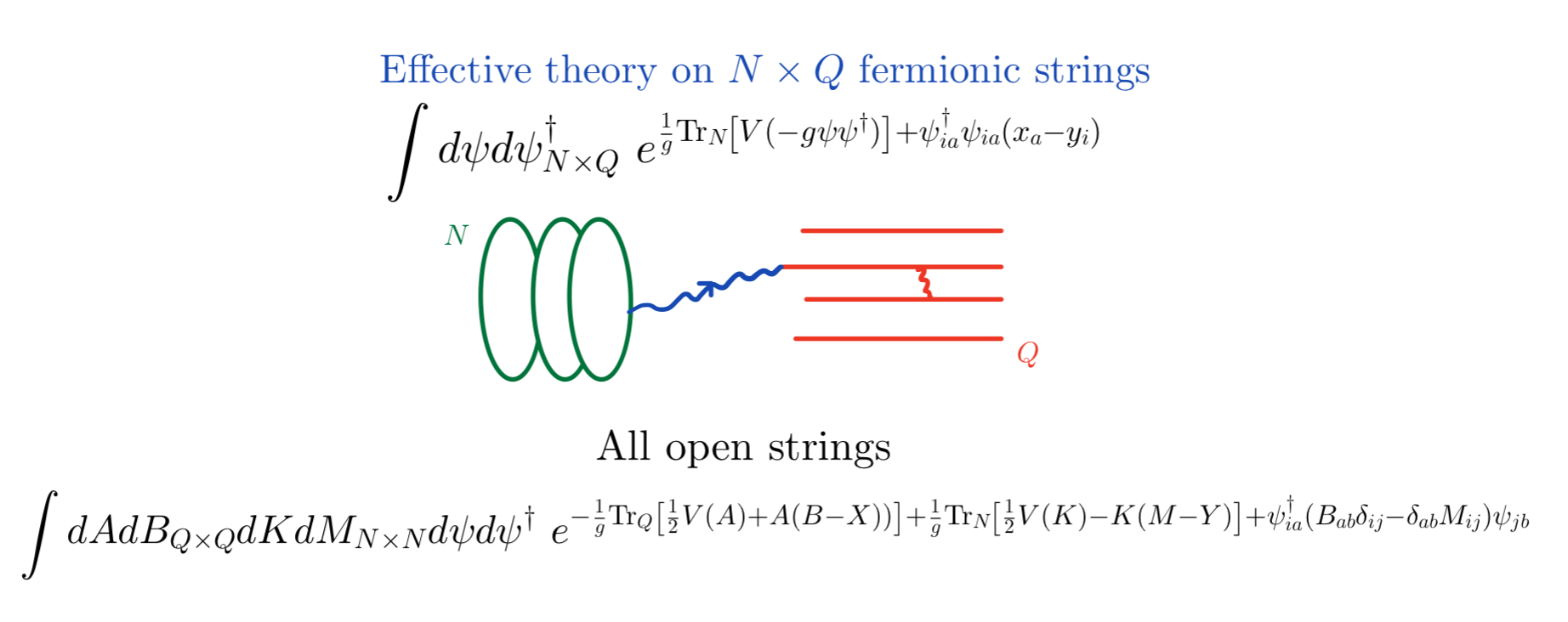}
\caption{\small{\textbf{More Open Strings}: The duality between the V- and F-type open string pictures can be made manifest by keeping track of the fermionic open strings stretched between the two types of branes. Here, $X$ and $Y$ appear on the same footing. By diagonalizing both $X_{ab}$ and $Y_{ij}$, which can always be done, we see that $(x_a-y_i)$ serves as a mass term for the fermions $\psi_{ia}$. We can interpret this as a proxy for the length of the string stretching between the $i$-th compact and $a$-th non-compact brane. } }
\label{fig:allopen}
\end{figure}

In string theoretic terms, the fermions represent the open strings between the $N$ compact and $Q$ non-compact branes. The derivation can therefore be recast as integrating back in or out various open string degrees of freedom. As shown in Fig. \ref{fig:allopen}, we could, for example, keep track of all open strings, leading to an integral over two matrices for the open strings on the $N$ branes, two matrices for the open strings stretched between the $Q$ other branes, and finally the integral over fermionic fields for the open strings stretched between the two. All matrix model identities can then be derived by integrating out some subset of open strings.  

With this bigger picture in mind, we now proceed to the concrete derivation. We first introduce the complex fermions $\psi_{ia}$ to rewrite the $Q$ determinants\footnote{We use Einstein summation convention throughout, unless noted otherwise.}.
\begin{equation}
    \prod_{a=1}^{Q} \det\left(x_a -M \right) = \det\left( X_{Q} \otimes \mb{I}_{N} - \mb{I}_{Q} \otimes M \right) = \int \prod_{c=1}^{Q} \prod_{k=1}^{N}d\psi_{ck} d\psi^{\dagger}_{ck} e^{\psi^{\dagger}_{ia} \left(X_{ab}\delta_{ij}-\delta_{ab}M_{ij} \right) \psi_{jb}},
\end{equation}
where $x_a$ are the diagonal elements of the $Q \times Q$ matrix $X_{ab}$\footnote{The first equality is most easily seen in our diagonal basis, $X_{ab}=x_a \delta_{ab}$. The $NQ \times NQ$ matrix $(X_{Q} \otimes \mb{I}_{N} - \mb{I}_{Q} \otimes M)$ decomposes into $Q$ diagonal blocks, with each of the $N \times N$ blocks being given by $(x_a \mb{I}_{N}-M)$. The determinant then factorizes into the product of the determinants of each block.}. Plugging this into Eq.~(\ref{eq:ZXY}), we find that we can now do the integral over $M$ since it appears linearly. It implements a delta-function constraint\footnote{We have been somewhat cavalier in our discussion of the appropriate integration contours. See \cite{Hashimoto_2005} for details.}, allowing us to do the remaining integral over $K$.
\begin{align}
      Z(X,Y) & =  \frac{1}{Z_{N}}\int dK d\psi d\psi^{\dagger} e^{\frac{1}{g} \Tr_{N} \left( V(K)+KY \right)+\psi^{\dagger}_{ia} X_{ab} \psi_{ib}} \int dM e^{- \frac{1}{g}M_{ij} \left( K_{ji}- g \psi^{\dagger}_{ia}\psi_{ja}\right) } \label{eq:fermsin}\\
     & = \int dK d\psi d\psi^{\dagger} e^{\frac{1}{g} \Tr_{N} \left( V(K)+KY \right)+\psi^{\dagger}_{ia} X_{ab} \psi_{ib}} \delta \left( K_{ji}- g \psi^{\dagger}_{ia}\psi_{ja}\right) \label{eq:deltafnK}\\
    & =  \int d\psi d\psi^{\dagger} e^{\frac{1}{g} \Tr_{N} V\left[(-g \psi \psi^{\dagger}) \right]+\psi^{\dagger}_{ia}\left( X_{ab} \delta_{ij} - \delta_{ab} Y_{ij} \right) \psi_{jb}} \label{eq:onlyferms},
\end{align}
where $(\psi\psi^{\dagger})_{ij} \equiv \psi_{ia} \psi^{\dagger}_{ja} = - \psi^{\dagger}_{ja} \psi_{ia} $. The above manipulations are identical to those in \cite{goel2021string}\footnote{We would like to thank Herman Verlinde for several discussions on these points. The work \cite{goel2021string}, in fact, provided the initial impetus to derive such a duality, in the search of a description valid for the interesting regime $Q>>N>>1$ studied in that paper. He and his collaborators were also aware of such a matrix model duality.}. Recall that before introducing the $Q$ other branes, the integral over $M$ enforced $K_{ij}=0$. In other words, there were simply no transverse open string degrees of freedom along that direction. The new delta function $\delta(K_{ji}-g \psi^{\dagger}_{ia}\psi_{ia})$ in Eq.~(\ref{eq:deltafnK}) shows that branes $i$ and $j$ can now interact via the open strings attached to a common, non-compact brane $a$. As pointed out in \cite{exact}, we can think of the bosonic variable $K_{ij}$ as a meson built from the fermionic open strings stretched between the two type of branes. As we will see shortly, this will also be true for $A_{ab}$. 

In the final line, we have exactly integrated out the open strings on the $N$ compact branes. The fermionic integral in Eq.~(\ref{eq:onlyferms}) represents the effective theory of the open strings stretched between the two stacks. In terms of the diagonal entries of $X, Y$, we can write it as
\begin{equation}
    Z[X,Y]=\int d\psi d\psi^{\dagger} e^{\frac{1}{g} \Tr_{N} V\left[(-g\psi \psi^{\dagger}) \right]+\psi^{\dagger}_{ia}\psi_{ia}(x_a-y_i)}.
\end{equation}
The appearance of $(x_a-y_i)$ as the mass term for the fermions suggests an interpretation as the length of the string connecting branes $a$ and $i$. In fact, our A-model picture will be in line with this identification.

We now want to introduce a set of $Q\times Q$ bosonic variables. As noted in \cite{goel2021string}, we can use the following `color-flavor transformation'\footnote{This terminology is borrowed from similar manipulations in the quantum chaos literature, see for example Chapter 6 of \cite{quantumChaos}.} to trade a trace over powers of the $N\times N$ matrix $\psi_{ia}\psi^{\dagger}_{ja}$ for a trace over powers of the $Q \times Q$ matrix $\psi^{\dagger}_{ia}\psi_{ib}$. 
\begin{equation} \label{eq:colorflavor}
     \Tr_{N} \left[ (\psi\psi^{\dagger})^{k} \right] = (-1)^{2k-1} \Tr_{Q} \left[ (\psi^{\dagger}\psi)^k \right].
\end{equation}
This simply follows from the cyclicity of the trace and the anticommutation rules for the fermions. It is the origin of the sign flip in front of the potential, in going from one matrix model to another. The 'color-flavor' transformation was introduced by Zirnbauer \cite{Zirnbauer1996} in the context of circular ensembles. It also plays a key role in the work of Altland and Sonner \cite{AltlandSonner2020} relating chaos and topology change in quantum gravity. The interesting follow-up work \cite{Altland2022} further made an explicit connection to Dijkgraaf-Vafa-like matrix models.   

At this point, we can now perform the same manipulations which led to the fermionic integral, but in reverse. We can integrate in an exact\footnote{To compare with the one-matrix model of \cite{exact}, we can first integrate out $K$ when $p=2$. In the Hubbard-Stratanovich construction they use, the equations of motion read $M_{ji}= g \psi^{\dagger}_{ia}\psi_{ja}$. In other words, the bosonic matrix entries can be viewed as collective fields for the fermions only on-shell.  This is the sense in which we can think of our derivation as an \textit{exact} collective field rewrite.} collective field $A_{ba}=-g \psi^{\dagger}_{ia}\psi_{ib}$ via 
\begin{align}
      Z(X,Y) & =  \int d\psi d\psi^{\dagger} e^{-\frac{1}{g} \Tr_{Q} V\left[(-g_s \psi^{\dagger}\psi ) \right]+\psi^{\dagger}_{ia}\left( X_{ab} \delta_{ij} - \delta_{ab} Y_{ij}\right) \psi_{jb}} \left( 1 = \int dA \delta(A_{ba}+g \psi^{\dagger}_{ia}\psi_{ib}) \right)\\
    & =  \frac{1}{Z_{Q}}\int dA d\psi d\psi^{\dagger} e^{-\frac{1}{g} \Tr_{Q} \left( V(A) + AX \right)-\psi^{\dagger}_{ia}Y_{ij}\psi_{jb}} \int dB e^{- \frac{1}{g}B_{ab} \left( A_{ba}+ g \psi^{\dagger}_{ia}\psi_{ib}\right) } \\
     & = \frac{(-1)^{NQ}}{Z_{Q}} \int dA dB e^{-\frac{1}{g} \Tr_{Q} \left( V(A)+A(B-X) \right)} \prod_{i=1}^{N} \det \left(y_i-B \right).
\end{align}
In going to the second line, we simply used the constraint to express the action in terms of $A$. Integrating out the fermions now gives $N$ determinant insertions, where the $y_i$ are the entries of $Y_{ij}$. In fact, by identical arguments, we can prove a slightly more general statement (which we will not use here) that includes insertions of traces of $K$ and $A$.
\begin{eqnarray}
   & &\frac{1}{Z_{N}} \int dK dM_{N \times N} e^{+ \frac{1}{g} \Tr_N \left( V(K) - K(M-Y) \right)} \prod_{\alpha=1}^{l} \Tr_{N}(K^{n_\alpha}) \prod_{a=1}^{Q} \det(x_{a}-M)  \nonumber \\
   &=& \frac{(-1)^{l+NQ}}{Z_{Q}}\int dA dB_{Q \times Q} e^{- \frac{1}{g} \Tr_Q \left( V(A) + A(B-X) \right)}  \prod_{\alpha=1}^{l}\Tr_{Q}(A^{n_\alpha}) \prod_{i=1}^{N} \det(y_{i}-B) , 
\end{eqnarray}
where the factor of $(-1)^l$ arises from the color-flavor transformation, Eq.~(\ref{eq:colorflavor}). 
This concludes the demonstration of the equality of the integrals in Eq.~(\ref{eq:source/det duality}), or the more general relation above.

There are two additional perspectives on this system of branes which we have yet to mention. The first corresponds to keeping track of all open string degrees of freedom. This can be done by simply integrating in $A_{ab}$ and $B_{ab}$ already at the level of Eq.~(\ref{eq:fermsin}).
\begin{align}
    Z(X,Y)_{N,Q} = & \int (dK dM)_{N \times N} \thinspace (dA dB)_{Q \times Q} \thinspace d\psi d\psi^{\dagger}_{N \times Q} \nonumber \\
     & e^{+\frac{1}{g}\left(\frac{1}{2} V(K)-K(M-Y) \right) -\frac{1}{g}\left(\frac{1}{2} V(A)+A(B-X) \right) + \psi^{\dagger}_{ai}\left( B_{ab} \delta_{ij}-\delta_{ab} M_{ij}\right) \psi_{jb}}. \label{eq:allstrings}
\end{align}
Note the prefactors in front of the potential terms. To obtain this expression, we used the fact that under the integral sign, the integral over $M$ and $B$ are enforcing delta function constraints, along with the color-flavor transformation. This enables us to share half of the potential between $K$ and $A$, making the (almost) symmetry between the two stacks of branes manifest. 

We mention one final presentation, this time entirely bosonic. We can integrate out the fermions in Eq.~(\ref{eq:allstrings}) to obtain the effective theory of open strings connecting the same type of branes:
\begin{align}
    Z(X,Y)_{N,Q} = & \frac{1}{Z_{Q}Z_{N}}\int (dK dM)_{N \times N} (dA dB)_{Q \times Q} e^{+\frac{1}{g} \Tr_{N} \left(\frac{1}{2} V(K)-K(M-Y) \right) -\frac{1}{g} \Tr_{Q} \left(\frac{1}{2} V(A)+A(B-X) \right)} \nonumber \\
    & \quad \times \det\left(B \otimes \mb{I}_{N}-\mb{I}_{Q}\otimes M\right).
\end{align}
The interactions arise entirely via the term $\det\left(B \otimes \mb{I}_{N}-\mb{I}_{Q}\otimes M\right)$. It shows how each type of brane looks like a determinant insertion for the other with both $X$ and $Y$ appearing as sources. 

To close this section, we highlight that such manipulations do generalize to more familiar settings of holography. For example, the authors of \cite{komatsuOCO} considered $Q$ insertions of half-BPS determinant operators in free $\mc{N}=4$ SYM. For $Q<<N$ (where we can ignore their backreaction on the geometry), these correspond to the insertion of giant gravitons, which are D3' branes wrapping an $S^3$ of the $S^5$, whose worldvolume intersects the boundary of $AdS$ at the point where the determinant operator is inserted. At $g_{YM}=0$, the action of the six scalars is quadratic. By performing almost identical steps, one can integrate out the original $N \times N$ scalars - tantamount to integrating out the original $N$ D3 branes which had sourced the AdS geometry. The fermionic integral is understood as the open strings stretching between the $N$ D3s and the $Q$ D3's. One can integrate in a $Q \times Q$ matrix and subsequently integrate out the fermions. The resulting integral over $Q \times Q$ bosonic degrees of freedom can then be understood as the open string theory on the giant gravitons. An open string field theory interpretation of the latter was proposed in \cite{komatsuOCO}. We discuss such ideas further in Section \ref{sec:discussion}.

\subsection{Exact brane wavefunctions} \label{sec:exactwfns}

Performing the open string path integral with determinant operator insertions computes the brane's wavefunction. In these topological models, we can do this exactly using the technique of bi-orthogonal polynomials. In this section, we find explicit expressions for the wavefunctions of the joint system of $N$ compact and $Q$ non-compact branes. Matching the results of both the V-type and F-type matrix integrals serves as an alternative proof of the exact equivalence of the two string descriptions. It also clarifies the physical interpretation of the brane system and connects nicely with the results of \cite{Aganagic_2005}. Indeed, we will find that the wavefunction of the two branes are related by a (Euclidean) Fourier transform, one of the main results of \cite{Aganagic_2005}, and further elaborated upon in \cite{Kimura_2014}\footnote{We thank T. Kimura for several important discussions on this section, based on his work in \cite{Kimura_2014}.}. For the most part, this section simply collects results and focuses on understanding the duality at the level of brane wavefunction. We relegate the detailed computations to appendix \ref{sec:appPolys}. 

We first construct the system of bi-orthogonal polynomials for our two-matrix model. This is a generalization of the orthogonal polynomial method used for single matrix integrals. Define two families of polynomials $P_{n}(x)$ and $Q_{n}(x)$. Their normalization is fixed such that the coefficient of the leading power equals one. They are required to satisfy the following bi-orthogonality relation. 
\begin{equation}
    \int dk dm e^{+\frac{1}{g}\left( V(k)-k m \right)} P_{i}(k) Q_{j}(m) = h_{i} \delta_{i j},
\end{equation}
where $h_i$ is a constant to be determined. In general, it is a difficult problem to find such polynomials explicitly. However, the fact that the potential for one of the matrices is trivial greatly simplifies the task. In these models, we can check the following polynomials meet these requirements.
\begin{align}
     P_{n}(k)= & k^{n}\\
    Q_{n}(m)= & [(g \partial_{z})^{n} e^{-\frac{1}{g}\left( V(z)-mz)\right)}]_{z=0},
\end{align}
with $h_{n}=(2\pi g) g^{n} n!$.
Furthermore, the two polynomials are related by the following integral transform:
\begin{equation} \label{eq:dualFZZTtoFZZT}
    P_{n}(y) = \frac{1}{2 \pi g} \int dk dm e^{+\frac{1}{g}\left( V(k)-k(m-y) \right)} Q_{n}(b),
\end{equation}
while the inverse transformation sends instead $P_{n} \rightarrow Q_{n}$.
\begin{equation}
    Q_{n}(x)= \frac{1}{2 \pi g} \int da db e^{-\frac{1}{g}\left(V(a) +a(b-x)\right)} P_{n}(b).
\end{equation}

Looking at the form of the exponent, we see these are essentially one-dimensional analogs of the V-type matrix model and its F-dual. This is no coincidence, as we will see it can be used to derive the open-open duality between the two.
The power of these methods comes from the fact that the exact correlators of determinant insertions can be written quite simply in terms of these polynomials. For example, a single non-compact brane  becomes
\begin{equation} \label{eq:singledualFZZT}
    \braket{\det(x-M)}_{N}=Q_{N}(x),
\end{equation}
while multiple insertions can be written in terms of a "Slater" determinant of single particle  'wavefunctions'.
\begin{equation}\label{eq:multdualFZZT}
    \Braket{\prod_{a=1}^{Q}\det(x_{a} - M)}_{N} = \frac{1}{\Delta(x)}\underset{a,b \in \{1,..,Q\}}{\det} \left[ Q_{a+N-1}(x_{b}) \right],
\end{equation}
where $\Delta(x) = \prod_{a< b} (x_a -x_b)$ is the Vandermonde determinant built out of the eigenvalues of $X$. In particular, this symmetrizes the expression under the interchange of any two branes $a$ and $b$. We highlight the shift by $N$ in the level of the orthogonal polynomial $Q_{a+N-1}$. The reduction to eigenvalues can famously be recast in terms of free fermions. The eigenvalue plays the role of the position coordinate of the fermion. In that picture, one might say that the $N$ eigenvalues making up the matrix model populate the first $N$ levels of the Fermi sea. The $Q$ determinants can then be thought of as populating the $Q$ levels above the Fermi surface. We will see that the source term plays a very similar role to determinant insertions, but without this extra shift. 

Before we can understand the role of the source term, we first consider determinant insertions of the other matrix. As might be expected, these are simply expressed in terms of the other polynomials $P_k$.
\begin{equation} \label{eq:singlFZZT}
    \braket{\det(y-K)}_{N}=P_{N}(y).
\end{equation}
If no $M$-branes nor source are present, the $k$-point correlator of $K$-brane insertions factorizes into the product of one-point functions, namely,
\begin{equation} \label{eq:multFZZT}
    \Braket{\prod_{a=1}^{k}\det(y_{a} - K)}_{N} = \frac{1}{\Delta(y)}\underset{a,b \in \{1,...,k\}}{\det} \left[ P_{a+N-1}(y_{b}) \right] = \prod_{a=1}^{k} y_{a}^N = \prod_{a=1}^{k} \Braket{\det(y_{a}-K)}_{N}.
\end{equation}
The third equality exploits the fact that the Vandermonde $\Delta(y)$ can  be written as $\det \left[ P_{a-1}(y_b)\right]$. The factorization can be easily seen directly from the matrix model
\begin{equation}
    \frac{1}{Z_{N}}\int dK dM e^{+ \frac{1}{g} \left( Tr V(K) - K M \right)} \prod_{a=1}^{k} \det(y_{a} - K) = \frac{1}{Z_{N}}\int dK e^{+ \frac{1}{g} \left( Tr V(K) \right)} \prod_{a=1}^{k} \det(y_{a} - K) \delta(K) = \prod_{a=1}^{k} y_a^{N}.
\end{equation}

Let us finally turn to the effect of the source. By using the Harish-Chandra-Itzykson-Zuber (HCIZ) formula \cite{harish1957differential,itzykson1980planar} twice to reduce the problem to eigenvalues, along with the orthogonal polynomial relation in Eq.~(\ref{eq:dualFZZTtoFZZT}), we find that 
\begin{equation} \label{eq:justsource}
    \frac{1}{Z_{N}}\int dK dM_{N \times N} e^{+ \frac{1}{g} \left( Tr V(K) - K(M-Y) \right)} = \frac{\det \left[  P_{i-1}(y_{j})\right]}{\Delta(y)}= 1.
\end{equation}
This can again easily be seen by simply doing the integral over $M$ enforcing $\delta(K)$. The important takeaway, however, comes from comparing this expression to Eq.~(\ref{eq:multFZZT}) - there is no shift by $N$ in the index of the orthogonal polynomials. Hence, we are not adding any new branes to the system, but rather we see that the original $N$ branes making up the matrix model can in fact be thought of as compact branes with $y_i=0$. In the language of free fermions, the source term does not populate additional levels above the Fermi sea. Its effect is to perform an integral transform of the single particle wavefunctions. However, the topological nature of the model is such that the locations $y_i$ don't enter the final answer, unless we perturb the system. 

As shown in the appendix, we can explicitly compute the path integral in the presence of both a source term for $K$ and determinant intersections for $M$. This can be interpreted as the wavefunction for the joint system of $N$ compact branes and $Q$ non-compact branes. It takes the form of a determinant of an $(N+Q)$ rank matrix,
\begin{eqnarray} \label{eq:Zwithboth}
   & \frac{1}{Z_N} \int dK dM_{N\times N} e^{+\frac{1}{g}\Tr \left( V(K)-K(M-Y)) \right)} \prod_{a=1}^{Q}\det(x_{a}-M) & \\ \nonumber
   & = & \\ \nonumber
   & \frac{1}{\Delta(x) \Delta(y)} \underset{\substack{
i,j \in \{1,\dots,N\} \\
a,b \in \{1,\dots,Q \} }}\det \left[ \begin{array}{c|c} 
    P_{i-1}(y_{j}) & Q_{i-1}(x_{b}) \\ \hline
    P_{a+N-1}(y_{j}) & Q_{a+N-1}(x_{b}) 
    \end{array}\right]. & 
\end{eqnarray}

We recognize the first diagonal blocks as the wavefunction of just the $N$ compact branes and the second one as that of the non-compact ones. The off-diagonal blocks stem from interactions between the two, mediated by fermionic strings connecting the two types of branes. We can also obtain this expression directly from the $Q \times Q$ matrix integral since it is symmetric under the interchange of $(N, Q)$ and $(y_i, x_a)$. This furnishes yet another further proof of the duality, strictly using bi-orthogonal polynomials.

\section{Feynman Graph Duality between V- and F-type Duals}\label{sec:graphduality}

One of the striking consequences of open-closed-open triality is that the individual Feynman diagrams of the two (open string) gauge theory descriptions should be graph dual to one another. In this section, we show how the manipulations deriving one open string picture from the other indeed implement a dynamical\footnote{`Dynamical' is generally a poor choice of words in the context of topological string theory, but this argument is expected to generalize to {\it bona fide} string theories with propagating modes. Indeed, this has been exploited for ${\cal N}=4$ Super Yang-Mills in \cite{brown2011complex, octagon, komatsuOCO}.} graph duality \cite{Joburg}. To do so, we represent the various steps pictorially in terms of Feynman diagrams. 

In order to simplify the discussion, we will consider the case of $p=2$, where we can reduce the two-matrix model to a Gaussian one-matrix integral. This will avoid having to keep track of two types of vertices, corresponding to the two matrices. Our discussion thus reduces to a graphical interpretation \cite{Joburg} of the derivation first presented in \cite{exact}, generalized to include the second external matrix $Y_{ij}$. 
Choosing $V(K)= K^2/2$, our starting point reads 
\begin{equation} \label{eq:GaussianStart}
   Z(X,Y) =  \frac{1}{Z_{N}} \int dM e^{- \frac{1}{2g} \Tr \left( M-Y \right)^2} \prod_{a=1}^{Q} \det\left( x_a-M\right).
\end{equation}

\begin{figure}[h!]
     \centering
         \includegraphics[width=0.4\textwidth]{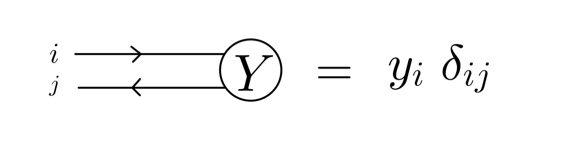}
        \caption{\small{\textbf{Diagrammatic representation of the tadpole term.}} }
        \label{fig:Ytadpole}
\end{figure}

Before we consider the effect of the determinants, let us make a remark on the role of the additional source term $Y$. From the form of the potential, it is clear that $Y$ is simply the one-point function of $M$:
\begin{equation}
    \Braket{M_{ij}}_{\text{no dets}}= Y_{ij}.
\end{equation}
Expanding the square in the exponent, there is a 'tadpole' term, linear in the source $Y$. On top of the usual double-line propagator for $M$, we therefore need an extra vertex to encode the coupling $\Tr( M Y)$. We represent this diagrammatically in Fig.\ref{fig:Ytadpole}. 
It is then clear that the source simply assigns the eigenvalue $y_i$ to each (compact) brane $i$. Each closed index-loop containing $k$ occurrences of $Y$ therefore gives a weight $\Tr_{N \times N}(Y^{k})$.

The determinant insertions in Eq.~(\ref{eq:GaussianStart}) play an analogous role for the non-compact branes. Using $ \ln \det \left( S \right) = \Tr \ln S $ and expanding the logarithm around large $x_{a}$, the determinant insertions generate a potential for $M$.
\begin{eqnarray} \label{eq:potentialgen}
   Z(X,Y) & = &  \frac{\det_{Q\times Q}(X)^N}{Z_{N}} \int dM e^{- \frac{1}{2g} \Tr \left( M-Y \right)^2 + \sum_{a=1}^{Q} \Tr \ln \left( 1- \frac{M}{x_{a}} \right)}  \\
   & = & \frac{\det_{Q\times Q} (X)^N}{Z_{N}} \int dM e^{- \frac{1}{g} \left( \frac{1}{2}\Tr \left( M-Y \right)^2 + \sum_{k=1}^{\infty} {\bar t}_{k} \Tr M^{k} \right)}, 
\end{eqnarray}
where we have introduced the Miwa "times"
\begin{equation} \label{eq:miwa}
    \frac{1}{g}{\bar t}_{k} \equiv \frac{1}{k} \Tr_{Q \times Q} \left( X^{-k} \right). 
\end{equation}
This is perhaps the simplest manifestation of open-closed duality. Taking the matrix potential to encode the closed string background, it shows exactly how the insertion of branes backreacts on the geometry. The relation in Eq.~(\ref{eq:miwa}) provides a dictionary between the open string moduli $x_{a}$ and (what will be) the closed string couplings ${\bar t}_{k}$. 

\begin{figure}[h!]
     \centering
         \includegraphics[width=0.8\textwidth]{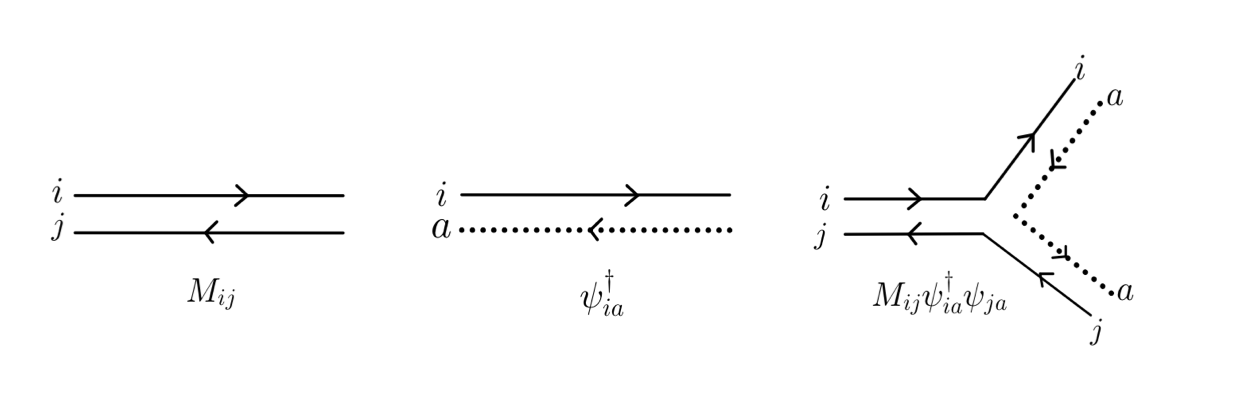}
        \caption{\small{\textbf{Feynman rules for the joint $M,\psi,\psi^{\dagger}$ system:} Edges with double solid lines represent propagators for $M_{ij}$, while the mixed index lines are for fermions.}}
        \label{fig:frMpsi}
\end{figure}

\begin{figure}[h!]
     \centering
         \includegraphics[width=0.6\textwidth]{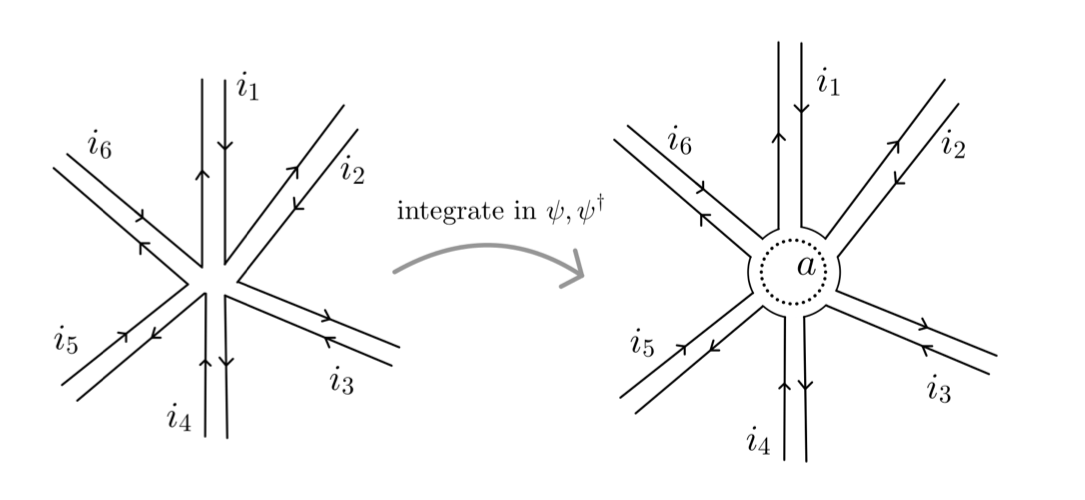}
        \caption{\small{\textbf{Step 1:} Rewriting the determinants in terms of fermionic integrals reintroduces the holes on the worldsheet corresponding to the string ending on the non-compact branes.}}
        \label{fig:vertexpuff}
\end{figure}

\begin{figure}[h!]
     \centering
         \includegraphics[width=0.8\textwidth]{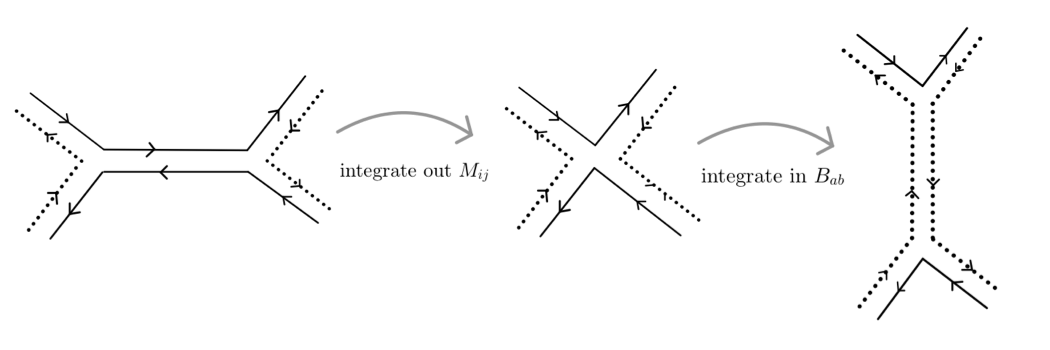}
        \caption{\small{\textbf{Steps 2 \& 3:} Integrating out the strings on the compact branes gives a quartic interaction vertex for the fermions. Integrating in the open strings on the non-compact branes resolves this vertex in `the other channel'. This exchanges each edge of the original V-type Feynman diagram with its dual edge.}}
        \label{fig:steps2&3}
\end{figure}

Recall that the first step in our derivation of Eq.~(\ref{eq:source/det duality}) rewrites these determinant insertions in terms of a fermionic integral over $N \times Q$ Grassmann-valued variables $\psi^{\dagger}_{ia},\psi_{ia}$. 
\begin{eqnarray}
   Z(X,Y) & = &  \frac{1}{Z_{N}} \int dM d\psi^{\dagger} d\psi e^{- \frac{1}{2g} \Tr \left( M-Y \right)^2 + \psi^{\dagger}_{ia} \left( X_{ab} \delta_{ij} - \delta_{ab}M_{ij} \right) \psi_{jb}}
\end{eqnarray}
The Feynman diagram rules of the $M,\psi, \psi^{\dagger}$ system are depicted in Fig.\ref{fig:frMpsi}, supplemented by the tadpole term in Fig. \ref{fig:Ytadpole}.
The cubic vertex $ M_{ij} \psi^{\dagger}_{ia} \psi_{ja}$ introduced in this rewriting essentially `splits' each leg of an interaction vertex into two fermion propagators. As shown in Fig. \ref{fig:vertexpuff}, this resolves the vertex in terms of a closed `$a$'-index loop. Physically, {\it we are reintroducing the holes in the worldsheet corresponding to the string ending on brane $a$}. In other words, we are re-expressing the complicated background encoded in the $\bar{t}_{k}$ in terms of a string moving in the simpler background with all $\bar{t}_{k}=0$, but including the $Q$ non-compact branes. 
Step 2 proceeds by integrating out the bosonic matrix $M_{ij}$, leaving us simply with the fermionic open strings stretched between the two sets of branes. This can be straightforwardly done, since $M$ appears at most quadratically. This generates quartic vertices for the fermions as shown in the middle picture of Figure \ref{fig:steps2&3}. In the process we have eliminated all edges with two solid lines.
\begin{eqnarray}
  Z(X,Y) & = & \int  d\psi^{\dagger}d\psi e^{\psi^{\dagger}_{ia}\psi_{ia} \left(x_{a} -y_{i}\right) + \frac{g}{2} \sum_{i,j=1}^{N}\sum_{a,b=1}^{Q} \psi^{\dagger}_{ja}\psi_{ia}\psi^{\dagger}_{ib}\psi_{jb} }.
\end{eqnarray}

\begin{figure}[h!]
     \centering
      \begin{subfigure}[b]{0.45\textwidth}
       \centering
         \includegraphics[width=\textwidth]{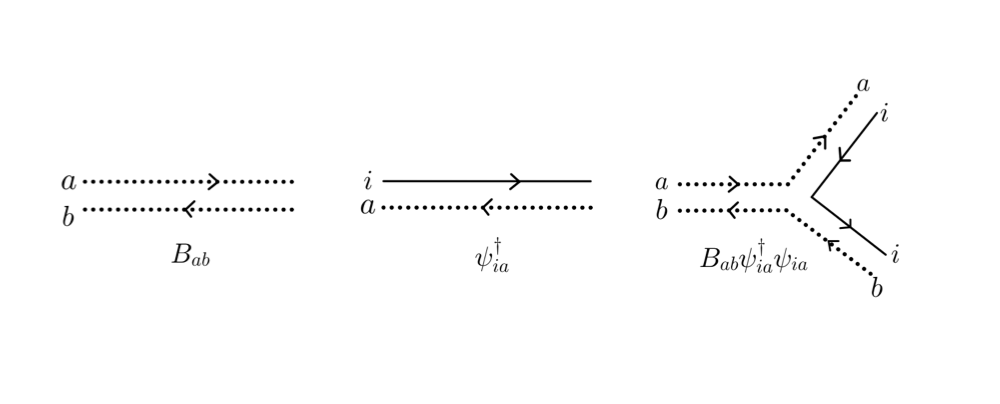}
         \caption{Feynman Rules for $B,\psi,\psi^{\dagger}$}
         \label{fig:Bpsifr}
     \end{subfigure}
     \begin{subfigure}[b]{0.45\textwidth}
         \centering
         \includegraphics[width=\textwidth]{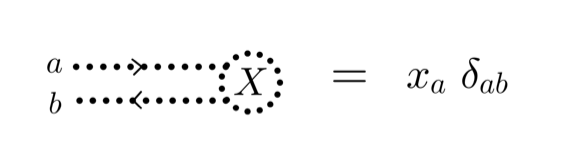}
         \caption{$X$-tadpole}
         \label{fig:Xtadpole}
     \end{subfigure}
      \caption{\small{\textbf{Feynman Rules relevant for Steps 2 \& 3}.}}
       \label{fig:frsteps23}
\end{figure}

The second Hubbard-Stratanovich transformation of step 3 integrates in the $Q \times Q$ matrix $B_{ab}$. 
\begin{eqnarray}
  Z(X,Y) & = & \frac{1}{Z_Q} \int dB_{Q \times Q} d\psi^{\dagger}d\psi e^{\frac{1}{2g} \Tr (B-X)^2 + \psi^{\dagger}_{ia} \left(B_{ab}\delta_{ij}-\delta_{ab}Y_{ij} \right)\psi_{jb} } .
\end{eqnarray}
This decouples the fermions, trading the direct four-fermion interaction for a Yukawa coupling to $B_{ab}$ - see Fig. \ref{fig:frsteps23} for the relevant Feynman rules. As can be seen in Fig.\ref{fig:steps2&3}, one might say it trades the `s-channel' interaction between the fermions, mediated by $M_{ij}$, for an exchange of $B_{ab}$ in the `t-channel'. From a graph perspective, it is clear this replaces each edge of the original Feynman diagram with its dual edge. In the process, the holes with the dashed lines (colour index $a$) have expanded while those of the solid lines (colour index $i$) have shrunk.

In Step 4, this shrinking is complete when we integrate out the fermions. We have now obtained a Feynman diagram in which 
each original face (closed solid line) has been replaced with a new (dotted-line) vertex. 
\begin{eqnarray}
  Z(X,Y) & = & \frac{(-1)^{NQ}}{Z_Q} \int dB_{Q \times Q} e^{\frac{1}{2g} \Tr (B-X)^2} \prod_{i=1}^{N} \det(y_i -B)
\end{eqnarray}

\begin{figure}[h!]
     \centering
         \includegraphics[width=0.6\textwidth]{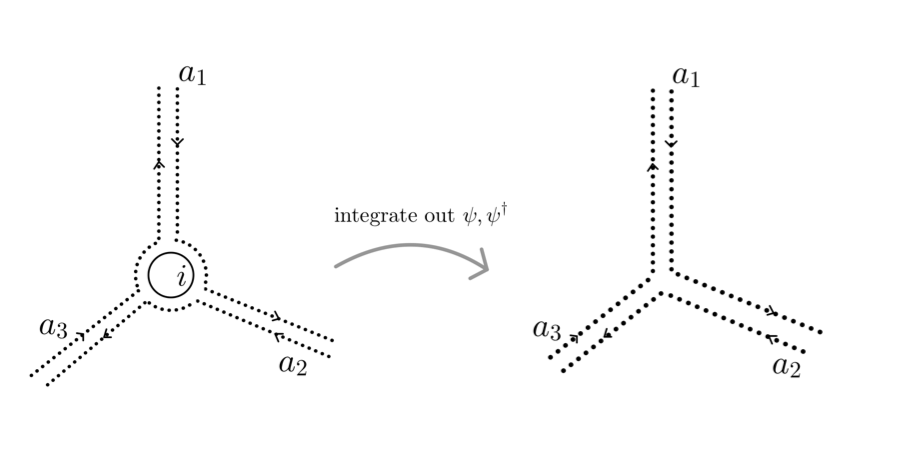}
        \caption{\small{\textbf{Step 4:} Integrating the fermions out shrinks the holes corresponding to the string ending on the compact branes. The original faces of the V-type description are thus mapped to vertices in the F-type dual.}}
        \label{fig:step4}
\end{figure}

Physically, the holes in the worldsheet corresponding to the $N$ compact branes have now disappeared, generating a potential for $B$ by the exact same mechanism described in Eq.~(\ref{eq:potentialgen}) - see Fig.\ref{fig:step4}. Each vertex is now weighted by the closed string couplings given by the second set of Miwa times 
\begin{equation}
   \frac{1}{g} \bar{s}_{k} =  \frac{1}{k} \Tr_{N \times N} \left(Y^{-k} \right) .
\end{equation}
Going back and forth between the two descriptions corresponds to various holes opening or closing up on the worldsheet.  The faces of the Feynman diagram in one description go over to the vertices in the other, and vice-versa.

\section{Strebel Differentials and V/F-type Dualities}

In this section, we use the Strebel parametrization of the moduli space of punctured Riemann surfaces $\mc{M}_{g,n}$ to elucidate the mechanisms by which open strings reconstruct closed ones \cite{freefieldsIII}. The special meromorphic Strebel quadratic differential on the closed string worldsheet will play the starring role. It encodes the complex structure of a compact Riemann surface into the combinatorial data of ribbon graphs. It thus provides a invaluable tool to understand how precisely closed strings emerge from gauge theory Feynman diagrams. 

\begin{figure}[h!]
    \centering
  \begin{subfigure}[b]{0.3\textwidth}
         \centering
         \includegraphics[width=\textwidth]{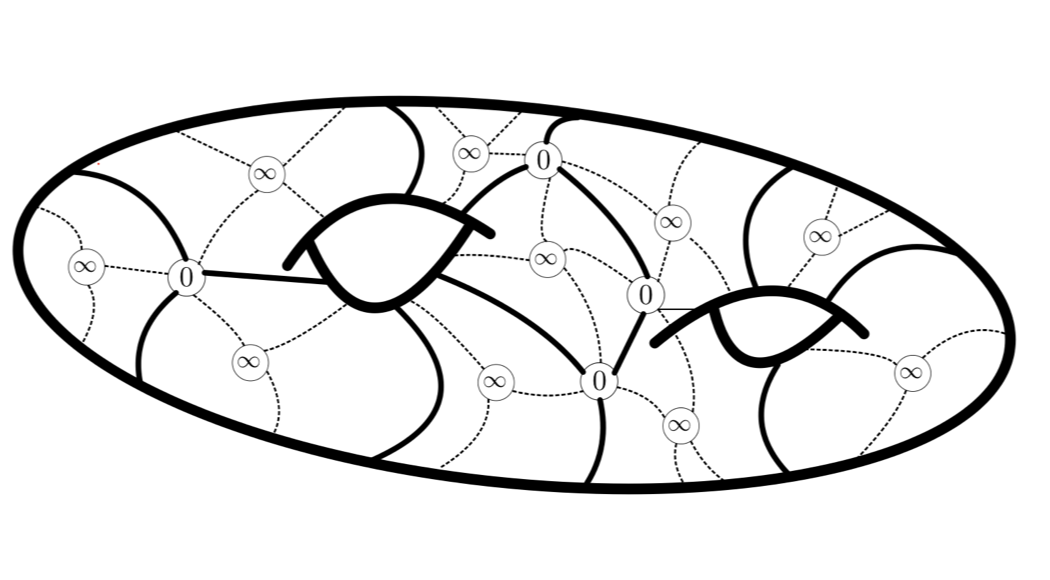}
         \caption{Closed string worldsheet}
         \label{fig:strebelgraphclosed}
     \end{subfigure}
     \begin{subfigure}[b]{0.3\textwidth}
         \centering
         \includegraphics[width=\textwidth]{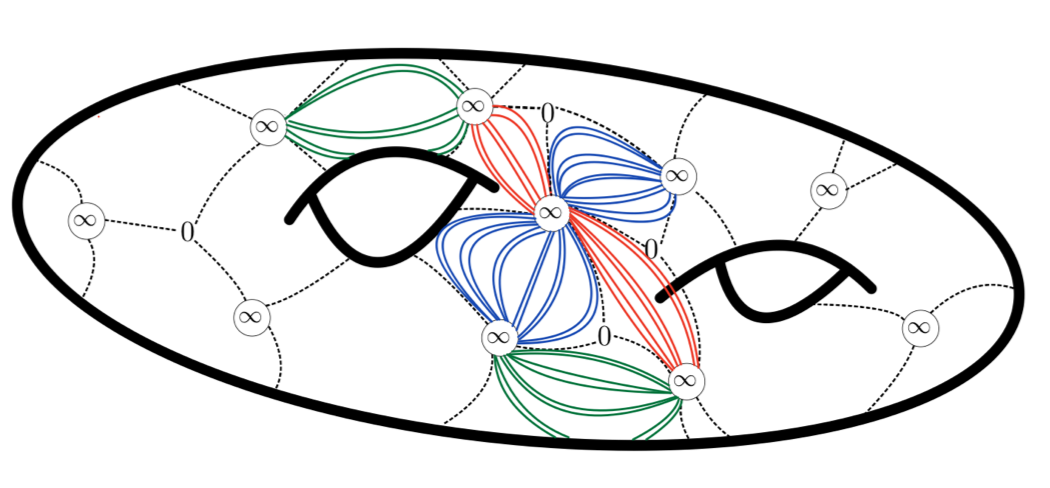}
         \caption{V-type reconstruction}
         \label{fig:vtypestrebel}
     \end{subfigure}
      \begin{subfigure}[b]{0.3\textwidth}
       \centering
         \includegraphics[width=\textwidth]{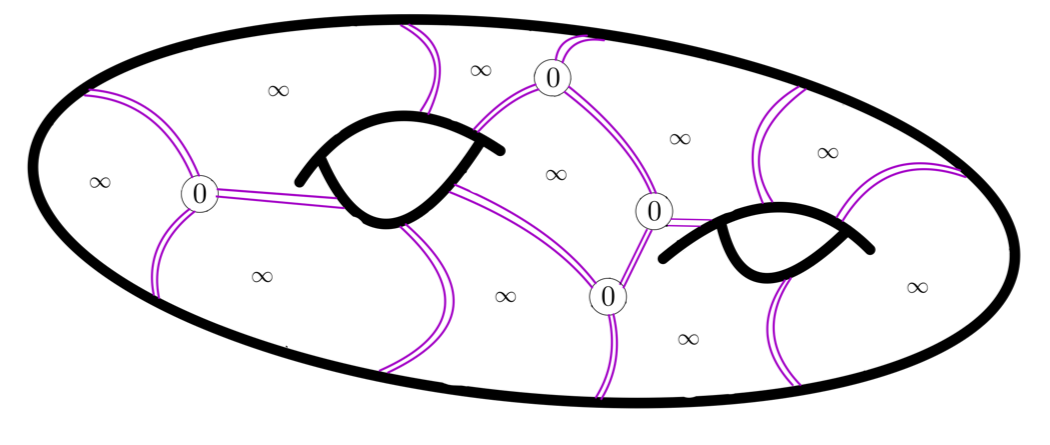}
         \caption{F-type reconstruction}
         \label{fig:ftypestrebel}
     \end{subfigure}
        \caption{\small{\textbf{A genus 2 Riemann surface with marked points and its characteristic horizontal and vertical trajectories}: a) The Strebel differential on a compact Riemann surface defines a family of critical curves. The critical horizontal trajectories connect zeroes of the differential. b) The proposal of \cite{freefieldsIII} is that the poles correspond to the insertion of the V-type gauge invariant operators, while (families of) vertical trajectories correspond to the 't Hooft double-line graphs. c) In an F-type duality, the critical horizontal trajectories would correspond to the spines of a open string field theory diagram.}}
        \label{fig:strebelgraph}
\end{figure}

String field theory often takes recourse to this parametrization \cite{ZwiebachProof}. We will see, for example, how a cubic vertex in F-type duality corresponds to the gluing of three open strings along their midpoint \cite{WittenOSFT}. Strebel graphs also form the basis of Kontsevich's proof of Witten's conjecture on $\psi$-class intersection theory on the moduli space of curves \cite{KontsevichAiry}, i.e. 2d topological gravity. Certain cohomology classes on $\mc{M}_{g,n}$ can be simply expressed in terms of graph data. Strebel differentials were proposed as a central ingredient in understanding gauge-string duality in \cite{freefieldsIII} where they give a well defined map from individual ribbon Feynman graphs to closed string worldsheets. A refinement of this proposal was made by \cite{razamatGauss} where the so-called integer Strebel differentials played a particularly important role. This also featured in the earlier proposal of \cite{gopakumar2011simplest, gopakumar2013correlators} for the topological string dual to the Gaussian model. Further, recent investigations \cite{gaberdiel2021symmetric, ads5strebel, knighton} of the tensionless limit in $AdS_3$ (and also $AdS_5$) have also seen a concrete realisation of the proposal of \cite{freefieldsIII, razamatGauss}. The corresponding free field theories seem to naturally pick out discrete points on $\mc{M}_{g,n}$ which correspond to the integer Strebel differentials.

To keep this work as self-contained as possible, we present a, hopefully, pedagogical introduction to this powerful tool. The mathematics is quite simple, involving no more than basic complex analysis. We believe the payoff for this small detour to be well worth it. Throughout, we focus on those aspects which will ultimately allow us to see how the open strings of the gauge theory are pieced together to form the closed strings of the bulk. We will also crucially use this framework in our third paper where it will be important in understanding how the A-model closed string dual arises from the Gaussian matrix model (and generalizations thereof). 

At its core, the Strebel differential establishes a bijection between metrized ribbon graphs of genus $g$ and $n$ faces (roughly, the gauge theory Feynman diagrams) and a point on the decorated, punctured moduli space $\mc{M}_{g,n} \times \mb{R}_{+}^{n}$. Indeed, it naturally gives a concrete prescription for how to assemble the closed string worldsheet from a collection of semi-infinite strips of the kind that arise in a V-type duality. It does so by providing us with the various open complex charts on the Riemann surface, along with the transition functions on their overlap. 

\begin{equation*}
    \text{pt} \in \mc{M}_{g,n} \times \mb{R}_{+}^{n} \underset{\text{Strebel's Thm}}{\Longleftrightarrow } \phi(z) dz^2 \underset{\text{V/H Trajectories}}{\Longleftrightarrow} \text{Metrized ribbon graph w/ genus } g, n \text{ faces}
\end{equation*}

We will see here that both V- and F-type dualities are geometrized in terms of the so-called vertical and horizontal trajectories of the Strebel differential. These are families of (real) curves foliating the Riemann surface. We would like to emphasize that \textit{this argument goes beyond the restricted setting of topological string theory considered in the previous sections}. It therefore strongly suggests that open-closed-open triality is, in fact, a \textit{generic} feature of holography. 

We will review only the most salient features of the Strebel construction. A large part of the initial discussion here was already presented in \cite{freefieldsIII}, and we follow its presentation closely. A beautifully clear and accessible mathematical introduction can also be found in \cite{mulaseStrebel}, in particular their section 4 spells out the various complex charts and transition functions. The fundamental result we will rely upon is a fundamental theorem due to Strebel \cite{strebel1984}, extended by Harer to parametrize the moduli space of complex curves $\mc{M}_{g,n}$ \cite{harer1988cohomology}. We first define the Strebel differential, briefly state the theorem, and then go on to explain its importance for open-closed string duality. 

\subsection{The Strebel differential}

Consider a compact Riemann surface $\Sigma_{g,n}$, of genus $g$, with $n$ marked points (`punctures'), admitting inequivalent complex structures, i.e. with $\chi(\Sigma_{g,n})=2-2g-n<0$. 
A Strebel differential $q$ on $\Sigma_{g,n}$ is a meromorphic quadratic differential. In other words, in every chart $U \subset \Sigma_{g,n}$, with local coordinate $z$, it takes the form 
\begin{equation}
    q = q_{zz}(z) dz \otimes dz \equiv \phi(z) dz^2 ,
\end{equation}
where $\phi(z)$ is a meromorphic function on $U$. Under a holomorphic change of variables $z \rightarrow w=w(z)$, it transforms simply as 
\begin{equation}
    q = \phi(z) dz^2 = \tilde{\phi}(w) dw^2 \rightarrow  \phi(z) = \tilde{\phi}(w(z)) \left(\frac{dw}{dz} \right)^2 .
\end{equation}
Note that, away from its zeroes and poles, $q$ defines a locally flat metric on $\Sigma_{g,n}$,
\begin{equation} \label{eq:StrebelMetric}
    ds^2 = | \phi(z) | dz d\bar{z} .
\end{equation}
In this gauge for the worldsheet metric, the curvature of the surface is localized at \textit{both} the zeroes and poles of the  differential.  

We require that the only singularities of the Strebel differential, $q$, be double-poles at the location of the $n$  marked points, with a prescribed behaviour characterized by $n$ real positive numbers (`residues') $L_1,...,L_n$\footnote{These were called $p_i$ in \cite{freefieldsIII}, and $a_i$ in \cite{mulaseStrebel}. The notation $L_i$ here is borrowed from Kontsevich's work \cite{KontsevichAiry}. They can be viewed as measuring the circumference of the asymptotic closed string states around each pole, as measured with the metric defined in Eq.~(\ref{eq:StrebelMetric}). Equivalently, they measure the size of the single trace operator insertion at these points in a V-type dual.}. More precisely, in a local coordinate chart containing the $k$-th marked point, it takes the form\footnote{The factors of $2\pi$ are inserted here for convenience, simplifying later expressions. } 
\begin{equation}
    \phi(z) dz^2  \overset{\substack{\text{near k-th}\\ \text{marked point}}}{=} - \frac{L_k^2}{(2\pi)^2} \frac{dz^2}{z^2}
\end{equation}

Strebel's theorem is about the existence of a quadratic differential on $\Sigma_{g,n}$, fulfilling all the above requirements as well as a crucial one regarding the nature of its horizontal trajectories. In order to explain this and connect with ribbon graphs, we will now need the notion of vertical and horizontal trajectories for a quadratic differential.

\subsection{Vertical and Horizontal trajectories} \label{sec:verthoztraj}

To see the connection between gauge theory Feynman diagrams and closed string worldsheets, we need to define a particular family of curves which foliate the Riemann surface. We first analyze their local behaviour, treating separately the neighborhood of zeroes, poles and regular points of the Strebel differential. How these three cases are patched together globally determines the moduli of the Riemann surface.

\begin{figure}[h!]
     \centering
      \begin{subfigure}[b]{0.3\textwidth}
       \centering
         \includegraphics[width=\textwidth]{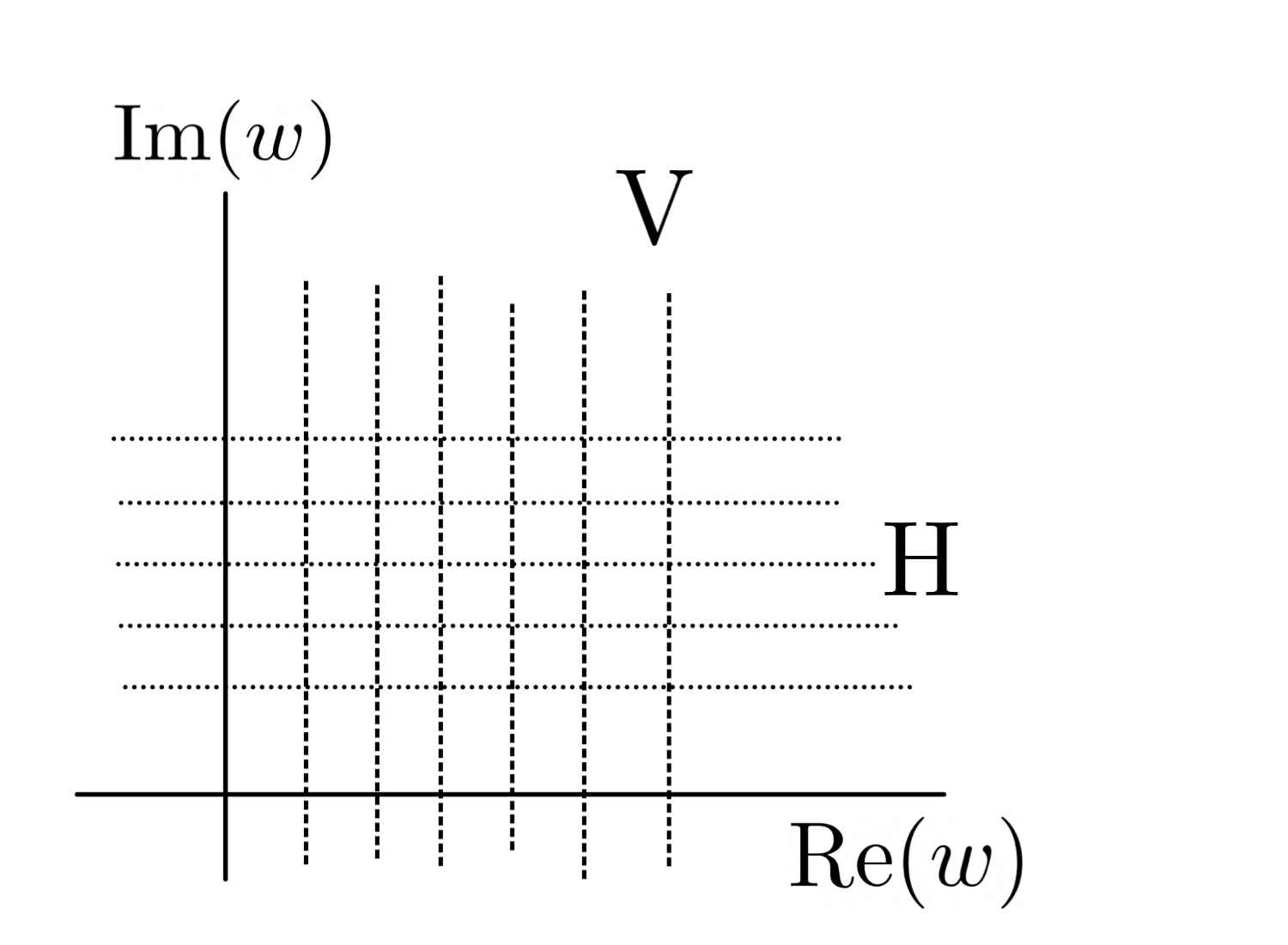}
         \caption{Near a regular point}
         \label{fig:nearregular}
     \end{subfigure}
     \begin{subfigure}[b]{0.3\textwidth}
         \centering
         \includegraphics[width=0.9\textwidth]{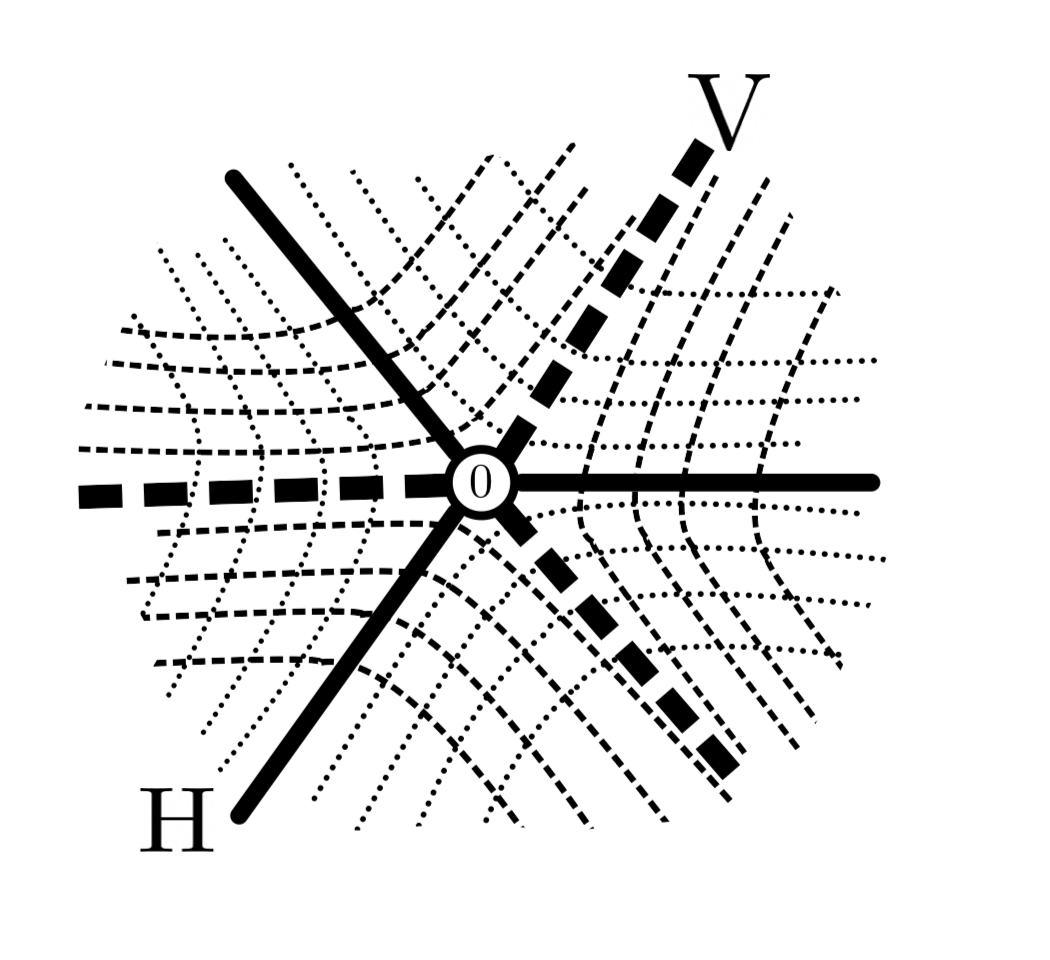}
         \caption{Near a $m$-th order zero (depicted $m=1$)}
         \label{fig:nearzero}
     \end{subfigure}
     \begin{subfigure}[b]{0.3\textwidth}
         \centering
         \includegraphics[width=0.9\textwidth]{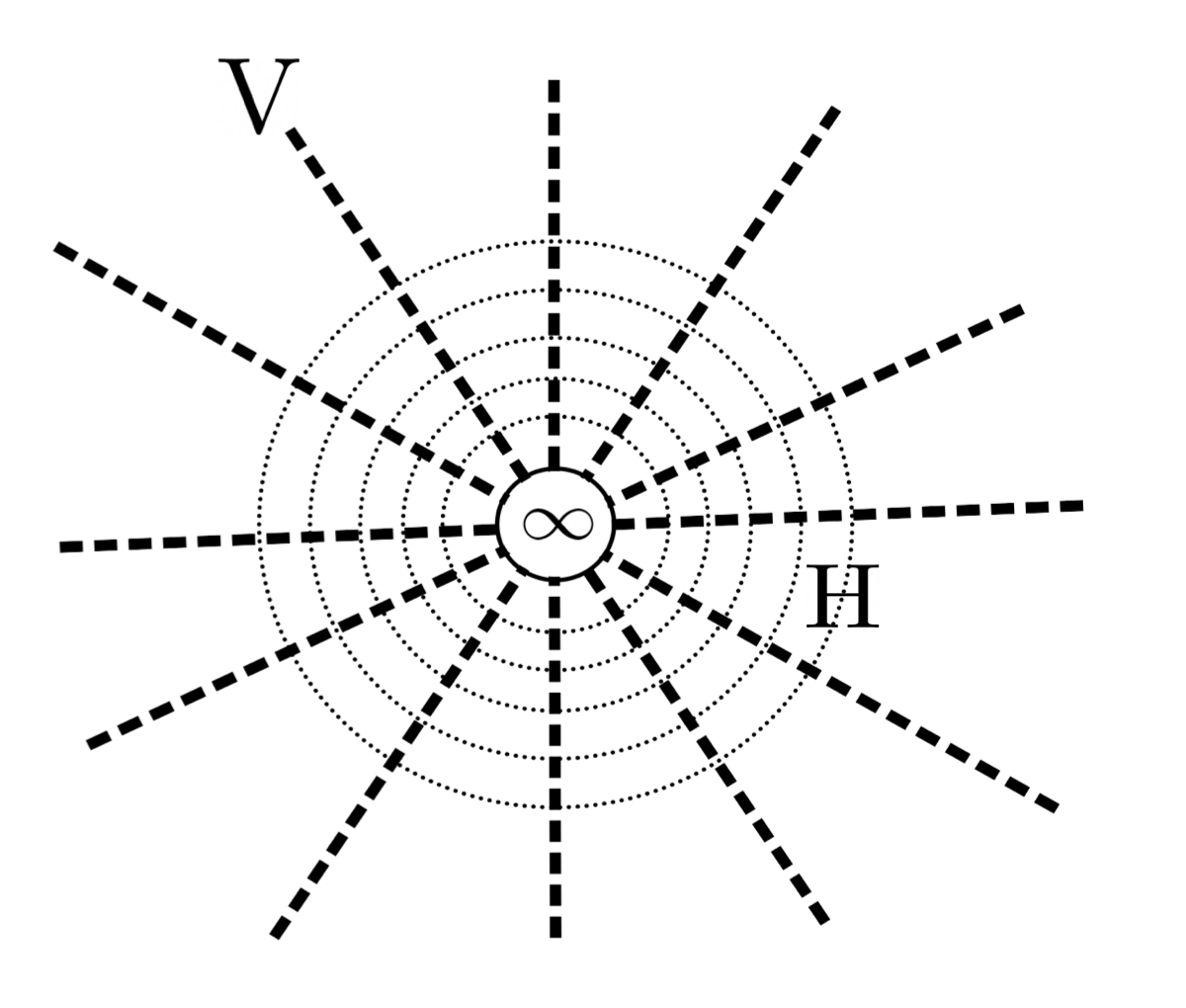}
         \caption{Near a double pole}
         \label{fig:nearpole}
     \end{subfigure}
        \caption{\small{\textbf{The three characteristic behaviours of vertical \& horizontal trajectories }: a) In the neighborhood of a regular point, we can defines coordinates $w$ such that the Strebel differential takes the simple form $q=dw^2$. The horizontal (dotted) and vertical (dashed) trajectories are straight lines running parallel to the real and imaginary axes, respectively. b) $(m+2)$ horizontal trajectories meet at an $m$-th order zero. The vertical trajectories are also half-lines emanating radially outward but are rotationally offset by angle $\pi/(m+2)$ relative to the horizontal ones. c) The horizontal trajectories are closed, and encircle the double pole. Any radially outgoing half-line is a vertical trajectory. }}
        \label{fig:strebelbehaviours}
\end{figure}

Vertical and horizontal trajectories can be defined for any quadratic differential and are (real) curves on $\Sigma_{g,n}$. We can parametrise them as $z_{V}(t)$ and $z_{H}(t)$, for $t$ in some specified interval of the reals. Their defining feature is that, along the trajectory, they satisfy
\begin{align}
    \phi\left( z_{V}(t) \right) \left( \frac{d z_{V}(t)}{dt} \right)^2 & < 0, \\
     \phi\left( z_{H}(t) \right) \left( \frac{d z_{H}(t)}{dt} \right)^2 & > 0 .
\end{align}
Note that the line element, defined as $d\tau=\sqrt{\phi(z)}dz$, is therefore real along horizontal trajectories, and purely imaginary along vertical ones. As we see below, the terminology arises from considering ${\mathbb C}$ where the horizontal and vertical trajectories of $q=dz^2$ are exactly lines parallel to the real and imaginary axes respectively (see Fig.~\ref{fig:nearregular}). On a general Riemann surface, the vertical and horizontal trajectories exhibit three distinct types of behaviours, depending on whether one is in the neighborhood of an ($m$-th order) zero, a double-pole or simply a regular point of the Strebel differential. 

\subsubsection*{a) Local behaviour near a regular point}

We begin by considering the case of a regular point on $\Sigma_{g,n}$.  In other words, in some local coordinate $z$, $\phi(z)$ is holomorphic and non-zero at $z=z_0$. Then, in a neighborhood of $z_0$, we can define so-called canonical coordinates $w$
\begin{equation}
    w(z)=\int^{z}_{z_{0}} \sqrt{\phi(z')} dz' ,
\end{equation}
such that the Strebel differential takes the very simple form $q=dw^2$. We can immediately write down the vertical and horizontal trajectories
\begin{align}
    w_H(t)= &\,  \alpha t + i  \beta ,\quad \alpha, \beta \in \mb{R}\\
    w_V(t) = & \, \alpha + i \beta t .
\end{align}
The horizontal trajectories therefore run parallel to the real axis, at various fixed $\text{Im}(w) = \beta$, while the vertical trajectories run parallel to the imaginary axis, label by the fixed values $\text{Re}(w)=\alpha$ - as shown in Fig~\ref{fig:nearregular}.  Near zeroes and poles of the Strebel differential, this picture will, however, need to be modified. 

\subsubsection*{b) Local behaviour near a zero}

Consider now an $m$-th order zero of the Strebel differential. We can always put it in the form 
\begin{equation}
    \phi(z)dz^2 = z^m dz^2 .
\end{equation}
The novel feature here is that $(m+2)$ vertical and horizontal trajectories will meet at the zero. To see this, note that radial half-lines, emanating from $z=0$ and parametrised via 
\begin{equation}
    z_{H}(t)=t e^{2\pi i \frac{k}{m+2}}, \quad \quad t>0,\qquad  k \in (0,1,..,m+1),
\end{equation}
are horizontal trajectories. The $(m+2)$ vertical trajectories are also radial half lines emanating from the zero, but offset with respect to the horizontal trajectories by an angle of $\frac{\pi}{m+2}$.
\begin{equation}
    z_{H}(t)=t e^{i \pi \frac{(2k+1)}{m+2}}, \quad \quad t>0, \qquad k \in (0,1,..,m+1).
\end{equation}
These horizontal and vertical trajectories are shown in Fig.~\ref{fig:nearzero} for the case of a simple zero ($m=1$). As can be seen in that figure, slightly away from the zero, the horizontal trajectories asymptote to the boundaries of the wedge defined by the intersection of two of these radial half-lines. The same holds for the vertical trajectories. 

\subsubsection*{c) Local behaviour near a double pole}

The behaviour near the marked points is particularly interesting. For a Strebel differential, we have specified it's form in the neighborhood of the $i$-th double-pole to be
\begin{equation}
    \phi(z) dz^2 = - \frac{L_i ^2}{(2\pi)^2} \frac{dz^2}{z^2}.
\end{equation}
One can quickly check that the horizontal trajectories are concentric circles, centered around the double pole:
\begin{equation}
    z_{H}(t) = re^{i \theta t}; \quad t\in (0,2\pi) \  \& \ r \in \mb{R},
\end{equation}    
while \textit{any} radial half line emanating from the pole
\begin{equation}
    z_{V}(t) = te^{i \theta} ,
\end{equation}
is a vertical trajectory. This tells us that near a double-pole, we have the geometry of a semi-infinite cylinder - see Fig. \ref{fig:nearpole}. From the closed string theory perspective, these will be the asymptotic closed string trajectories created by the vertex operators on the worldsheet located at the marked points (or poles). Note that the proper length of the circular horizontal trajectories, as defined via the line element $d\tau$ is in fact independent of $r$ as simply seen from
\begin{equation}
    \oint d\tau =  \frac{L_{i}}{2\pi i} \oint \frac{dz}{z} = L_i.
\end{equation}
The $n$ real numbers $L_i$ therefore correspond geometrically to the proper circumferences of these asymptotic closed strings, as measured using the Strebel metric. 

\subsubsection*{Global Structure: putting it all together}

\begin{figure}[h!]
     \centering
         \includegraphics[width=0.6\textwidth]{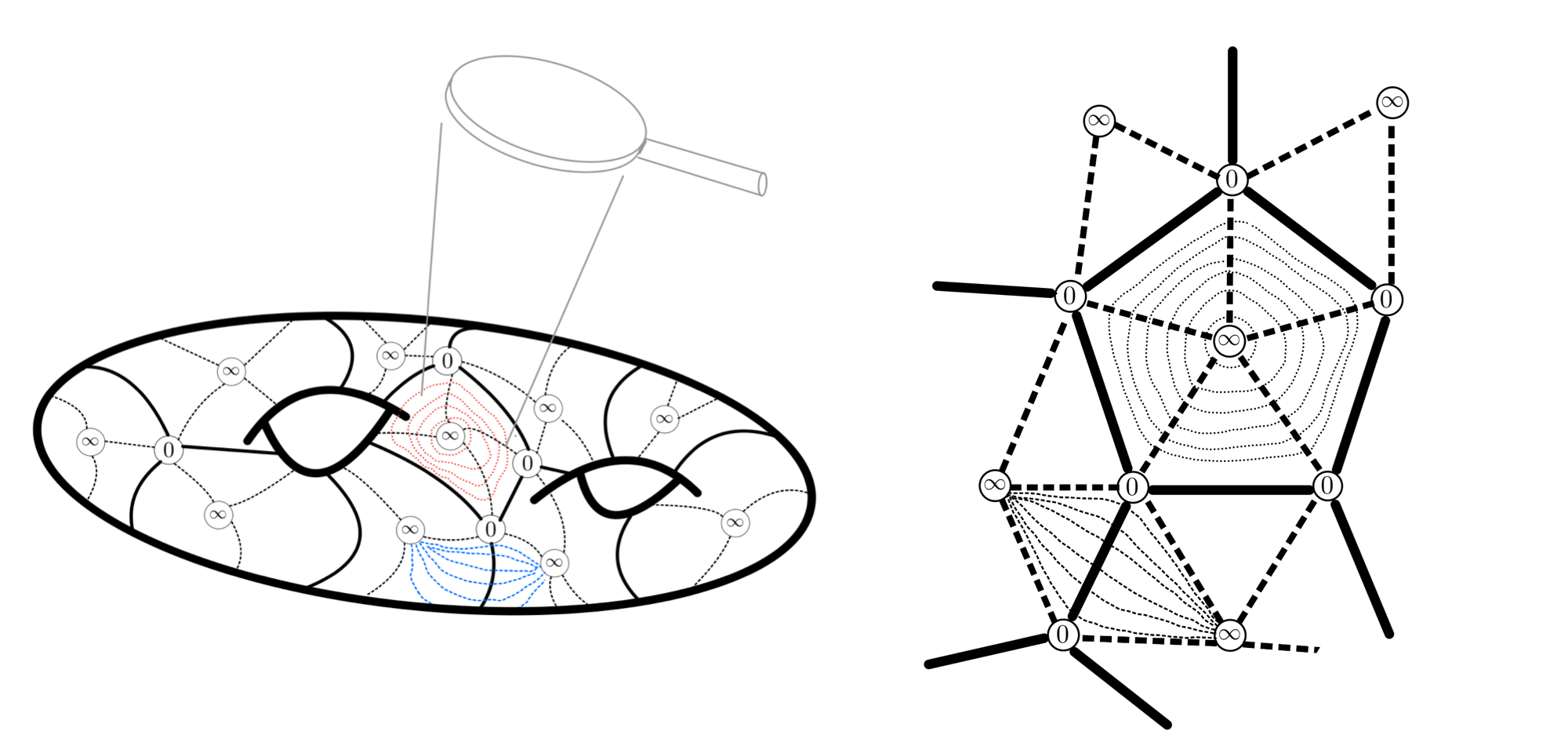}
        \caption{\small{\textbf{Global structure of the trajectories:} (Left:) In dotted red lines are the compact horizontal trajectories encircling each double-pole. They define open disc regions around each marked point, whose closure gives the critical graph of the Strebel differential. Blue dashed lines are vertical trajectories connecting poles. The thick dashed lines are special vertical trajectories connecting zeroes and poles. (Right:) Zooming in on a part of the worldsheet, we see how the overlapping regions reflecting how the local behaviour near regular points, poles and zeroes are connected.}}
        \label{fig:globalzoomin}
\end{figure}

We have now collected all the necessary ingredients to understand Strebel's theorem and the resulting Strebel map between the Riemann surface and ribbon graphs on the surface. 
While the local behaviour of horizontal and vertical trajectories is universal, their global behaviour is complicated. In general, horizontal trajectories, for instance, do not close. 
The remarkable property of the Strebel differential is that most of it's horizontal trajectories are closed and form disk shaped domains, which are separated by a graph on the Riemann surface. In fact, the Strebel differential is the unique meromorphic quadratic  differential (with specified residues $L_i$), whose non-compact horizontal trajectories form a set of measure zero.
This is the content of Strebel's theorem which thus gives a unique assignment of a quadratic differential to every point on ${\mathcal M}_{g,n}\times {\mathbb R}_+^n$ where the $L_1,...,L_n$ parametrise the $\mb{R}^{n}_{+}$.  

Pictorially, one 'grows' the punctures (double poles) into bigger and bigger holes, chewing away at the worldsheet until we are left just with a skeletal version of the surface. More precisely, the compact horizontal trajectories encircling each double pole define the open disc regions around each marked point. Their closure is given by the set of non-compact horizontal trajectories connecting two zeroes of the Strebel differential. The union of these non-compact horizontal trajectories define what is known as the critical (or Strebel) graph of the differential. This defines a ribbon graph on the Riemann surface. See Fig. \ref{fig:globalzoomin}. 
 
Each $m$-th order zero corresponds to a vertex of valency $(m+2)$ on this graph. Thus generically this Strebel graph has cubic vertices (simple zeroes). We can assign a length to (`metrize') each edge connecting vertices $i$ and $j$,
\begin{equation}
    l_{ij} = \int_{V_i}^{V_j} \sqrt{\phi(z)}dz ,
\end{equation}
which is indeed real, because we are integrating along the horizontal trajectory connecting the two vertices, and can be taken (by suitable orientation) to be positive. Each face of the graph contains exactly one marked point or double pole. Since the proper length of each circular horizontal trajectory encircling that $k$-th double-pole is given by $L_k$, we know that the sum of the lengths of the edges bounding the $k$-th face must also satisfy
\begin{equation}
    \sum_{(ij)\  \text{around face} \ k} l_{ij} = L_{k}.
\end{equation}

Variations of the edge lengths $l_{ij}$ at fixed $L_k \in \mb{R}_{+}^{n}$ are therefore variations of the Riemann surface's moduli. They move us within the $\mc{M}_{g,n}$ of the total $\mc{M}_{g,n} \times \mb{R}_{+}^{n}$ isomorphic to the set of all metric ribbon graphs of genus $g$ and $n$ faces. In fact, Kontsevich showed how to translate between the flat measure $\prod_{e \in \text{edges}} dl_{e}$ and a natural volume form on $\mc{M}_{g,n} \times \mb{R}_{+}^{n}$. While we refer the reader to  \cite{KontsevichAiry} for details, we can make a quick dimension counting argument as a sanity check. $\mc{M}_{g,n}$ is not quite a manifold, and instead consists of pieces of different dimension. The top-dimensional component has real dimension $(6g-6+2n)$. On the other hand, the generic Strebel graph will be trivalent, since zeroes of the differential will usually be simple\footnote{Higher order vertices arise only when some of the Strebel lengths $l_{ij}$ go to zero i.e. when zeroes coalesce and are thus of higher codimension in the moduli space.}. From $V-E+F = 2-2g$, a genus $g$ fatgraph with $n$ faces and trivalent vertices will have $6g-6+3n$ edges. The $(6g-6+2n) + n$ real lengths assigned to each edge therefore suffice to characterize one of the top dimensional cells of $\mc{M}_{g,n} \times \mb{R}_{+}^{n}$. As one sums over inequivalent critical graphs (with the same genus) and $n$ faces, one gets a simplicial decomposition of $\mc{M}_{g,n} \times \mb{R}_{+}^{n}$.

We see from the above construction that the faces of the critical or Strebel graph of the Strebel differential thus provides a \textit{polygonalization} of the Riemann surface. One can show that given the lengths, the complex structure is determined. This is done essentially by defining the natural complex cordinates locally in charts around zeroes, poles and patching them together with holomorphic transition functions. See \cite{mulaseStrebel} for more details.

\begin{figure}[h!]
     \centering
         \includegraphics[scale=0.2]{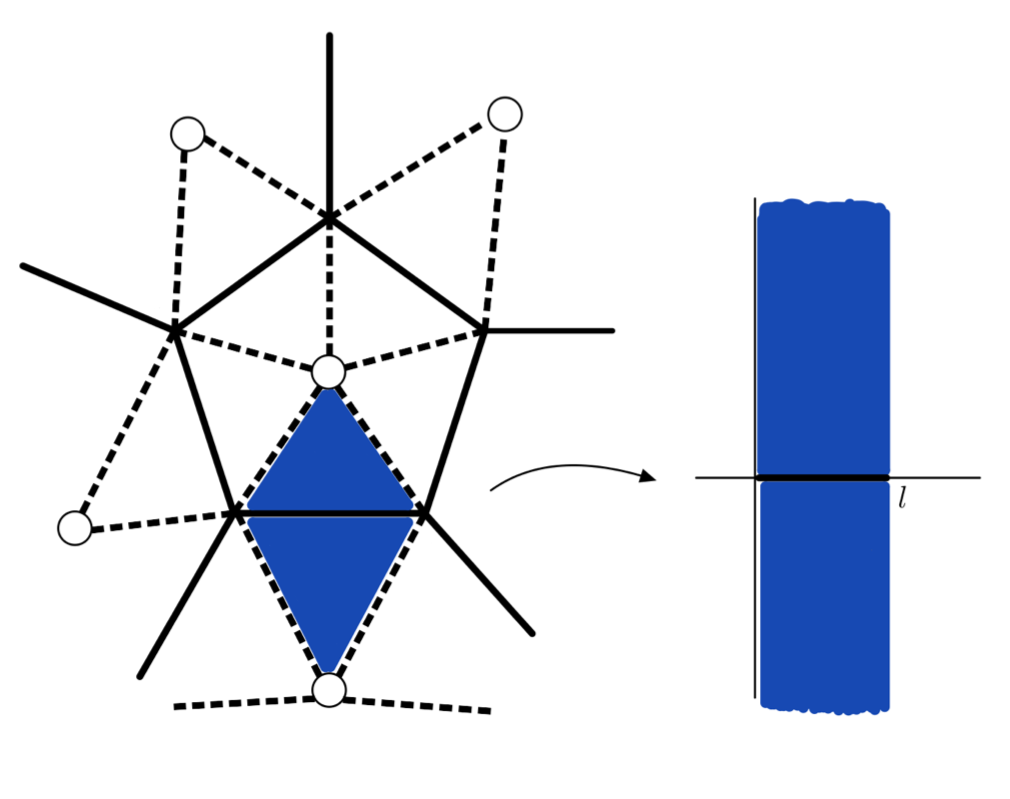}
        \caption{\small{\textbf{Infinite strips as basic building blocks:} Any point within the colored region can be conformally mapped to a point on the infinite strip. Each strip can be viewed as defining an edge dual to one of the critical graph. Thus the union of these strips defines a dual graph to the Strebel graph.}}
        \label{fig:strips}
\end{figure}

What will be particularly important for gauge-string duality is that the vertical trajectories allow us to define a canonical \textit{triangulation} which is {\it graph dual} to the Strebel graph. Focus on an edge of the Strebel graph, the two faces that share that edge (and their corresponding poles) as well as the two zeroes at the endpoints. Passing through each of these zeroes are vertical trajectories which begin or end at the two poles of the adjacent faces. These lines therefore connect Strebel graph vertices to the center of each face sharing the said vertex. These are depicted by thick dashed lines in Fig. \ref{fig:strips}, with the double-pole being represented by the white dot. These vertical trajectories thus bound a quadrilateral. Repeating for all the edges of the Strebel graph gives a dissection of the entire surface into a union of quadrilaterals with two poles as apex vertices (together with the two zeroes). Since the poles are an infinite proper distance away from any other point, each quadrilateral can be viewed as conformally equivalent to a single infinite strip. Equivalently, we can associate each such strip to an edge of the graph dual to the Strebel graph - the strip is transversal to the original edge. Adjacent pairs of strips are then glued together along the edges of the quadrilaterals. Multiple strips then come together at a zero (generically three). Alternatively, we can view (generically) three dual edges surrounding each vertex of the original Strebel graph. This gives the promised triangulation.



\subsection{Reconstructing closed string worldsheets from open strings: \\ V vs. F-type duality}\label{sec:VFconstr}

We have seen how to associate to a point on the (decorated) moduli space of punctured Riemann surfaces $\mc{M}_{g,n}\times{\mathbb R}^n_+$, a metrized ribbon graph. This was done through the critical graph of the unique Strebel differential. As just discussed, to each edge of the critical graph - or alternatively to each diamond shaped region - we associate an infinite strip. The width of the strip is simply the Strebel length of the edge. As we now discuss, gauge-string duality (specially, of the V-type) consists of gluing together these strips (identified now with the 't Hooft double line diagrams) to form the closed string worldsheet. In other words we use this procedure as a precise way by which the gauge theory Feynman diagrams build up the dual Riemann surface. We elaborate briefly on this proposal \cite{freefieldsadsII, freefieldsIII, Joburg}. We will also see how F-type open string descriptions can also be viewed in a similar vein with the critical graph itself now playing the main role. These two descriptions are graph dual to each other, as we have seen, consistent with the duality  of V- and F-type descriptions as discussed in Sec. \ref{sec:graphduality}.   



\subsubsection*{V-type duals}

\begin{figure}[h!]
     \centering
      \begin{subfigure}[b]{0.32\textwidth}
       \centering
         \includegraphics[width=\textwidth]{Figs/V_type_reconstruction.png}
         \caption{V-Dual Reconstruction}
         \label{fig:vtypereconstruction}
     \end{subfigure}
     \begin{subfigure}[b]{0.3\textwidth}
         \centering
         \includegraphics[width=\textwidth]{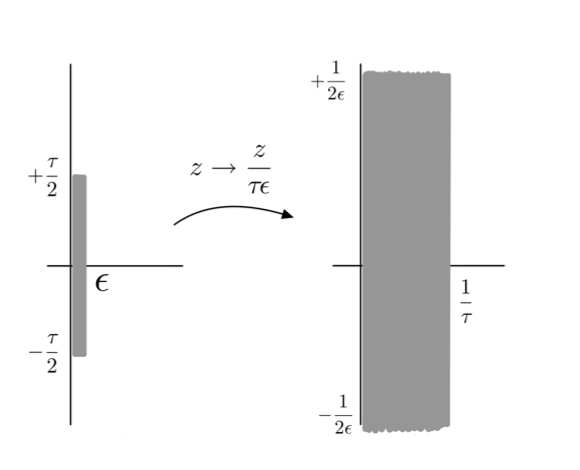}
         \caption{From worldlines to open string strips}
         \label{fig:WLtoString}
     \end{subfigure}
     \begin{subfigure}[b]{0.3\textwidth}
         \centering
         \includegraphics[width=\textwidth]{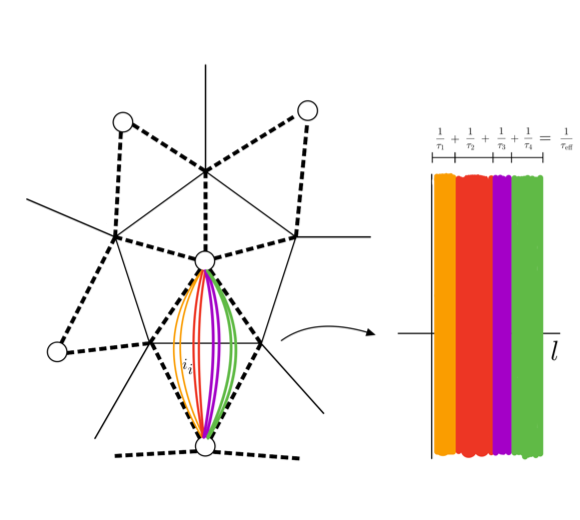}
         \caption{Gluing homotopically equivalent ribbons (~string bits) }
         \label{fig:adjoiningRibbons}
     \end{subfigure}
        \caption{\small{\textbf{V-type Duality}: a) In V-type duality, homotopically equivalent Wick contraction between vertices of the gauge theory Feynman diagram reconstruct diamond shaped regions on the closed string worldsheet. (\textit{the different colors are for clarity only}) b) Worldlines of the gauge theory Feynman diagrams, with length $\tau$, can be viewed as infinite open string strips of width $\sim 1/\tau$.  c) We "close up holes" by gluing adjacent ribbons sharing a matrix index $i$. The widths of the adjacent strips simply add, with the Strebel length $l = \frac{1}{\tau_{\text{eff}}}$}}
        \label{fig:VtypeDuals}
\end{figure}

In V-type duality, vertices of the gauge theory Feynman diagrams coincide with the marked points (insertions) of the dual closed string worldsheet. We will view internal vertices (in a perturbative expansion) as dual to additional insertions of the closed string vertex operators. The first step in reconstructing the worldsheet is to bunch together all homotopically equivalent propagators (arising from the free Wick contractions) between two gauge theory vertices \cite{freefieldsadsII}. This simply means that such edges can be deformed into each other without crossing any other edge or vertex. The resulting graph has been called a skeleton Feynman graph in this context. Perhaps the simplest way to phrase V-type open-closed duality is that each such bunch of ribbon graphs will reconstruct a diamond-shaped region on the closed string worldsheet - see Fig.\ref{fig:vtypereconstruction}. This "diamond" is the same one we discussed earlier in Fig. \ref{fig:strips}, 

Next, we associate a length to each such set of propagators. It was suggested in \cite{freefieldsadsII, freefieldsIII} that the Schwinger parametrisation of the propagator gives a natural length assignment. Thus, for instance, for a free massless propagator\footnote{See \cite{gursoy1,Gursoy2} for an interesting recent proposal to use the Schwinger parametrisation of the interacting gauge theory amplitudes to reconstruct the closed string geometry away from $AdS$.}, we can rewrite 
\begin{equation} \label{eq:Schwinger}
    \frac{1}{p^2} = \int_{0}^{\infty} d\tau e^{-\tau p^2}.
\end{equation}
Indeed the proper time $\tau$ is the parameter of the `worldline moduli space' in the first quantised interpretation of the propagator. For a large $N$ gauge theory (with all fields in the adjoint) each such double line can be assigned a strip of width $\frac{1}{\tau}$. The heuristics is shown in Fig.\ref{fig:WLtoString}, where an infinitesimal strip (since it is a worldline) of width $\epsilon$ and length $\tau$ is conformally mapped to one of length $\frac{1}{\epsilon} \rightarrow \infty$ and width $\frac{1}{\tau}$ giving rise to an infinite strip of fixed width. Identifying this strip with that of the previous section, we see that in V-type duality, (Euclidean) time on the worldsheet runs along vertical trajectories. The width of the strip corresponds to the size of the open string `bit'. We see that the skeleton graph is the graph dual to the critical Strebel graph. 

When we bunch together homotopically equivalent edges, they can be replaced with a single `effective' propagator. In our Schwinger parametrization, the inverse of the individual times add up, much like resistors set up in parallel in an electric circuit: $\sum_{i} \tau^{-1}_{i} = \tau^{-1}_{\text{eff}}$. This has a beautiful geometric interpretation: by adjoining the strips of adjacent ribbons, the widths simply add. One then assigns the Strebel length 
\begin{equation} \label{eq:inverseSchwinger}
     l_{ij} = \frac{1}{(\tau_{\text{eff}})_{ij}} .\qquad (\text{V-type duality})
\end{equation}
i.e. directly proportional to the width of the strip which is transverse. We can then rewrite the Schwinger parametrised amplitude of the field theory as an integral over the stringy moduli space $\mc{M}_{g,n}$.

It was pointed out in \cite{aharony2007remarks}
that the Schwinger proper time representation of field theory amplitudes does not preserve special conformal transformations (which is a symmetry of the free theory). Therefore, the dual worldsheet integrands constructed this way
do not make manifest all the symmetries of the $AdS$ dual geometry. It might therefore well be that there is a somewhat different assignment of the Strebel lengths to worldline parameters which preserves these symmetries. Indeed, in the case of (zero dimensional) matrix integrals, where we do not have the notion of a proper time at all, it was proposed by Razamat \cite{razamatGauss} that we should simply assign $l_{ij}=n_{ij}$ where $n_{ij}$ is the number of homotopically equivalent propagators (`string bits') which are glued together. Thus one picks out special points on $\mc{M}_{g,n}$ with integer Strebel lengths. This was also the basis for the proposal for string dual to the Gaussian matrix model in \cite{gopakumar2011simplest}. We will see how to make this more precise in \cite{DSDIII}. In the meanwhile, the tensionless limit of string theory on $AdS_3$ dual to the free symmetric orbifold CFT\cite{eberhardt2019worldsheet} has been shown to exhibit a localisation of its amplitudes on moduli space \cite{eberhardt2020deriving, dei2021free}. In fact, for correlators with large twist (a Gross-Mende or BMN-like limit) it was very explicitly shown in \cite{gaberdiel2021symmetric} that the points that the string worldsheet localises to are those with integer Strebel length. The integers correspond to the width of the strip measured by the number of wick contractions. The Strebel differential can also be very explicitly written down in terms of the Schwarzian of a covering map. We will discuss a very similar limit of large traces in the matrix model context at more length in our second and third papers \cite{DSDII,DSDIII}, along with its relation to the double-scaling limit.  



\begin{figure}[h!]
     \centering
      \begin{subfigure}[b]{0.32\textwidth}
       \centering
         \includegraphics[width=\textwidth]{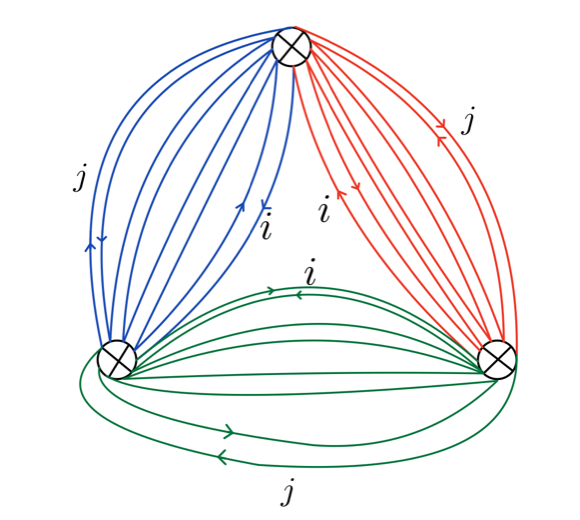}
         \caption{Feynman diagram of a V-type dual}
         \label{fig:3vertices}
     \end{subfigure}
     \begin{subfigure}[b]{0.3\textwidth}
         \centering
         \includegraphics[width=\textwidth]{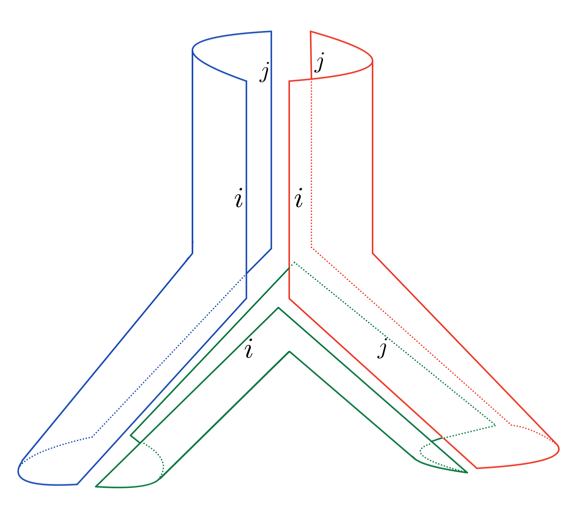}
         \caption{Three strips}
         \label{fig:3strips}
     \end{subfigure}
     \begin{subfigure}[b]{0.3\textwidth}
         \centering
         \includegraphics[width=\textwidth]{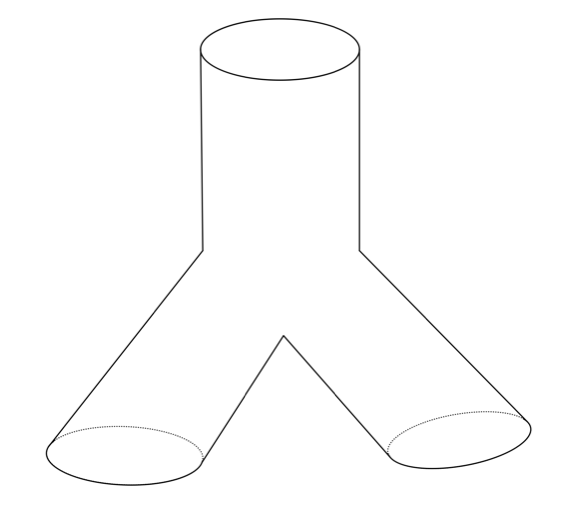}
         \caption{The resulting closed string worldsheet}
         \label{fig:3closedstrings}
     \end{subfigure}
        \caption{\small{\textbf{Gluing open string strips in V-type duality}:} a) One of the simplest Feynman diagrams of a V-type dual. We have colored homotopically equivalent ribbons the same color. b) These map onto the three strips glued together along faces. We have highlighted the two faces, labeled by the color indices $i$ and $j$. c) The resulting closed string worldsheet. Asymptotic closed string states correspond to vertices of the Feynman diagram of the V-type dual. }
        \label{fig:Vtype3pt}
\end{figure}

To summarise, replacing the homotopically equivalent edges of the gauge theory Feynman diagram by an effective propagator is just the first step of `closing up the holes'. Geometrically, it welds the thin strips together into a larger one of width proportional to the number of contractions. Once this is done, these in turn must also be glued together at the faces of the V-type diagrams (vertices of the Strebel graph). We show a simple example in Fig. \ref{fig:Vtype3pt}. 
This mechanism of closing up the holes is not that of a D-brane boundary state shrinking and being replaced by closed string insertions. 
Rather the holes in V-type duality are closed colour index (gauge invariant) loops formed by adjacent ribbons and building up a closed worldsheet. We have drawn attention to this fact by highlighting the index `i' in Fig.~\ref{fig:adjoiningRibbons} - see also Fig.~\ref{fig:Vtype3pt}.

\subsubsection*{F-type duals}

The F-type open string description, in general, is likely not to be a simple gauge theory but rather a full fledged open string theory\footnote{Before the advent of holography, Witten had interpreted the Feynman diagrams of Chern-Simons theory as an open string field theory\cite{WittenCSasOSFT}. In the context of the A-model topological strings, the open string field theory reduced to a gauge theory, very much like the matrix integrals in Sec. \ref{sec:twostacks} arose from the topological B-branes.}. The open-closed string duality, in this case, is one of D-brane boundary states closing up. This process can be viewed through the lens of Strebel differentials as well, essentially the way it is used in open string field theory \cite{ZwiebachProof}. Indeed, this is how Kontsevich employed the Strebel parametrisation of moduli space to capture closed string amplitudes (intersection numbers on $\mc{M}_{g,n}$) in terms of the combinatorial data of the Strebel lengths \cite{KontsevichAiry}.  

\begin{figure}[h!]
     \centering
      \begin{subfigure}[b]{0.32\textwidth}
       \centering
         \includegraphics[width=\textwidth]{Figs/F_type_reconstruction.png}
         \caption{F-type Dual Reconstruction}
         \label{fig:ftypereconstruction}
     \end{subfigure}
     \begin{subfigure}[b]{0.3\textwidth}
         \centering
         \includegraphics[width=\textwidth]{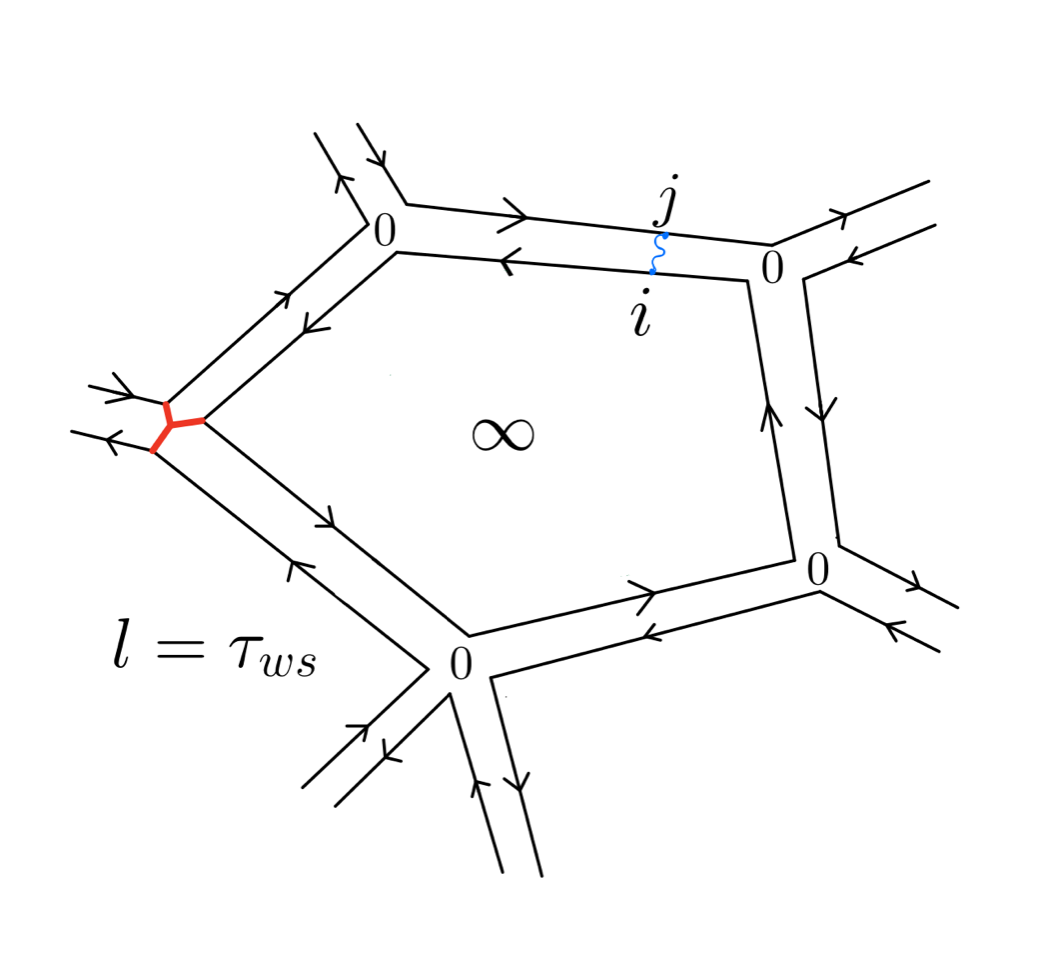}
         \caption{Edges as open string strips}
         \label{fig:Ftypeopenstrings}
     \end{subfigure}
     \begin{subfigure}[b]{0.3\textwidth}
         \centering
         \includegraphics[width=\textwidth]{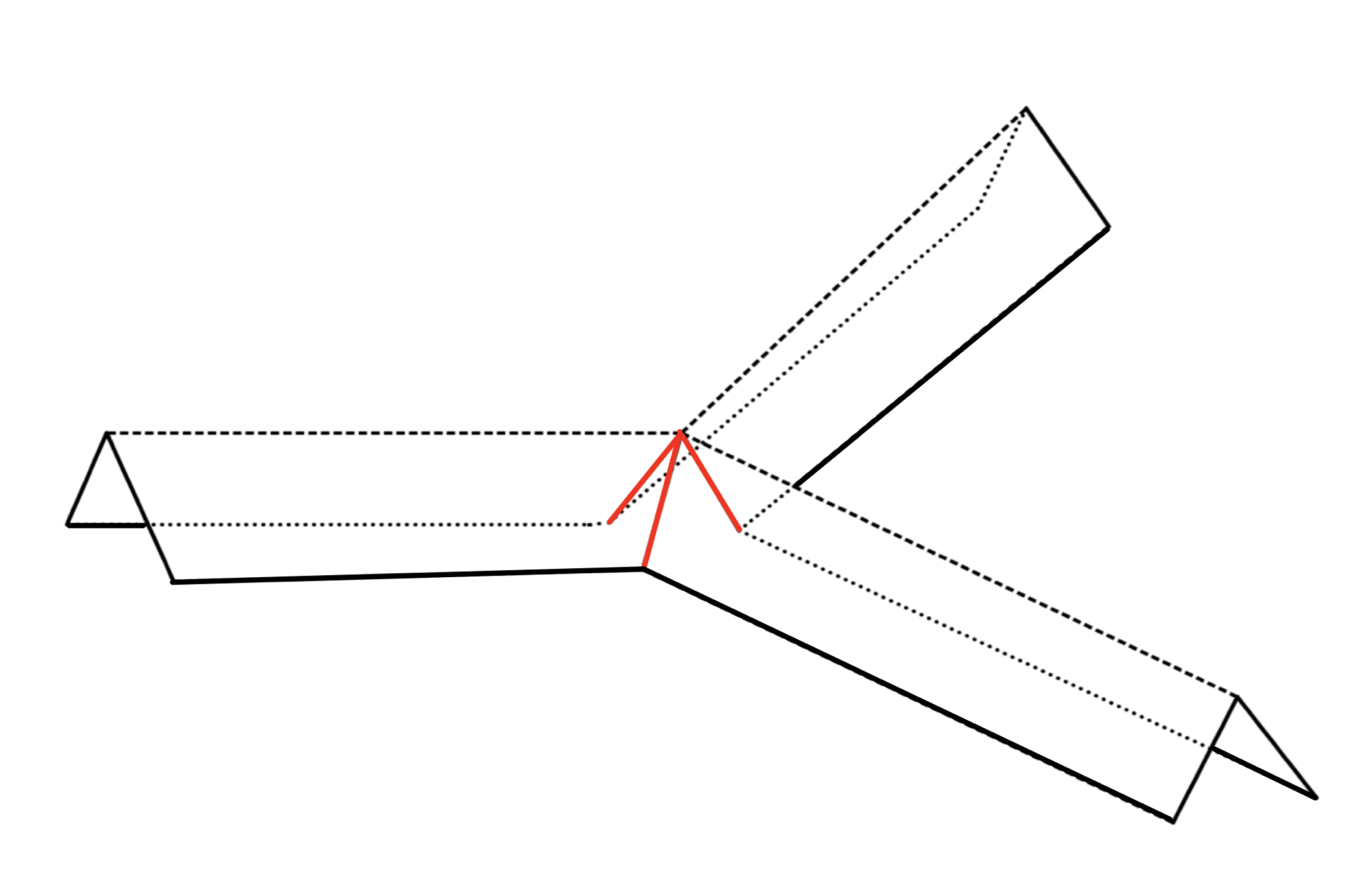}
         \caption{Gluing 3 open strings along their midpoint}
         \label{fig:midpoint}
     \end{subfigure}
        \caption{\small{\textbf{F-type Duality}: a) In F-type duality, the gauge theory Feynman diagrams correspond to the critical graph of the Strebel differential of the closed string worldsheet. b) The worldsheet-strips of open strings ending on branes make up the double-line ribbons of the gauge theory diagram. The amount of (euclidean) time evolution of that open string maps onto the Strebel length associated to that edge. c) Vertices of the Feynman diagram represent open string interactions. The open strings are glued along their midpoint, as in the cubic vertex of open string field theory.}}
        \label{fig:FtypeDuals}
\end{figure}

Thus we can view the open string theory diagrams (or in the special cases where it reduces to a matrix or gauge theory Feynman diagram) as precisely a `thickened' version of the critical graph of the Strebel differential of the dual closed string, see Fig. \ref{fig:ftypereconstruction}. There are Chan-Paton factors labeling each side of these ribbon graph edges, on which the open strings end. As we have explained above, the Strebel graph also assigns a length to each edge. 
This means we can think of each edge of the Feynman diagram as an open string strip of fixed width and length $\tau$. The length of each edge quantifies the (euclidean) time evolution of each open string. In F-type duality, worldsheet time thus runs along the horizontal trajectories. We therefore identify the Strebel lengths with the open string worldsheet time 
\begin{equation}
    (\tau_0)_{ij} = l_{ij} \qquad (\text{F-type})
\end{equation}
 as in Fig. \ref{fig:Ftypeopenstrings}. We note that this is the inverse relationship to that in V-type duality. 
 
The cubic vertices of the critical Strebel graph correspond to the cubic interactions of the open string field theory \cite{WittenOSFT} - the open-string equivalent of the "pair of pants" diagram.  
Thus to each open string worldsheet we can associate the fattened Strebel graph which can be viewed as including additional horizontal trajectories beyond the critical one. Recall that the horizontal trajectories within each face form concentric circles surrounding the double poles of the Strebel differential of the closed string dual. We can think of them as a cartoon of the hole shrinking to zero size. The D-brane boundary condition of the open string gets replaced by a (sum of) local vertex operator(s) at the puncture on the closed string worldsheet. See, for instance, Eq.~(3.6) of \cite{Gaiotto:2003yb} for how this might be quantified in the case of the Konstsevich model, an F-type dual to the $(2,1)$ minimal string. A similar picture was also realised in the duality of the topological closed string on the conifold to large $N$ Chern-Simons theory \cite{GV99}. In fact, in the linear sigma model derivation of this duality by Ooguri and Vafa \cite{OV2002worldsheet}, there was a similar occurrence of holes being replaced by closed string insertions that deform the background away from the free gauge theory (or singular conifold geometry).

To summarise, F-type duals are directly inheriting a Strebel parametrisation from their open string field theory descriptions which carry over to a closed string description when we view the holes as being replaced by closed string insertions. 
It would be good to have more tractable examples of F-type duality that go beyond the simple matrix model or topological gauge theories that have so far been considered. We make some comments about a possible open string F-type description for ${\cal N}=4$ super Yang-Mills in section~\ref{sec:discussion}.

\pagebreak

\part{Application: Closed String Duals to the One Matrix Model}


\section{The F-type Dual and the $c=1$ Theory}

We will now see a concrete application of the ideas of open-closed-open triality discussed at length in the first part of this paper. A special case of the duality between matrix models described in Sec. \ref{sec:twostacks} will allow us to connect observables of the general one matrix model (including, of course, the free Gaussian) to those of the two-dimensional $c=1$ string theory at self-dual radius. This will be via a duality to the so-called Imbimbo-Mukhi matrix integral \cite{ImbimboMukhi}. The latter was shown to capture correlation functions of tachyon vertex operators in the $c=1$ string at self-dual radius, to all orders in the string coupling. 

In the broader context of our three articles, the map to the $c=1$ theory provides a convenient stepping stone to check our proposed A- and B-model worldsheet duals. The $c=1$ string at self-dual radius is an extremely well studied noncritical string background. 
In particular, the theory at self-dual radius shows many indications of being a topological string theory and in fact, a pair of topological string descriptions (mirror to each other) have been shown to give rise to all the observables of this theory. It is this feature that we will exploit to argue that correlators of the one matrix model are the same as those of the A- and B-models with some appropriate backgrounds turned on. 

Indeed, Mukhi and Vafa had proposed \cite{MukhiVafa} the A-twisted $SL(2,\mb{R})_{1}/U(1)$ Kazama-Suzuki coset SCFT as an alternative closed string worldsheet description to the standard one of Liouville theory coupled to the compactified $c=1$ boson (the "tachyon"). The equivalence of the two essentially follows from a Wakimoto representation of the coset theory \cite{MukhiVafa, ashok2006topological}.  
Open-closed-open triality guarantees an all-genus identification between correlators of matrix traces and of tachyon vertex operators in the $c=1$ string theory (with a certain background turned on, as we will see). Using the relation to the twisted coset model, in turn, guarantees an exact agreement between correlators of single trace operators and those of particular physical vertex operators in the $SL(2,\mb{R})_{1}/U(1)$ theory. This is to all orders in the string coupling or genus expansion. 

Similarly, a B-model topological string description of the $c=1$ string at self-dual raidus was also proposed \cite{ghoshalMukhiLG, hananyOzPlLG}. This involves a topological Landau-Ginzburg theory with a singular superpotential $W(Z)=\frac{1}{Z}$ coupled to 2d topological gravity. Again it was argued that observables in this theory reproduce the tachyon correlators of the $c=1$ string theory. This also follows from the mirror symmetry between this LG description and the topological cigar theory
\cite{hori2001duality}. Again, this implies the equality of correlators of single trace operators with those of observables in the B-model theory (with a deformation of the superpotential). 
Once we have these agreements we will not need the intermediate crutch of the $c=1$ string theory any more. Thus we will concentrate in the future instalments of this work \cite{DSDII, DSDIII} to showing how one can directly derive the B-model and A-model string dual backgrounds respectively. 

In this section, we will first show the equivalence of the hermitian matrix model with the Imbimbo-Mukhi matrix model. We will then briefly discuss the relation of the latter to the $c=1$ string theory. As a concrete sanity check (and of normalisation factors etc.), we will exhibit a computation of the one point function $\Braket{\Tr M^{2n}}$ in the Gaussian matrix model, to all orders in $1/N$. We then compute the all-genus correlation function of one negative momentum tachyon in a background of tachyons of postive momentum $(+2)$ and find that they indeed agree genus by genus. 

\subsection{From the Hermitian one matrix model to the Imbimbo-Mukhi Model}

The  Imbimbo-Mukhi matrix model \cite{ImbimboMukhi}, which captures the $c=1$ string at self-dual radius, reads (see for example Eq.~(4.12) of \cite{mukhireview})
\begin{equation}\label{eq:IMpartfn}
    Z_{IM}(t_k,\bar{t}_k)= \det(X)^{-i \mu} \int dA_{Q \times Q} e^{+ i\mu \sum_{k=1} t_{k} \Tr (A^{k})  + i \mu \Tr (A X)  - (i \mu + Q) \Tr \log(A) }.
\end{equation} 
Here we have suggestively labeled the matrix $A$ and the source $X$ to connect with our earlier notation. We will see that we will need to further identify $N=+ i\mu$, playing the role of the inverse string coupling. The coefficients $\bar{t}_k$ - often called Miwa times  - are in turn defined in terms of the source matrix $X$.
\begin{equation} \label{eq:bartimes}
    i\mu \bar{t}_k \equiv \frac{1}{k} \Tr_{Q} \left( X^{-k}\right).
\end{equation}
We also note the appearance in the integral of another infinite set of times $t_k$. As we review later, $Z_{IM}(t_k,\bar{t}_k)$ is the generating function of tachyon correlators of both positive and negative momentum, which accounts for the two sets of times which appear in this model. 

We will now arrive at the above model as a  special case of the RHS of the matrix duality in Eq.~(\ref{eq:source/det duality}). Setting $Y=0$ and rewriting the coupling $g^{-1}=N$ in Eq.~(\ref{eq:source/det duality}) gives us the relation
\begin{eqnarray} \label{eq:IMStart}
   & \frac{1}{Z_{N}} \int dK dM_{N \times N} e^{+ N \Tr \left( V(K) - KM \right)} \prod_{a=1}^{Q} \det(x_{a}-M) & \nonumber \\ 
     & =  \frac{1}{Z_{Q}}\int dA dB_{Q \times Q} e^{- N \Tr \left( V(A) + A(B-X) \right)}  \det(B)^{N}. &
\end{eqnarray}
We start with the RHS of Eq.~(\ref{eq:IMStart}) and note that $A$ and $B$ couple only via their product. Therefore, we may define a new matrix\footnote{We thank Herman Verlinde for this insight, as well as discussions on related transformations applied to the $G,\Sigma$ collective field description of the SYK model.}  \begin{equation}
    C = AB .
\end{equation}
The Jacobian of this transformation relates $dC$ to $dB$ via
\begin{equation}
    dC = \det\left[ \frac{\partial C_{ab}}{\partial B_{cd} }\right] dB = \det[A_{db} \delta_{ac}] dB = \det(A)^Q dB , 
\end{equation}
and we can trivially rewrite the remaining integrand in terms of $C$ using $\det(B)^{N}= \frac{\det(C)^{N}}{\det(A)^N}.$

The RHS of Eq.~(\ref{eq:IMStart}) (divided by a factor of $\det(X)^N$) now factorizes as the product of two one-matrix models
\begin{align}
     \frac{ \det(X)^{-N} }{Z_{Q}} \int dA_{Q\times Q} e^{-N \Tr \left(V(A)-AX \right)-(N+Q)\Tr \log(A)} \times \int dC_{Q\times Q} e^{-Q \frac{N}{Q}\Tr C + N \Tr \log(C) } .
\end{align}
We note that the matrix integral over $C$ is nothing but the Penner matrix model (up to a constant) as can be seen by  shifting $C \rightarrow 1-\tilde{C}$, with a coupling constant set by the ratio $Q/N$ \cite{harer1986euler, penner1988perturbative}.

We also see that up to an irrelevant normalisation factor the $A$ integral is precisely that of the Imbimbo-Mukhi model $Z_{IM}(t_k, \bar{t}_k)$
in Eq.~(\ref{eq:IMpartfn}) when we take 
\begin{equation}
    V(A)= \sum_{k=1} (-t_{k}) A^{k},  
\end{equation}
and make the identification $N = i \mu$.


On the other side, we see that the LHS of Eq.~(\ref{eq:IMStart}) can be viewed as a generating function for correlators of traces of $M$ (after dividing by a factor of $\det(X)^N$). This can be seen as usual by exponentiation of the determinants using $\log \det(1-M/x_a) = \Tr_{N} \log (1-M/x_a)$ and a Taylor expansion of the logarithms around large $x_a$ as in Eq.~(\ref{eq:potentialgen}).  
We see the appearance of the times $\bar{t}_k$, as in Eq.~(\ref{eq:bartimes}). 

Thus we have established that 
\begin{eqnarray}\label{eq:IM=gauss}
  &\frac{1}{Z_N} \int dK dM_{N \times N} e^{+ N \Tr \left( V(K) - KM - \sum_{k=1}^{\infty} \bar{t}_k  M^{k}\right)} &\nonumber \\ 
 & = \frac{(-1)^{Q^2}e^{-NQ}\det(X)^{-N}}{\sqrt{Z_{Q}}} \int dA_{Q\times Q} e^{-N \Tr \left(V(A)-AX \right)-(N+Q)\Tr \log(A)} \times Z_P, & 
\end{eqnarray}
with the potential $V$ as in Eq.~(\ref{eq:potentialgen}).
We note that the degree of the potential determines which times $t_k$ are turned on. Similarly, the rank $Q$ of the source matrix 
$X$ determines how many independent couplings $\bar{t}_k$ are turned on. 

In particular, if we further specialise
to a quadratic potential $V(K)= \frac{1}{2}K^2$
then that implies that only one of the times $t_k$ is turned on,  
$t_k=-\frac{1}{2}\delta_{k,2}$. In this case we can do the Gaussian integral over $K$ and be left with a relation between two one matrix integrals.
\begin{eqnarray}\label{eq:IM=1HM}
  &\frac{1}{\sqrt{Z_{N}}}\int  dM_{N \times N} e^{-N \Tr \left( \frac{1}{2}M^2 - \sum_{k=1}^{\infty} \bar{t}_k  M^{k}\right)} &\nonumber \\ 
 & = Z_{P} \frac{(-1)^{Q^2}e^{-NQ}}{\sqrt{Z_{Q}}}\det(X)^{-N} \int dA_{Q\times Q} e^{-N \Tr \left(\frac{1}{2}A^2-AX \right)-(N+Q)\Tr \log(A)}  . &  
\end{eqnarray}
where $Z_P$ is the partition function of the Penner model, whose free energy computes the Euler characteristic of the moduli space of Riemann surfaces with at least one marked point $\chi(\mc{M}_{g,s})$:
\begin{equation}
   \log Z_{P} = \sum_{g} N^{2-2g}\sum_{s \geq 1} \left(\frac{N}{Q} \right)^{2-2g-s} \chi\left(\mc{M}_{g,s} \right).
\end{equation}

Eq.~(\ref{eq:IM=1HM}) is the main result of this section. 
We have found a relation between the generating function of correlators in the one Hermitian matrix model and those in the Imbimbo-Mukhi matrix model. Note that we can consider an arbitrary potential and correlators therein by considering nonzero background values of $\bar{t}_k$. 
In the next subsection we review how the Imbimbo-Mukhi matrix model captures correlators in the $c=1$ string theory at self-dual radius. We then exploit the relation of this latter string background to various topological string theories to verify our proposed duality. 


\subsection{Imbimbo-Mukhi model and the $c=1$ string at self-dual radius} \label{sec:IMc=1}

The $c=1$ string under discussion here, is a non-critical string which propagates on a space-time with an exponential potential in one of the two directions, corresponding to a cosmological term whose coefficient is the cosmological constant $\mu$. On the worldsheet, this is captured by a Liouville CFT. The other target-space direction $X$ is described by a free boson CFT. The boson can be compactified to be a circle of radius $R$ \cite{grossKlc=1}, and here we choose $R=1$ in appropriate units, corresponding to the self-dual value under the target-space duality transformation. The main observables of this theory are an infinite discrete set of modes of a single massless scalar field in the 2d spacetime (misnamed the `tachyon'). The tachyon modes $T_n$, when the spacetime direction $X$ is a circle at self-dual radius, are labeled by their momentum $n\in \mb{Z}$. This momentum is conserved in correlation functions.

This is one of the few string theory backgrounds which is exactly solved to all orders in perturbation theory in the coupling $1/\mu^2$. The solution is elegantly expressed in terms of recursion relations of an infinite-dimensional $W_{\infty}$ symmetry algebra. 
The scattering of the physical states, or equivalently tachyon correlators in euclidean signature, was fully computed by Moore, Plesser and Ramgoolam \cite{mooreRangPlessScattering, mandal1991interactions} (there is a large body of work on the $c=1$ string theory including its scattering amplitudes, see \cite{ginsparg1993lectures, klebanov1991string} for reviews). 
Dijkgraaf, Moore and Plesser \cite{Dijkgraaf:1991qh}
exploited the free fermionic dual of this theory and the resultant $W_{\infty}$ constraints to obtain a generating function for all (connected) tachyon correlators, $F(t_k,\bar{t}_k)$ at the self-dual radius\footnote{The appearance of free fermions is no coincidence, as this two-dimensional string theory is dual to the singlet sector of the (Euclidean) $c=1$ matrix \textit{quantum mechanics}. The eigenvalues of this matrix description can be recast in terms of free fermions in a certain potential.}. The $t_k$ couple to tachyons of positive momenta, and similarly the $\bar{t}_k$ to negative momenta. In \cite{ImbimboMukhi}, Imbimbo and Mukhi showed that $F(t_k,\bar{t}_k)$ could be written as the free energy of the matrix model given in Eq.~(\ref{eq:IMpartfn})\footnote{It should be noted that a two matrix model very similar to our V-type model on the LHS of Eq. (\ref{eq:IM=gauss}), was proposed in \cite{BonoraXiong2matrix}, motivated by integrability, as a generating function of $c=1$ correlators. This was done independently of the Imbimbo-Mukhi model but now we see they are related. We also note that the Imbimbo-Mukhi model with a background momentum $(+2)$ can be shown to be equivalent to the Kontsevich-Penner model of \cite{ChekhovMakeenkoKP}. We will discuss this matrix model in \cite{DSDIII}, in connection with our A-model closed string dual. Finally, there is another interesting matrix model description of the $c=1$ string theory in terms of a so-called normal matrix model \cite{AlexandrovNormal}. This has even featured in a complex model version of open-closed-open triality \cite{brown2011complex}. It would be interesting the understand the connection to this latter model a bit better.}
\begin{equation}
    Z_{IM}(t_k,\bar{t}_k) = e^{\mu^2 F_{IM}(t_k,\bar{t}_k)}.
\end{equation}
In other words, connected correlation functions of tachyon vertex operators in the `big phase space' (i.e. with all the times turned on or in an arbitrary background with tachyon momenta turned on), can be computed via 
\begin{equation}
    \Braket{\prod_{i=1}^{n_+} k_{i}T_{k_i} \prod_{j=1}^{n_{-}} {k_{j}}T_{-k_{j}} } = \prod_{i=1}^{n_+}  \frac{\partial}{\partial t_{k_{i}}} \prod_{j=1}^{n_{-}}   \frac{\partial}{\partial \bar{t}_{k_j}} F_{IM}(t_k,\bar{t}_k).
\end{equation}
The usual correlators of the $c=1$ string at self-dual radius are recovered by setting $t_k=\bar{t}_{k}=0$ after taking the derivatives. 
What will be important for us is that computations of correlators in a general one hermitian matrix model potential correspond to turning on background ${\bar t}_k$ in addition to the non-zero $t_2$ already mentioned above. 

Thus, single trace correlators for a single matrix $p$-th degree potential is given in terms of tachyon correlators in a background with a finite set of times being non-zero.
\begin{equation}
   \Braket{\frac{1}{N k_1 } \Tr M^{k_{1}}...\frac{1}{N k_n } \Tr M^{k_{n}}}_{general}= \left. \Braket{\prod_{j=1}^{n_{-}} T_{-k_{j}}} \right|_{\bar{t}_{k<p}\neq  0, t_{2}\neq0}.
\end{equation}
Since this is an all-genus statement and holds for all $n$-point functions, we can, in fact, state an operator dictionary between matrix traces and tachyon vertex operators of negative momentum 
\begin{equation} \label{eq:traces=tachyons}
     \frac{1}{N k} \Tr M^{k} \Longleftrightarrow T_{-k}.
\end{equation}
This reflects the fact that the equality of correaltors also holds in the background of arbitrary negative momentum tachyons, together with that for the positive momentum mode $(+2)$. From momentum conservation, the positive momentum  background deformation is necessary for correlators of purely negative momentum tachyons to not vanish identically. 

\subsection{Matching the one-point function at all genus} \label{sec:allordercheck}

As a sanity check of this rather surprising correspondence, we now show, by explicit and direct computation, that $\Braket{\frac{1}{N}Tr M^{2n}}$ in the Gaussian matrix model agrees to all orders in the genus expansion with the one point function of negative momentum tachyons $\Braket{2n T_{-2n}}_{t_2}$\footnote{The observation that these specific correlators are equal had, in fact, been obtained in unpublished work by the first author, in collaboration with S. Mukhi, in 1995. However, the reason for this matching was a mystery at the time. Open-closed-open triality now provides the underlying justification. In fact, it proves the stronger statement that all $n$-point correlators must agree as we have seen, and also away from the `free' Gaussian case.}. The $t_2$ subscript refers to the background of positive momentum tachyons being turned on.  In our companion paper \cite{DSDII} describing the B-model string dual, we do a further explicit check of the genus zero three and four-point functions. 

We begin with the $c=1$ string computation and later show how the matrix model reproduces it, using orthogonal polynomial techniques. The $W_{\infty}$ constraints allow us to express the tachyon correlators in terms of the $W$-currents. We will not need the most general correlators of this theory, so we write down the expression which gives rise to all correlation functions with a single tachyon of negative momentum, in the background of arbitrarily many tachyons of positive momentum\cite{Dijkgraaf:1991qh}. This is expressed as a contour integral \footnote{We include an implicit factor of $(2\pi i )^{-1}$ in  the contour integral $\oint$. } 
\begin{equation}
    n\Braket{T_{-n}} = \frac{1}{n+1} \left(\frac{1}{i \mu } \right)^{n+1} \oint dz W^{n+1}(z) ,
\end{equation}
where the W-currents are given in terms of the free fermions by 
\begin{equation}
    W^{n+1}(z) \equiv \bar{\psi}(z) \partial_{z}^{n+1} \psi(z).
\end{equation}
The free fermion variable encode the couplings $t_k$ to the tachyons $T_k$ of positive momentum through the bosonization relation 
\begin{eqnarray}
  \psi(z) & = & e^{i \mu \phi(z)} \\
  \phi(z) & = & \log z + \sum_{k=1}^{\infty} \frac{1}{k} t_{k} z^{k}
\end{eqnarray}
Differentiating appropriately in the various $t_k$ gives the correlators of $T_{-n}$ with the tachyons $T_k$. These have an expansion in inverse powers of $i \mu$, which is the genus expansion of the string theory. 
Restricting to the case at hand with only $t_2$ turned on, and using the identification $i \mu =N$, the exact correlator can be written as 
\begin{equation} \label{eq:1ptc=1}
    2n \Braket{T_{-2n}} = \frac{1}{N^{2n+1}} \frac{1}{2n+1} \oint dz z^{-N}e^{- \frac{N}{2}t_2 z^2 } \partial_{z}^{2n+1} \left( z^{N} e^{+\frac{N}{2}t_2 z^2 }\right) .
\end{equation}

We now turn to the matrix model computation. Choosing $V(K)=t_2K^2$, we can explicitly integrate out $K$ to obtain the standard Gaussian one-matrix model
\begin{equation}
    \int dM_{N \times N} e^{-\frac{N}{2 t_2} \Tr M^2}.
\end{equation}
We want to check the agreement between Eq.~(\ref{eq:1ptc=1}) and the the Gaussian matrix average 
\begin{equation}
    2n\Braket{T_{-2n}} = \left( \frac{2 t_2}{N} \right)^{n} \frac{\int d \tilde{M} e^{- \Tr \tilde{M}^2} \frac{1}{N} \Tr \tilde{M}^{2n}}{\int d \tilde{M} e^{- \Tr \tilde{M}^2}} .\label{eq:gaussianeq}
\end{equation}
Here we have rescaled $M$ to isolate the dependence on $t_2$ and simplify subsequent equations. After reducing the problem to eigenvalues, these integrals can be evaluated, exactly in $N$, using orthogonal polynomials $F_{n}(\lambda)$ (section 2.2 of \cite{Zubernotes} provides a quick but clear introduction to these methods). These are normalized  $F_{n}(\lambda)= \lambda^{n}+ \mc{O}(\lambda^{n-1})$ and satisfy the orthogonality relation 
\begin{equation}
    \int d \lambda e^{- V(\lambda)} F_{m}(\lambda) F_{n}(\lambda) = h_{m} \delta_{m n} .
\end{equation}
For the Gaussian with $V(\lambda)=\lambda^2$, these are simply rescaled Hermite polynomials $F_n(\lambda)=2^{-n}H_{n}(\lambda)$, with $h_m = 2^{-m} m! \sqrt{\pi}$.
Using the fact that the integrand is symmetric amongst all eigenvalues, and rewriting the denominator in Eq.~(\ref{eq:gaussianeq}) in terms of a product of the $h_i$, the expectation value takes the simple form 
\begin{equation} \label{eq:preCD}
    \Braket{\frac{1}{N} \Tr \tilde M^{2n}} = \frac{1}{N} \sum_{k=0}^{N-1} \frac{1}{h_k} \int dx e^{-x^2} F_{k}(x)F_{k}(x) x^{2n}.
\end{equation}
We can now use the Christoffel-Darboux formula to perform the sum over $k$:
\begin{equation}
     \sum_{k=0}^{N-1} \frac{1}{h_k}  F_{k}(x)F_{k}(x) = \frac{1}{h_{N-1}}\left( F'_{N}(x) F_{N-1}(x)-  F_{N}(x) F'_{N-1}(x)\right).
\end{equation}
Inserting this into Eq.~(\ref{eq:preCD}), writing $x^{2n}= (2n+1)^{-1}\partial_x  x^{2n+1}$, integrating by parts, and finally using the relation $(2x - \partial_x)F'_{k}(x)=2k F_{k}(x)$, we arrive at
\begin{equation}
    \Braket{\frac{1}{N} \Tr \tilde M^{2n}} =  \frac{1}{N} \frac{1}{h_{N-1}} \frac{2}{2n+1} \int dx e^{-x^2} x^{2n+1} F_{N}(x) F_{N-1}(x) .
\end{equation}
We now resort to two integral representations of the Hermite polynomials to rewrite
\begin{eqnarray}
  F_{N}(x) & = & 2^{-N} (-1)^N e^{x^2} \frac{1}{2 \sqrt{\pi}} \int (is)^N e^{i s x-s^2/4} \\
  F_{N-1} & = & 2^{-(N-1)} (N-1)! \oint \frac{e^{2 u x -u^2}}{u^{N}}.
\end{eqnarray}
The integral over $x$ can be done by trading powers of $x$ for derivatives $\partial_u$.  The result is a $\delta$-function enforcing $s=2i u$. We are thus left with a single integral over $u$
\begin{equation}
     \Braket{\frac{1}{N} \Tr \tilde M^{2n}}  = \frac{1}{N} \frac{1}{2n+1} \frac{1}{2^{2n}} \oint du \frac{1}{u^{N}}e^{-u^2} \partial_{u}^{2n+1} \left( u^{N} e^{+ u^2 } \right),
\end{equation}
which indeed agrees with Eq.~(\ref{eq:1ptc=1}), upon rescaling $u = \sqrt{N t_2/2}z$ and including the prefactors as in Eq.~(\ref{eq:gaussianeq})

\subsection{Double-Scaling: Generalized Kontsevich model as OSFT} \label{sec:GKM}

While most of our arguments thus far have been devoted to making sense of the closed string duals to matrix models \textit{away} from the double-scaling limit, we make a small detour here to highlight an application of open-closed-open triality to the $(p,1)$ minimal string. As first shown in \cite{daul1993rational}, certain multi-critical regimes of a two-matrix model are dual to the $(p,q)$ minimal models coupled to two-dimensional topological gravity. In particular, by choosing an appropriate $p$-th order potential $V_{p}(K)$ and double-scaling the $N \times N$, V-type $K,M$-matrix model, we focus on the $(p,1)$ minimal string theory. Determinant insertions in this double-scaled matrix integral correspond to the addition of FZZT branes (see \cite{nakayama2004liouville} for a review of branes in the Liouville CFT). In this section, we show that the F-type dual to this system is the generalized Kontsevich model, reviewed in \cite{dijkgraaf1990notes}.  

The authors of \cite{exact} already showed this result for the $p=2$ case. They obtained the cubic Kontsevich model of \cite{KontsevichAiry} starting from the double-scaled $N \times N$ Gaussian matrix model with $Q$ FZZT brane insertions. They did so by first deriving the F-type $Q \times Q$ matrix model, and translating the large $N$ (not $Q$) double-scaling of the V-type description. Until then, the exact relationship between the double-scaled models dual to the minimal string theories and Kontsevich's had remained rather mysterious. We now understand it as a manifestation of open-closed-open triality.

The interpretation of the cubic Kontsevich model in terms of open string field theory came from the work of \cite{Gaiotto:2003yb} who argued that the full cubic open string field theory on $Q$ FZZT branes in the $(2,1)$ minimal string theory localized to Kontsevich's matrix integral. The results of \cite{exact} were extended in \cite{Hashimoto_2005} to the $(p,1)$ string, albeit using slightly different methods \footnote{Our result here is the generalization of their Eq.~(2.21) to $Q>1$.}. They first computed the expectation value of $Q$ determinant insertions in terms of orthogonal polynomials. Then, they double-scaled that expression, and showed it to be equivalent to the (higher) matrix Airy function 
\begin{equation}
    \int dZ_{Q \times Q} e^{\Tr\left( \frac{Z^{p+1}}{p+1} + Z \tilde{X}\right)},
\end{equation}
which thus plays the role of an open string field theory of $Q$ FZZT branes in the $(p,1)$ minimal string. 

We show instead that translating the large $N$ double-scaling directly to the Imbimbo-Mukhi matrix model reduces to the $p>2$-generalizations of the Kontsevich model. In order to engineer a $p$-th order critical point, we need to choose a specific potential. We will follow \cite{Hashimoto_2005}, taking 
\begin{equation}
    V_{p}(z)= \sum_{k=1}^{p} \frac{1}{k} (z+1)^{k}. 
\end{equation}
We now take this potential in the Imbimbo-Mukhi matrix integral,
\begin{equation}
    Z_{IM} \propto \int dA_{Q\times Q} e^{- \frac{1}{g} \Tr \left(V_{p}(A)-AX \right)-(N+Q)\Tr \log(A)} 
    \end{equation}
and send both $N$ and $g^{-1}$ to infinity, while rescaling the matrix $A$ as well as the source $X$. More precisely, we send $\epsilon \rightarrow 0$ such that
\begin{align}
    g = & \frac{1}{N} =\epsilon^{p+1},\\
    A = & -1+ \epsilon Z, \\
    X = & \epsilon^{p} \tilde{X}.
\end{align}
A simple Taylor expansion of the $A$-integral action in power of $\epsilon$ gives 
\begin{equation}
   S(A) \rightarrow  -\frac{1}{\epsilon} \Tr \tilde{X} + -i \pi Q^{^2} + \Tr\left( \frac{Z^{p+1}}{p+1} + Z \tilde{X}\right) + \mc{O}(\epsilon).
\end{equation}
We will ignore constant prefactors, as well as the normalization of the integral. The divergent $\Tr \tilde{X}$, on the other hand, can be traced back to the fact that determinant insertions alone generally do not have a well-defined double scaling. We need to include a classical potential contribution. Indeed, note that
\begin{equation}
  \frac{1}{\epsilon} \Tr \tilde{X} = \frac{1}{g} \Tr X
\end{equation}
This divergence is therefore remedied by considering instead the double scaling limit of 
\begin{equation}
    e^{+\frac{1}{g}\Tr X} \braket{\prod_{a=1}^{Q} \det(x_{a}-M)}
\end{equation}
In fact, in our choice of the potential, we had implicitly absorbed a linear potential term for $M$ via $K \rightarrow K+1$, see Eq.~(2.6) of \cite{Hashimoto_2005}. Summarizing, we have found that 
\begin{equation}
  e^{+\frac{1}{g}\Tr X} \braket{\prod_{a=1}^{Q} \det(x_{a}-M)} \rightarrow \int dZ e^{\Tr\left( \frac{Z^{p+1}}{p+1}+ Z \tilde{X}\right)} \times (\text{Penner Model}).
\end{equation}

There is an interesting deformation of this result, which adds a term linear in $Z$ to the action of the generalized Kontsevich model. It involves a slightly different double-scaling which relaxes the relation $g N =1$ to 
\begin{equation}
    g N = 1 - \epsilon^{p} \mu
\end{equation}
thus keeping the term linear in $Z$ in the expansion of $\Tr \log(A)$ and giving 
\begin{equation}
  e^{-\frac{1}{g}\Tr X} \braket{\prod_{a=1}^{Q} \det(x_{a}-M)} \rightarrow \int dZ e^{\Tr\left( \frac{Z^{p+1}}{p+1}+ Z \tilde{X} + \mu Z \right)}
\end{equation}
Section 6 of \cite{Gaiotto:2003yb} explains how this realizes a non-zero bulk cosmological constant. From a string field theory perspective, it captures a simple deformation of the closed string background through an open-closed vertex linear in the string field.

Open-closed-open triality thus provides further evidence that the generalized Kontsevich model arises as the open string field theory on $Q$ FZZT branes in the $(p,1)$ minimal string. More importantly, it clarifies the role of the double-scaling limit, and advocates that the theory makes sense, even without taking such a limit. Again, this is rather surprising from the minimal string perspective, where it was understood as a continuum limit of the discretization of the string worldsheet. Instead, we see that the generalized Kontsevich models sit in a particular subsector of deformations of the $c=1$ string theory at self-dual radius. The order of the potential $p$ in the Kontsevich model reflects the positive momentum tachyon backgrounds of momentum $\leq p$ in the $c=1$ theory.

\section{Verifying the Simplest Gauge String Duality}\label{sec:Amodelpersp}

What we have shown in the previous section is how correlators in the one hermitian matrix model are the same as those of the Imbimbo-Mukhi matrix model which captures tachyon correlators in the $c=1$ string. We can then use the (A- and B-model) topological string presentations of the $c=1$ string and its amplitudes 
(but now with some backgrounds turned on) to conclude the equality of correlators, as advertised in the beginning of the paper. As stressed there, we will use this indirect route only as a stepping stone towards an eventual direct derivation of both A- and B-models. 

In this section, we will first review the map from the $c=1$ string to the A-twisted Kazama-Suzuki coset. 
We will then outline some of the ingredients in the derivation which will be elaborated on in \cite{DSDIII}. We will next turn to the B-model Landau-Ginzburg reformulation of the $c=1$ string and again briefly explain its relation to the B-model derivation of our second paper.


\subsection{The A-model string dual}


The $c=1$ string theory with the boson compactified at self-dual radius has many simplifications which indicate that it can be described by a topological string theory very much like the other minimal model string theories (for the latter, see for e.g. \cite{dijkgraaf1991topological}). The first concrete description which realises this idea was put forward 
in \cite{MukhiVafa} by Mukhi and Vafa. They proposed a A-model twisted $N=2$ worldsheet SCFT which is equivalent to the original bosonic worldsheet theory of the $c=1$ string. This relies on the close correspondence between twisted $N=2$ worldsheet theories and the bosonic string. The precise $N=2$ background is that of the supersymmetric Kazama-Suzuki coset $SL(2,\mathbb{R})_1/U(1)$ which is a cigar like geometry. For a general (susy) level $k$ of the $SL(2,\mathbb{R})$, this coset has central charge\footnote{Recall that the SUSY $\mathfrak{sl}(2, \mathbb{R})_k$ is equivalent to the bosonic $\mathfrak{sl}(2, \mathbb{R})$ at level $(k+2)$ together with three free fermions. The coset by the SUSY ${\rm U}(1)$ removes one free boson and one free fermion. Thus the central charge of this supersymmetric coset is the same as that of the bosonic coset but with an effective level of $(k+2)$.} 
\begin{equation}
c=\frac{3(k+2)}{k} .  
\end{equation}
Precisely at $k=1$, we get a central charge $c=9$. This is the `critical' central charge of the A-twisted theory, being also the central charge of an $N=2$  Calabi-Yau three fold background\footnote{The two dimensional black hole background \cite{witten2dBlackHole, MSW, Elitzur:1990ubs, DVV2dblackholeprop } is also a cigar geometry which, however, involves a different embedding of the $U(1)$ as well as a different level $k$ of the $\mathfrak{sl}(2, \mathbb{R})_k$, corresponding to a critical dimension of 26.}.   

In \cite{MukhiVafa}, a matching of the physical cohomology of the coset theory and that of the $c=1$ string theory was carried out. However, 
after the realisation of the importance of spectrally flowed representations of $\mathfrak{sl}(2, \mathbb{R})$, for strings on $AdS_3$ \cite{MOOgI}, a more complete analysis was carried out by Ashok, Murthy \& Troost\cite{ashok2006topological}. The equivalence of the physical cohomologies of the coset model and the $c=1$ string has thus been established. In particular, \cite{ashok2006topological} showed that all the physical states in the coset description descend from spectrally flowed $j=1/2$ discrete representations $D^{(w)}_{j=1/2}$ of $\mathfrak{sl}(2,\mathbb{R})$,  (where $w$ is the amount of spectral flow). 
As we have seen, traces in the gauge theory are related, using the open-closed-open triality to the Imbimbo-Mukhi matrix model, to negative momentum tachyons, as in Eq.~(\ref{eq:traces=tachyons}). We thus arrive at the following identification\footnote{In our conventions, $\phi$ is the cigar radial direction and $X$ the compact angular direction. Then the positive momentum tachyon modes correspond to the vertex operators $T_{+k} = c e^{-\frac{(k-2)}{\sqrt{2}}\phi}e^{+i\frac{k}{\sqrt{2}}X}$. }:
\begin{equation}
    \frac{1}{N k} \Tr M^{k} \Leftrightarrow T_{-k} \Leftrightarrow \mathcal{V}_{k} = c e^{-\frac{(k-2)}{\sqrt{2}}\phi}e^{-i\frac{k}{\sqrt{2}}X}.
\end{equation}
This dictionary combined with our open-closed-open triality relation translates into an equality of $n$-point correlators, to all genus. 
\begin{equation}
    \langle\frac{1}{N k_1}{\rm Tr}M^{k_1}\ldots \frac{1}{N k_n}{\rm Tr}M^{k_n} \rangle_g = \langle\mathcal{V}_{k_1}\ldots \mathcal{V}_{k_n}\rangle_g.
\end{equation}

We now give a brief idea of how this equality of correlators can be made manifest. In other words, to derive this equality by tautologising it. This will be fleshed out in \cite{DSDIII} and the present discussion only gives a broad picture. The starting point is the observation that the $\mathcal{V}_{k}$ are essentially the same vertex operators considered in the tensionless worldsheet theory dual to the symmetric orbifold in the $AdS_3/CFT_2$ correspondence \cite{eberhardt2020deriving, eberhardt2019worldsheet, dei2021free}. In fact, these operators are in the $D^{(k)}_{j=1/2}$ spectrally flowed representations of $\mathfrak{sl}(2,\mathbb{R})$ which played a crucial role. The operators locally create a branching of order $k$ (the amount of spectral flow) at their insertion. 
The Ward identities of the level one $SL(2,\mathbb{R})$ worldsheet theory can be used to show that these correlators localize to discrete points on the moduli space \cite{eberhardt2020deriving, Eberhardt:2020akk, Knighton:2020kuh}. The only contributions come from points which admit a holomorphic covering to the Riemann sphere $\mathbb{P}^{1}$, with the specified branchings at the insertions. 

These Ward identities also apply to the correlators of the twisted coset model. The only difference here is that the physical vertex operators are not labelled by an additional boundary position label as in the $AdS_3$ case. In some sense, they are all mapped to the same point in the target space -  the boundary ($\infty$) of the cigar geometry. In fact, this makes the topological structure of the target space like that of a $\mathbb{P}^1$. Thus, there is a cycle structure of $(k_1)...(k_n)$ above the branchpoint at $\infty$ corresponding to these external insertions.
We saw that the quadratic potential of the Gaussian translated into a tachyon background of momentum $(+2)$. 
The corresponding vertex operators are in the twice spectrally flowed sector and thus create branchpoints of order $(+2)$. Momentum conservation dictates that there are $\frac{|k|}{2}$ of them, where  $|k|=\sum_{i=1}^{n} k_i$. This leads to another branchpoint, say at $1$,  with cycle structure  
$(2)^{|k|/2}$. Finally, the interactions from the Liouville wall at the strong coupling tip of the cigar leads to another set of branchings. Note these are present even in the absence of the tachyon background.
This leads to a third branching in the strong coupling region at $0$ - the tip of the cigar. Since the target space is topologically a ${\mathbb P}^{1}$, for the purpose of these twisted correlators, the product of these three permutations must multiply to the identity. This reflects a trivial monodromy around the three branchpoints. 
Thus we need to have a holomorphic covering map from the worldsheet to ${\mathbb P}^{1}$ which is branched precisely over three points and with the above cycle structures. The worldsheet correlators dual to the expectation values of traces in the matrix model must therefore localize to special points on the moduli space which admit such covering maps.  

This is, in fact, exactly what was predicted for correlators in the Gaussian matrix model in \cite{razamatGauss, gopakumar2011simplest, gopakumar2013correlators} following the work of \cite{DiFrancescoItzykson, koch2010matrix}. The latter had pointed to the special significance for Gaussian correlators of the so-called Belyi maps which are holomorphic coverings branched over exactly three points. This had been related in  \cite{gopakumar2011simplest, gopakumar2013correlators} to the integer Strebel length points on moduli space using the results of \cite{mulaseStrebel}. We will see in our third paper how the above worldsheet picture closes the circle and gives a {\it bona fide} worldsheet explanation of this localisation on moduli space.

\subsection{The B-Model string dual}

The $c=1$ string at self-dual radius has a topological B-model description as well, in terms of a topological Landau-Ginzburg theory. This description was proposed and developed in \cite{ghoshalMukhiLG,ghoshalMukhiIm, hananyOzPlLG}. Again, this is analogous to the similar description for the $c<1$ string theories (see, for instance,  \cite{dijkgraaf1991topological,dijkgraaf1990notes}), albeit more subtle. All these descriptions are in terms of a single superfield $Z$ governed by a LG superpotential $W(Z)$. 
The observables of this theory form the so-called chiral ring, consisting of polynomials in the field $Z$, modulo the relation $dW(Z)=0$. The topological LG-description of the $c=1$ string was first proposed as an analytic continuation of the so-called $A_{k+1}$ minimal models (with $W(Z)=Z^{k+2}$) continued to to $k=-3$ (in the same spirit that the dual KS-model was originally viewed as an $SU(2)/U(1)$ coset at level $k=-3$). This suggests a $1/Z$ superpotential.

The authors of \cite{ghoshalMukhiLG,ghoshalMukhiIm,hananyOzPlLG} reproduced various tachyon correlators of the $c=1$ for the theory with $W(Z)=\frac{1}{Z}$. They identified positive momentum tachyons with positive powers of the field $Z$,
\begin{equation}
    T_{k>0} \Leftrightarrow Z^{k-1},
\end{equation}
with $T_{0}$ as the cosmological constant operator. Turning on a background of positive momentum tachyon modes amounts to deforming the superpotential as follows: 
\begin{equation}
    W(Z) \rightarrow W(Z,t) = \frac{1}{Z} + \sum_{k>0} t_{k} Z^{k-1}.
\end{equation}
The times $t_k$ are to be identified with those appearing in section \ref{sec:IMc=1}. In particular, for the momentum $(+2)$ background dual to the Gaussian matrix model, we recover the superpotential $W(Z)=Z^{-1}+t_{2}Z$, as advertised in Fig.\ref{fig:Gaussianproposal}. 

To complete the verification of the proposed operator dictionary, we need to re-express the negative momentum tachyons in this language. In the unperturbed theory (i.e. with all $t_{k}=0$), they simply correspond to negative powers of $Z$, $T_{-k} \Leftrightarrow Z^{-k-1}$. However, when we turn on the deformation, this mapping becomes more complicated. For all intents and purposes of this section, we can identify 
\begin{equation}
\frac{1}{Nk} \Tr M^{k} \Leftrightarrow T_{-k} \Leftrightarrow  {\mathcal T}_{-k}(Z) \equiv \left( \frac{\partial}{\partial Z}  W(Z,t)^{k}\right)_{-},
\end{equation}
where the minus-subscript means we only keep negative powers of $Z$ in the final expression (see Eq.~(2.3) of \cite{ghoshalMukhiIm}). 
The initial calculations of \cite{ghoshalMukhiLG, hananyOzPlLG} did not explicitly couple the topological Landau-Ginzburg theory to gravity. Instead, the residue calculus of \cite{vafatopLG} for pure matter topological LG models was supplemented by a particular set of contact terms. As first explained by Losev in \cite{losev1993descendants}, these can be understood as remnants of a gravity sector which has been `integrated out'. Losev had further proposed precise equivalent pure matter operators whose correlators would match those of gravitational descendants. In any case, with an appropriate definition of the ${\mathcal T}_k(Z)$, we have the equality
\begin{equation}
    \langle\frac{1}{N k_1 }{\rm Tr}M^{k_1}\ldots \frac{1}{N k_n }{\rm Tr}M^{k_n} \rangle_g = \langle\mathcal{T}_{-k_1}\ldots \mathcal{T}_{-k_n}\rangle_g.
\end{equation}

Once again, the idea will be to make this equality of correlators manifest. 
In \cite{DSDII}, we will do so via the  Eynard-Orantin machinery of topological recursion for matrix models\footnote{We note that topological recursion methods have recently been employed for computing correlators of resolvents in the matrix ensemble dual to JT gravity \cite{SSSJTmatrix}.  We will discuss the relation more in \cite{DSDIII}.}.  The large $N$ loop equations of the Gaussian matrix model give rise to the spectral curve
\begin{equation}
   t_2 y^2 - x y + 1 = 0 
\end{equation}
where $y$ plays the role of the resolvent. 
Up to a linear redefinition of $(x,y)$ this is nothing other than the (complexified) Wigner semi-circular distribution. 
This curve can be uniformized in terms of a complex coordinate $z$ 
\begin{equation}
    x(z) = \frac{1}{z}+t_2 z \quad \quad y(z) = z
\end{equation}
The branchpoints of the spectral curve satisfy $dx(z)=0$. We would like to identify $x(y)$ - obtained from solving $x$ in terms of $y$ from the spectral curve - with the superpotential\footnote{A very similar proposal was also made in \cite{Aganagic_2005} and we will return to this point in section \ref{sec:discussion}.}. This gives $W(Z)=Z^{-1}+ t_2 Z$ for the closed B-model string dual to the Gaussian matrix model, as promised in Fig.\ref{fig:Gaussianproposal} and verified from the $c=1$ string description. 

While this is suggestive, we can do more. 
In fact, given a solution of the topological recursion relations, one can write the generating function of 
correlators in terms of integrals of characteristic classes over multiple  copies (colours) of moduli space \cite{eynard1bp,eynard2bp}. The number of colours corresponds to the number of branchpoints of the spectral curve, i.e. the number of edges of the eigenvalue distribution. In the case of the single cut solution of the one hermitian matrix model, this is just two. For this case, we will be able to give an explicit expression for the operators (in terms of characteristic classes) on this bicoloured moduli space which correspond to the operators 
$\frac{1}{k}{\rm Tr}M^k$. 

We propose to view the expression for the closed string operators, in terms of characteristic classes on $\mc{M}_{g,n}$, as the result of having integrated out the matter sector of LG theory and expressed everything in terms of the remaining fields of 2d gravity. To arrive at this picture, we identify the branchpoints of the spectral curves with the critical points of the LG-superpotential as suggested above. Eynard's colored moduli space becomes the (compactified) B-model moduli-space of constant maps, satisfying $dW=0$\footnote{For a quadratic superpotential, the identity operator is the only primary and there is only one branchpoint. The gravitational descendants are then simply the $\psi$-classes of Witten-Kontsevich intersection theory. In this case, Eynard-Orantin topological recursion can be shown to be equivalent to the recursion relations of Dijkgraaf, Verlinde and Verlinde \cite{dijkgraaf1991topological}. This case also arises as a BMN like-limit of large operators in the Gaussian correlator. One then effectively zooms in on the edges of the eigenvalue distribution.}. 




\pagebreak

\section{Discussion}\label{sec:discussion}

In lieu of a conclusion, we discuss some further points for exploration that are suggested by our broad approach and results. 

\begin{itemize}
    \item \textbf{Connection to string bits \& the BMN limit}\\
The picture for V-type open-closed duality in \ref{sec:VFconstr}, connects nicely to previous suggestions of a string bit model \cite{Thorn:1991fv, Klebanov-susskind}. However, the difference is that here we are not in light cone gauge but in the different Strebel gauge. In any case, each ribbon or double line of the gauge theory Feynman diagram corresponds to the worldsheet of a `string bit'. The insertion of, say, $\Tr M^{k}$ creates an asymptotic string state consists of $k$ string bits, see Fig.\ref{fig:stringbits}.

\begin{figure}[h!]
     \centering
         \includegraphics[width=0.5\textwidth]{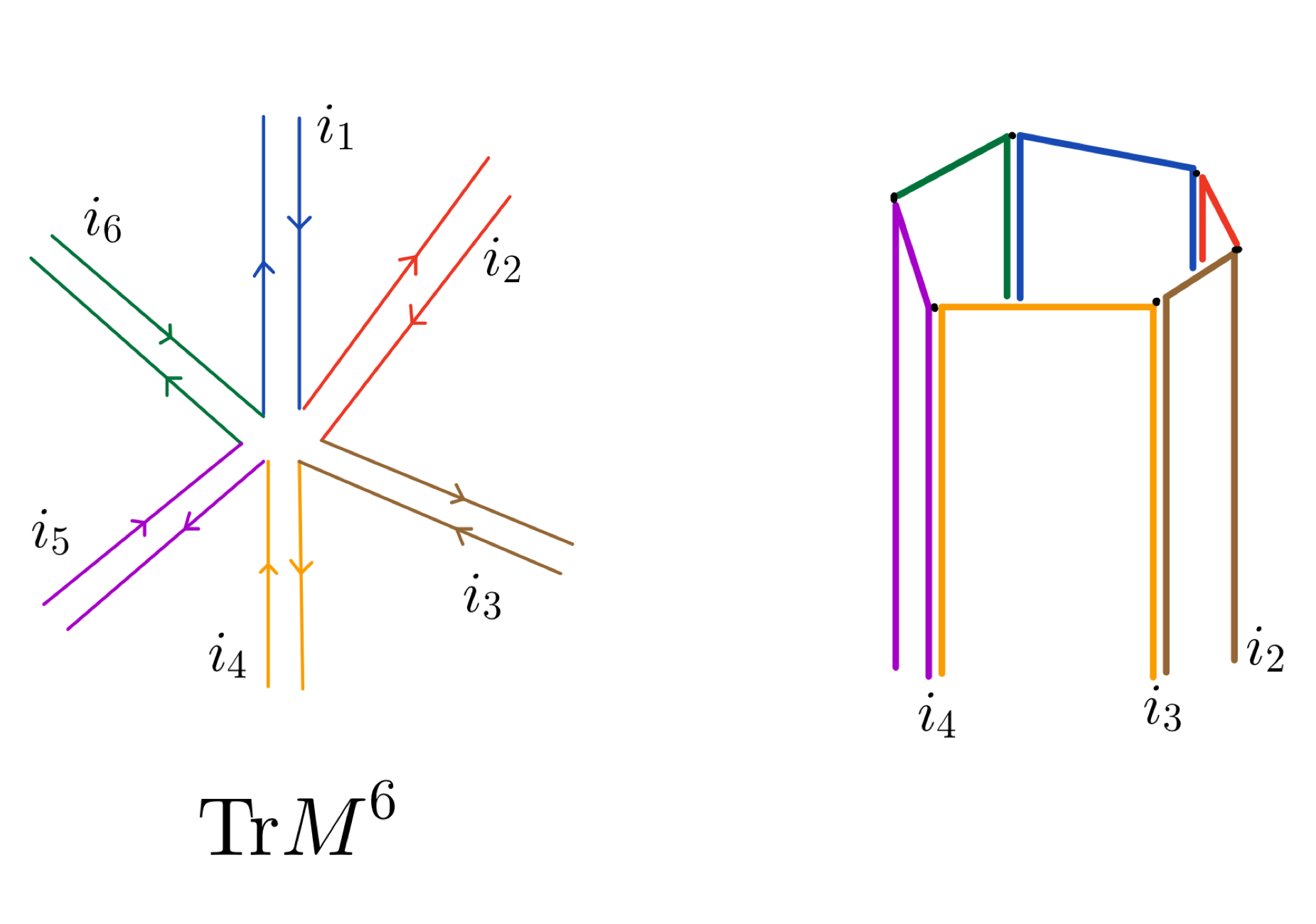}
        \caption{\small{\textbf{String bit picture in V-type duality:} Each open strip originates as a ribbon of the gauge theory Feynman diagram and can be viewed as the worldsheet of a string bit. Single trace operators creating asymptotic closed string states consist of as many string bits as there are matrix elements in the trace.}}
        \label{fig:stringbits}
\end{figure}

In the $0$-dimensional gauge theories considered here, each string bit is assigned equal weight, which is simply the trivial propagator of the matrix model. This is the string bit interpretation of Razamat's prescription \cite{razamat2010matrices} for an integer Strebel length proportional to the number of Wick contractions or propagators. 
More precisely, these assigned lengths correspond to the width of the flat panels (shown in Fig. \ref{fig:stringbits}) making up the cylindrical worldsheet of the asymptotic closed string. 
In analogy to a bicycle chain, each closed string consists of finitely many rigid links. It can therefore assume only a restricted set of configurations. This is fundamentally the origin of the localization of the string path integral onto discrete points on the moduli space which we see manifest in the A-model description. In a BMN-like (or Gross-Mende) limit, where we take the number of operators within each trace to be large, the string consists of many, many such bits. It can now take on practically any shape and we cover the moduli space densely. In our companion papers, we will examine the BMN limit explicitly, both from an A- and B-model perspective.

We can also ask how the above BMN-limit is related to the standard double-scaling limit. Traces of large (even) numbers of matrices are dominated by their largest eigenvalues. Hence, these correlators probe the vicinity of the edges of the large $N$ eigenvalue distribution. But we know that this is exactly what double-scaling also achieves. We will see this more explicitly in \cite{DSDII, DSDIII}.

One might wonder in what sense the string-bit discretization is related to the `old matrix model' picture of the worldsheet discretization by Feynman diagrams. There, one obtained a discrete tessellation by considering the graph dual to the Feynman diagram. All the curvature is then located at vertices (or equivalently, faces of the original diagram). In the Strebel gauge for the metric Eq.~(\ref{eq:StrebelMetric}), the curvature is instead localized to \textit{both} the zeroes and the poles of the Strebel differential and thus shared amongst vertices and faces. In some rough sense, topological strings in Strebel gauge function as an analogue of light cone gauge.


\item \textbf{An A-model picture of the matrix model?} 

We would like to suggest a picture of the matrix models of Sec. \ref{sec:twostacks}, in terms of the open strings of the D-branes in our A-Model dual (see for e.g.  \cite{Ribault:2003ss,fotopoulos2005d}). This is therefore to be understood in a probe-brane limit. 
\begin{figure}[h!]
     \centering
      \begin{subfigure}[b]{0.4\textwidth}
         \centering
         \includegraphics[width=\textwidth]{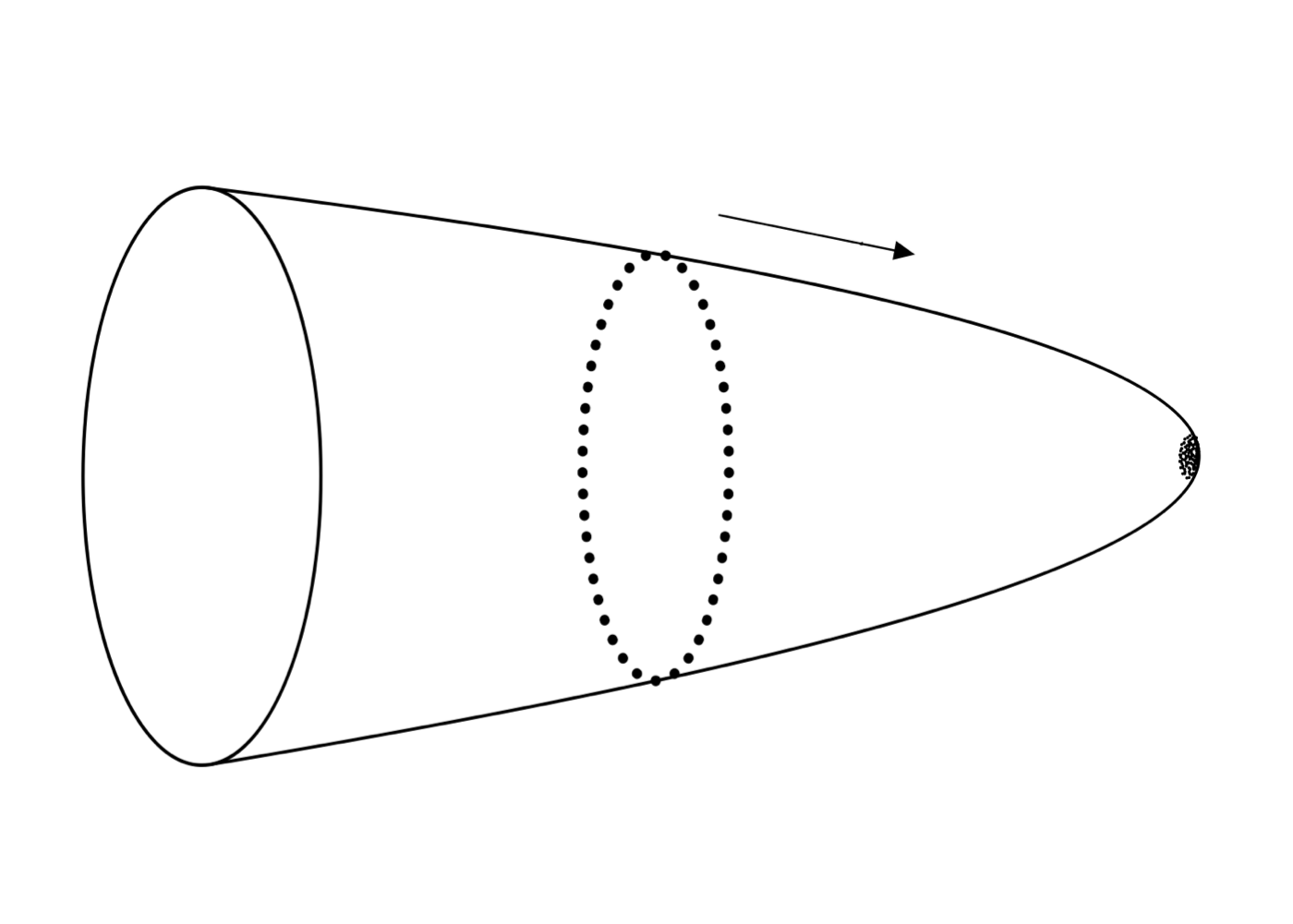}
         \caption{"Angular D1'" (D0-brane)}
         \label{fig:angularD1}
     \end{subfigure}
     \begin{subfigure}[b]{0.4\textwidth}
         \centering
         \includegraphics[width=\textwidth]{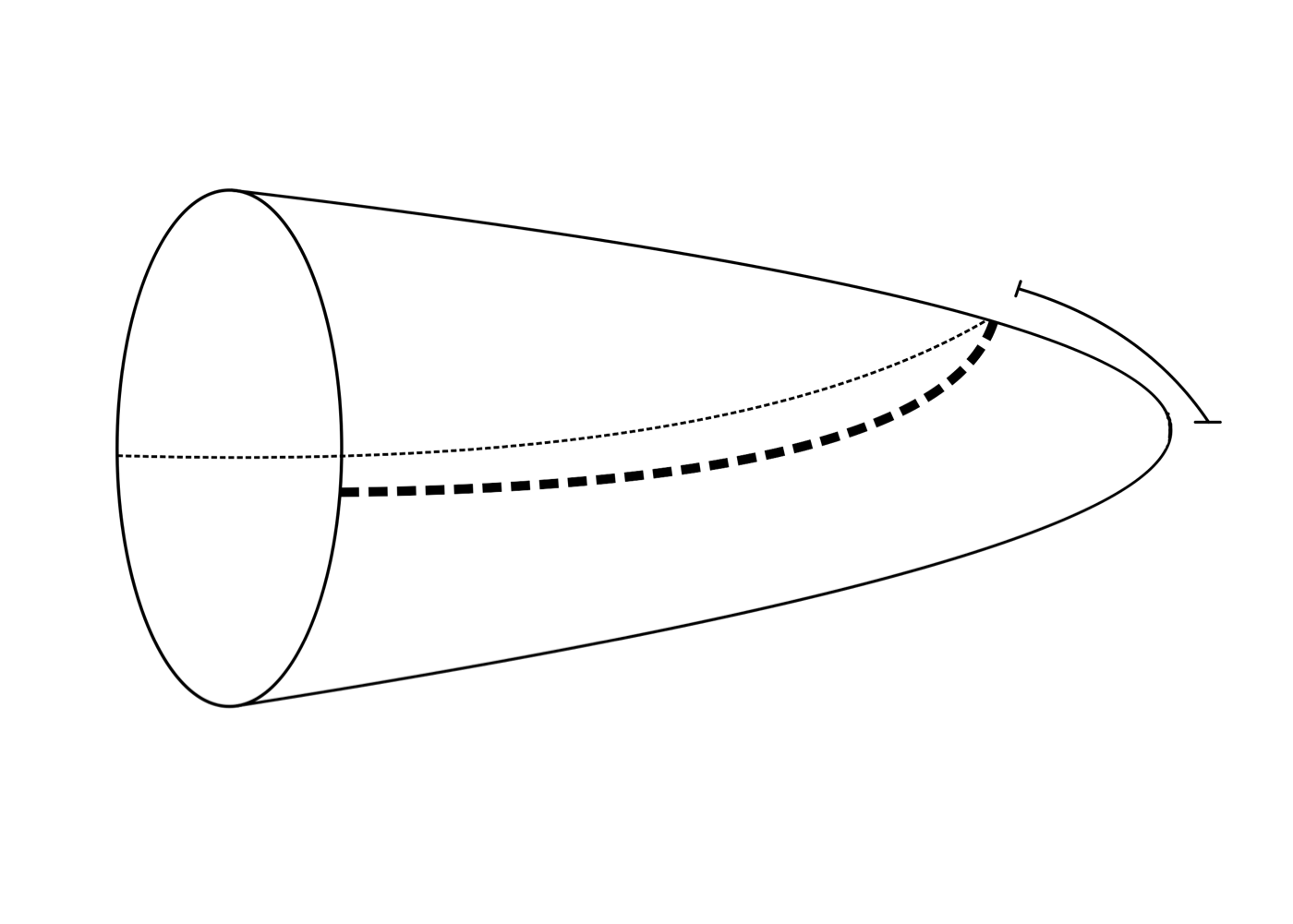}
         \caption{Radial D1-brane}
         \label{fig:radialD1}
     \end{subfigure}
        \caption{A-model Branes: The two relevant stable branes on the cigar geometry defined by the $SL(2,\mb{R})_1/U(1)$ coset include the D0-branes localized at the tip, and the D1-branes extended in the radial direction. It is helpful to think of the D0's as a D1 branes extended in the angular direction, which for energetic reasons have shrunk to the tip.}
        \label{fig:AngandRadD1s}
\end{figure}
We view the compact branes of Sec. \ref{sec:twostacks} as D-branes wrapping the angular $S^1$ of the cigar geometry. They shrink to the tip, so as to minimize their energy - see Fig.\ref{fig:angularD1}. These are the D0 branes discussed in \cite{Ribault:2003ss,fotopoulos2005d}. The fact that they have no moduli is reflected on the matrix model side in that the partition function of the $N \times N $ matrix model does not depend on the source $Y$ (if no other determinant insertions are present). These are the analogue of ZZ branes in the usual minimal string theories. Note that these compact branes naturally live in the strong coupling region. We believe this to be a rather generic feature of V-type duals. It gels well with our intuition from ZZ branes in Liouville theory, which are also localized behind the Liouville wall. We generically expect these V-type duals to be the natural description for strongly coupled (large 't Hooft coupling) open string theories.

The non-compact branes are the radial D1 branes and correspond to the FZZT branes from the point of view of the Liouville direction. As shown in Fig. \ref{fig:radialD1}, they come in from infinity until they reach a minimal radial distance from the tip.  We propose to identify the eigenvalues $x_a$ of the external source $X_{ab}$ with this minimal distance of the D1 to the tip. Semi-classically, it is clear that these branes want to remain in the weak coupling region, of large $x_a$, so as to minimize their length. This feature was also seen in the Kontsevich model, where the free energy was given in a  perturbation theory around large source terms (see Thm. 1.1 of \cite{KontsevichAiry}). While we have not been able to make a precise argument yet, we expect that F-type duality is the natural description for weakly coupled (small 't Hooft coupling) open string theories.

\begin{figure}[h!]
\centering
\includegraphics[width=0.5\textwidth]{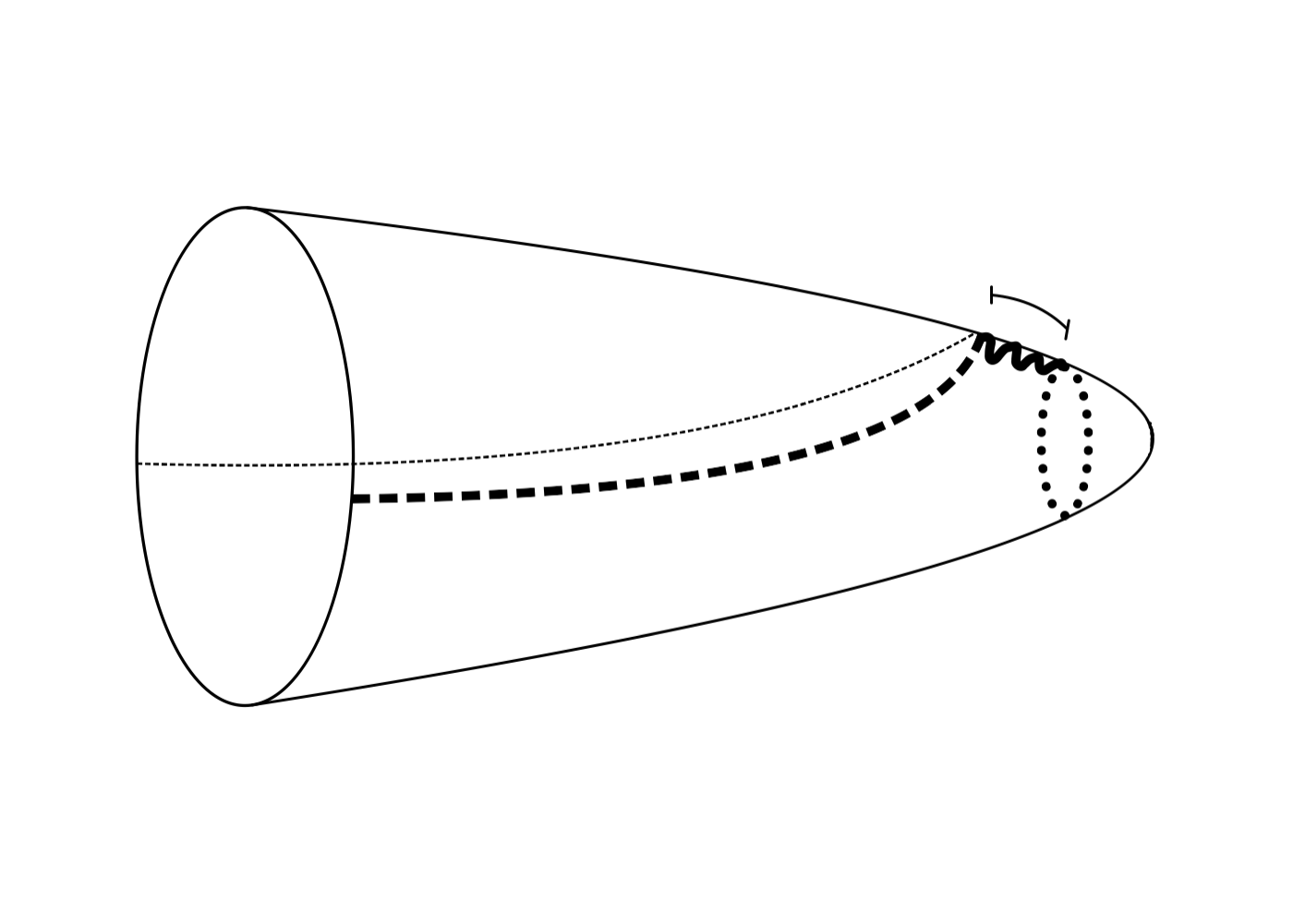}
\caption{\small{\textbf{D1-D1' Boundstate}: We propose the following interpretation of the matrix model duality in terms of a set  of radial and angular D1 branes, and the various open strings between them. The fermionic open strings, denoted here by the wavy line connect radial and angular D1s. In a way, they pull the D0's away the tip, making them into fully fledged D1 branes (and vice versa the D0 branes pull the D1 branes in towards the tip). We view their mass term $(x_a - y_i)$ as measure of the minimal distance between the two branes, as indicated by the small interval symbol above. Cf. Fig. \ref{fig:2stacks}}}
\label{fig:D1D1bound}
\end{figure}

The fermionic theory describes the open strings between the compact and non-compact branes. In section \ref{sec:twostacks}, we proposed that the mass, $(x_a-y_i)$, of the fermions $\psi_{ia}$ should provide a proxy for the length of the string. From a semi-classical A-model perspective, it appears that these fermionic strings can pull in the compact branes a distance $y_i$ away from the tip, as depicted in Fig. \ref{fig:D1D1bound}. The former D0 branes are now honest angular D1 branes, with moduli $y_i$. This would explain the dependence, on both the $x_a$ and $y_i$, of the exact brane wavefunctions computed in Sec. \ref{sec:exactwfns}. For now, these are suggestive interpretations. It should, however, be possible to test these predictions by methods similar to those used in \cite{fotopoulos2005d} to probe the spectrum of open strings stretched between two radial D1-branes. 

 \item \textbf{Relation to the Dijkgraaf-Vafa conifold proposal}
 
    Dijkgraaf and Vafa proposed the Gaussian matrix model (or more generally a one Hermitian matrix model) as an open string dual to the closed string B-model on the deformed conifold (and its generalisations) \cite{DijkgraafVafa02,DVongeom02}. As we review in appendix \ref{sec:holCSarg}, the open string field theory on $N$ D-branes wrapping a $\mathbb{P}^1$ in the resolved conifold geometry reduces to a matrix integral with a quadratic potential. The resulting matrix model's spectral curve, of the form $f(x,y)=0$, encodes the emergent closed string geometry. More precisely, the dual is believed to be a non-compact Calabi-Yau, embedded in $\mathbb{C}^{4}$ by the algebraic relation
    \begin{equation} \label{eq:cy3target}
    uv-f(x,y)=0.
    \end{equation}
    For the Gaussian, an appropriate choice of variables indeed gives the equation for the deformed conifold.  
    The dimensional reduction does not capture local operators in the open-string gauged $\beta \gamma$ system on the $\mathbb{P}^1$, which is an intermediate step in the derivation of the matrix integral from the OSFT. Hence, correlators in the Gaussian matrix model should only capture a subset of observables of the dual closed string on the deformed conifold. This is indeed what Costello and Gaitto find \cite{CostelloGaiottoTH}. In \cite{Budzik:2021fyh}, Budzik and Gaiotto identify the precise subset of correlators computable by the matrix model. 
    
    How does the B-twisted sigma-model with target space given by Eq.~(\ref{eq:cy3target}) relate to the Landau-Ginzburg description we have explored in this paper? In \cite{Aganagic_2005} -  see Eq.~(5.7) - it was suggested that the B-model with LG superpotential $W(y)$ corresponds to a target space geometry via 
    \begin{equation}\label{eq:CY-superpot}
        uv-(x-W(y))=0 .
    \end{equation}
    This was motivated by the LG/Calabi-Yau correspondence for strings propagating on ADE singularities \cite{MartinecManual,vafa1989catastrophes}. 
    In our case with $W(y)=\frac{1}{y}+t_2y$, we see on substitution into Eq.~(\ref{eq:CY-superpot}), that (for $y\neq 0$), we have the equation
    \begin{equation}
        uvy-(xy-t_2y^2)=1 .
    \end{equation}
    After a redefinition of variables $uy \rightarrow u$ and a shift of $(x,y)$ this is the equation of a deformed conifold\footnote{In \cite{OV2002worldsheet, okuda2004d}, the superpotential $W(Z)=1/Z$ was proposed as the LG description of the deformed conifold (i.e. without the background momentum deformation which gives the linear term). This appears to be somewhat in tension with our result. It would be desirable to clarify the exact relationship with these proposals.}. 
    
\item \textbf{Relation to JT Gravity}   

    Matrix models of two-dimensional gravity have enjoyed a recent resurgence, prompted in large part by the reformulation of Jackiw-Teitelboim gravity as a matrix integral in \cite{SSSJTmatrix}. There are three important differences between such  models and the open-string descriptions we have discussed here. Firstly, those theories are studied in the double-scaling limit. Secondly, as reviewed in \cite{CJohnsonReview}, the matrix there corresponds to a representative Hamiltonian, drawn from an ensemble. The dimension of the Hilbert space dictates the size of the matrix. Finally, the emergent two-dimensional gravity is that of the target space, and not the worldsheet of a string. However, the authors of \cite{EVerlindeJTstring} applied the topological string point of view advocated here to the matrix models of JT gravity, and recast the Kodaira-Spencer \cite{KodairaSpencer} string field as a 'third-quantized' universe field theory.

    \item \textbf{Some speculations on an F-type dual to IIB strings on $AdS_5 \times S^5$}

Open-closed-open triality suggests there should exist both a V- and F-type dual to the same closed string theory. It immediately begs the question as to what the `other' open string description in the well-known holographic correspondences might be. In particular, one might ask what an F-type dual to IIB string theory on $AdS_5 \times S^5$ could look like. 

As discussed above, the known examples of F-type duality seem to arise as weakly coupled open string theories on non-compact branes. In \cite{witten2004perturbative}, Witten proposed a twistor string description of perturbative amplitudes in $\mathcal{N}=4$ super Yang-Mills theory. In fact, it shares striking structural similarities with the simple topological strings we have considered in this paper.
For example, the twistor string description also descends from a B-model open string field theory on $N$ space-filling branes, which again reduces to a holomorphic Chern-Simons theory - for a very nice review, see \cite{cachazo2005lectures}. Reproducing the full set of interactions in ${\mathcal N}=4$ SYM requires the introduction of additional $D1$ branes. As shown in \cite{mason2008heterotic}, the open-string field theory on the D1 branes reduces to a gauged $\beta \gamma $ system, while the strings joining the compact and non-compact branes are again fermionic. Integrating these out is tantamount to the insertion of determinants (and anti-determinants). 

Heckman and Verlinde \cite{Heckman:2011ju,Heckman:2011qu,VerlindeHeckAmplitudesGaussian} have, in fact, derived a very interesting matrix model descending from this holomorphic Chern-Simons description. The natural observables in F-type duals are not the single trace operators familiar from V-type gauge-string duality. This might well be the underlying reason for the rather complicated expressions given in, for instance, \cite{chicherin2015correlation} to describe (chiral) stress-energy tensor correlation functions in $\mathcal{N}=4$ SYM. It also perhaps provides a rationale for the natural twistor-string description of the closed string worldsheet dual to free  $\mathcal{N}=4$ SYM proposed by the first author and M. Gaberdiel \cite{gaberdiel2021worldsheet,gaberdiel2021string}. We hope to make this set of ideas more concrete in future work. It would also be interesting to connect this with the open string description on giant gravitons as studied by \cite{komatsuOCO} - see comments in Sec. \ref{sec:twostacks}.


\item \textbf{Embedding in AdS/CFT}

    The perspective taken in this paper follows the conventional holographic framework of open/closed string duality. As exhibited in Fig. \ref{sec:bigpicture}, many of our results embed nicely in the broader context of AdS/CFT. 
    The Gaussian matrix model (and the related free matrix quantum mechanics) encodes a particular half-BPS sector $\mathcal{N}=4$ SYM \cite{erickson2000wilson, drukker2001exact, Lin:2004nb, berenstein2004toy,itzhaki2005large,pestun2012localization,okuyama2018connected}. In a certain sense, the worldsheet theories we propose here are dual to this subsector and should presumably find a nice embedding in the full bulk string theory as well. A natural class of observables, suggested by this embedding, are then ${\rm Tr}_R(e^M)$ (in different representations $R$). The Gaussian model has also been used to calculate certain 1/8 BPS Wilson loops in \cite{giombi2010correlators,giombiKomatsu2018exact}, leaving open the possibility to extend our methods to less supersymmetrically protected sectors of the correspondence. 
    
    Returning, however, to the half BPS sector, Fig. \ref{fig:biggerpicture} sketches a more systematic way in which the Gaussian matrix model captures a topological subsector of the 2d chiral algebra associated to $\mathcal{N}=4$ SYM \cite{bonettiRastelli2018}. We have already commented above on the twisted holography program of \cite{CostelloGaiottoTH} and how the matrix model correlators capture a further subset of this chiral algebra subsector. Since the latter is holographically dual to the B-model string theory on the deformed conifold $SL(2,\mathbb{C})$, we expect an embedding of our B-model description in that of \cite{CostelloGaiottoTH}. For the most part, \cite{CostelloGaiottoTH} frames the duality in terms of the spacetime Kodaira-Spencer field theory, which plays the role of the bulk closed string field theory. It would be very interesting to understand this duality from an A-model perspective as well as from the worldsheet, which would more easily connect to our results here. 

    In fact, the close resemblance between the group manifold $SL(2,\mathbb{C})$ and $AdS_3 \times S^3$ might give some clues as to the appearance of the (twisted) coset $SL(2,\mathbb{R})/U(1)$ at level one. The level one ${\mathfrak sl}(2, {\mathbb R})$ theory took center stage in the derivation of  $AdS_3/CFT_2$ in the tensionless limit \cite{eberhardt2019worldsheet, eberhardt2020deriving}. 
    The dual symmetric product orbifold correlators also exhibits a very similar connection to integer Strebel differentials as in the Gaussian matrix model \cite{gaberdiel2021symmetric}. Thus it is natural to try and relate our A-model picture to that of tensionless $AdS_3 \times S^3$.
      
\end{itemize}

\section*{Acknowledgments}
We greatly benefited from insightful discussions with many people, including David Gross, Igor Klebanov, Juan Maldacena, Steve Shenker, Erik Verlinde, Herman Verlinde, Spenta Wadia and other participants of the very stimulating HirosiFest@Caltech.

R.G. acknowledges enlivening discussions wth Sumit Das, Abhijit Gadde, Gautam Mandal, Shiraz Minwalla, Onkar Parrikar, Sandip Trivedi and other members of the String group at TIFR, during a seminar on this work. He also thanks ASICTP, Trieste, for hospitality while completing the final stages of this work. He would like to thank Matthias Gaberdiel, Pronobesh Maity and Ashoke Sen for many discussions on related matters.  He thanks the people of India for their continuing support for basic sciences. His work is particularly supported by the project RTI4001 of the Dept. of Atomic Energy, Govt. of India, as well as by a J. C. Bose Fellowship of the DST-SERB. 

E.M. would particularly like to thank Bruno Balthazar, Emil Martinec and Herman Verlinde for collaborations in the very early stages of this work, as well as the fruitful questions from the high energy groups at MIT and Stanford after early presentations of these results. E.M. is also grateful for extensive conversations with, and many useful suggestions regarding the material presented here from, Anthony Ashmore, Daniel Brennan, Clay C\'ordova, Taro Kimura, Manki Kim, Jorrit Kruthoff, Yuri Lensky, Daniel Ranard, Pavel Wiegmann and Barton Zwiebach. He would also like to thank Bertrand Eynard, Nicolas Orantin and Chiu-Chu Melissa Liu for correspondence. E.M. is supported by a Kadanoff fellowship at the University of Chicago, and would like to acknowledge the support of the Jones Endowment for Physics Research.

\pagebreak

\appendix

\section{Localization to the 2-matrix model: A B-model perspective} \label{sec:holCSarg}
In order to interpret the matrix model equivalence in Eq.~(\ref{eq:source/det duality}) in terms of an open-open string duality between V- and F-type descriptions, we first need to show in what way these simple matrix integrals descend from a system of branes. We will sketch a rough picture thereof in this section. The following argument is due to Dijkgraaf \& Vafa in \cite{DijkgraafVafa02}, see also \cite{marino2004houches} for a more pedagogical derivation. 
We begin from a B-Model topological string perspective. A proposal for the possible A-model picture is given in section~\ref{sec:discussion}.

Consider the following non-compact Calabi-Yau threefold $\mc{X}$, consisting of two holomorphic line bundles over the sphere $\mathbb{P}^1$ \footnote{The Calabi-Yau condition $c_1(\mc{X})=0$ is satisfied for this choice of bundle since $c_1(\mc{X}) = c_1(\mc{O}(-a))+ c_1(\mc{O}(a-2))+ c_1(\mathbb{P}^1)=(-a)+(a-2)+2=0$.}
\begin{equation}
    \mc{O}(-a) \oplus \mc{O}(a-2) \rightarrow \mathbb{P}^1.
\end{equation}
We now consider wrapping $N$ compact D2-branes\footnote{Note that in topological string theory, a D$p$ brane is conventionally defined to a have a $p$-dimensional worldvolume, and not a $p+1$-dimensional one, as in the non-topological context.} on the $\mb{P}^1$. From \cite{WittenCSasOSFT}, we know that the open string field theory on these $N$ branes localizes exactly to the holomorphic Chern-Simons theory
\begin{equation}
    S_{\text{OSFT}}= \frac{1}{2g_s} \int_{\mc{X}} \Omega \wedge \Tr_{N} \left( \mc{A} \wedge \bar{\partial } \mc{A} + \frac{2}{3}\mc{A} \wedge \mc{A} \wedge \mc{A} \right),
\end{equation}
where $\Omega$ is the holomorphic three-form on $\mc{X}$. The open string degrees of freedom, stretched between the $N$ compact branes, are in the gauge field $\mc{A}$. As is standard, we decompose $\mc{A}$ into longitudinal and transverse pieces $\mc{A}_{\mu \in z,0,1}=(A_{//},\Phi_0,\Phi_1)$. Here $z$ is the coordinate on the $\mb{P}^1$, while $0,1$ denote the two normal directions. 

Start with the case of $a=0$, i.e. an $ \mc{O}(0) \oplus \mc{O}(-2) \rightarrow \mathbb{P}^1$ bundle, $\Phi_0$ is simply a scalar, while $\Phi_1$ is a $(0,1)$ form. Plugging in the above decomposition of $\mc{A}$ into the holomorphic Chern-Simons action, we find that it simplifies to
\begin{equation}
    S_{\text{OSFT}} = \frac{1}{g_s} \int_{\mb{P}^1} \Tr_{N} \Phi_1 \bar{D}_{A_{//}} \Phi_0,
\end{equation}
where $\bar{D}_A{_{//}} = \bar{\partial} + [A_{//},-]$. This may be viewed as a gauged $\beta \gamma$ system. 

It turns out that for more complicated bundles, i.e. the cases of $ \mc{O}(-a) \oplus \mc{O}(a-2) \rightarrow \mathbb{P}^1$ with $a \neq 0$, we need only add an appropriate superpotential to this action (see sections B.1 and B.2 of \cite{SuperpotforCS} for an explanation as to how exactly this encodes deformations of the complex structure). 
\begin{equation}
 S_{\text{OSFT},a\neq 0} =  \frac{1}{g_s} \int_{\mb{P}^1}  \Tr_{N} \big(\Phi_1 \bar{D}_{A_{//}} \Phi_0+ \omega  W(\Phi_0)  \big),
\end{equation}
where $W(\Phi_0)$ depends on the precise bundle, and $ \omega$ is a normalized volume $(1,1)$ form on $\mb{P}^1$. Varying the action with respect to $\Phi_1$ sets 
\begin{equation}
     \bar{D}_{A_{//}}\Phi_0=0.
\end{equation}
On the sphere, since $\Phi_0$ is a scalar, this translates to the stronger statement that $\Phi_0$ is a constant matrix which we will call $K$ 
\begin{equation}
    (\Phi_0)_{ij} = K_{ij}.
\end{equation}
Varying $\Phi_0$ instead gives 
\begin{equation}
    \bar{D}_{A_{//}} \Phi_1 = W'(\Phi_0)\, \omega .
\end{equation}
Since $\Phi_0$ is constant, we see that  $ * \bar{D}_{A_{//}} \Phi_1 $ must also be constant
\begin{equation}
    (* \bar{D}_{A_{//}} \phi_1)_{ij} = M_{ij}.
\end{equation}

This suggests that the open string field theory on the $N$ compact branes can be reduced to  
\begin{equation}
   Z_{OSFT} = \frac{1}{Vol U(N)}\int dK dM e^{\frac{1}{g_s} \Tr_{N} \left( V_p(K)-K M \right) },
\end{equation}
where $p$ denotes the order of the potential for $K$ (to be clear, $V_p(K)$ should be identified with the superpotential $W$ above). A more careful treatment is certainly needed to make this argument precise. The main takeaway here is that the two matrices describe the two transverse directions of the open strings, analogously to the six real scalars in the standard holographic example of $\mc{N}=4$ SYM.

Now that we have reduced the full open string field theory to a $0$-dimensional integral, it is worth getting better acquainted with these matrix models. First off, we highlight that $M$ appears only linearly in the action via the interaction term $\Tr K M$. We can therefore perform the integral over $M$, which gives rise to - after a suitable choice of contour - a $\delta$-function enforcing $K_{ij}=0$.  
\begin{equation}
     Z_{OSFT} = \frac{1}{\text{Vol}  U(N)} \int dK e^{\frac{1}{g_s} \Tr_N V(K)} (2\pi g)^{N^2} \delta(K) =  \frac{(2\pi g)^{N^2} e^{\frac{1}{g_s} \Tr_N V(0)} } {Vol (\ U(N))}.
\end{equation}
Note that we can add a constant to the superpotential to set $V(0)=0$. We will use this freedom to simplify some later expressions. If the superpotential further satisfies $V'_p(0)=0$, then $\Braket{M_{ij}}=0$ \footnote{There is no real justification for such saddle point equations, directly at the level of matrix entries. We present them here so as to build some intuition for the model.}. Indeed, the equations of motion read $M_{ij}=V'_{p}(K)_{ij}$ and $K_{ij}=0$. Note that we could also consider a non-zero background for $M_{ij}$ by turning on a source for $K$
\begin{equation}
    Z_{N}(Y)= \frac{1}{\text{Vol} (U(N))}\int dK dM e^{\frac{1}{g_s} \Tr_{N} \left( V_p(K)-K(M-Y) \right) }
\end{equation}
so that now the variation of the action with respect to $K_{ji}$ gives $M_{ij}=V'_{p}(K)_{ij} +Y_{ij}$. If $V'_p(0)=0$, this gives $\Braket{M_{ij}}=Y_{ij}$. However, if no other operator insertions are present, $Z_{N}(Y)$ is in fact independent of $Y$. This can be seen in two ways. Firstly, the simple linear appearance of $M$ in the action means that we could simplify shift $M$ by $Y$ and recover the `bare' partition function. Else, we could do the integral over $M$, which again enforces the constraint $K_{ij}=0$, so that the term proportional to $\Tr KY$ simply vanishes. As we see in Sec. \ref{sec:twostacks}, things change drastically in the presence of other branes (i.e. determinant insertions for $M$). In fact, the eigenvalues of $Y_{ij}$ will become open string moduli for the $N$ compact branes. This will again have an interesting interpretation from the A-model side, see Section \ref{sec:discussion}.  
Finally, note that when $p=2$, we can first do the integral over $K$ instead. This gives the Gaussian one-matrix model.

\section{Bi-orthogonal polynomials \& their duality} \label{sec:appPolys}

\subsection{Construction of the polynomials}

For the two-matrix model, we can construct bi-orthogonal polynomials satisfying 
\begin{equation} \label{eq:orthpoly}
    \int dk db e^{+\frac{1}{g}\left( V(k)-km \right)} P_{i}(k) Q_{n}(j) = h_{i} \delta_{i j },
\end{equation}
normalized such that $P_j(k) = k^{j}+ \mc{O}(k^{j-1})$ and $Q_{j}(m)=m^j+ \mc{O}(m^{j-1})$. 
Following \cite{Hashimoto_2005}, we can write these bi-orthogonal polynomials in our case fully explicitly
\begin{align}
     P_{n}(k)=k^{n}\\
    Q_{n}(m)=[(+g \partial_{z})^{n} e^{-\frac{1}{g}\left( V(z)-zm)\right)}]_{z=0}.
\end{align}
Here we are assuming that we have normalized the potential (i.e. chosen an appropriate "integration" constant), such that $V(0)=0$.

We can show these satisfy Eq.~(\ref{eq:orthpoly}) as follows:
\begin{align}
     \int dk dm e^{+\frac{1}{g}\left( V(k)-km \right)} P_{n}(k) Q_{j}(m) \\
     & = \int dk dm e^{+\frac{1}{g}\left( V(k)-km \right)} [k^n][(g \partial_{z})^{j} e^{-\frac{1}{g}\left( V(z)-mz)\right)}]_{z=0}\\
     & = \int dz \int dk dm e^{+\frac{1}{g}\left( V(k)-km \right)} [k^m]\delta(z) [(g \partial_{z})^{j} e^{-\frac{1}{g}\left( V(z)-mz)\right)}] \\
     & = \int dz \int dk dm e^{+\frac{1}{g}\left( V(k)-km \right)} [k^m] [(+g \partial_{z})^{j} \delta(z)] [e^{-\frac{1}{g}\left( V(z)-mz)\right)}].
\end{align}
Now we can do the integral over $m$
\begin{equation}
    \int dm e^{\frac{1}{g}m(k-z)}= (2 \pi g ) \delta(k-z)
\end{equation}
and with this resulting delta-function, we can also trivially do the integral over $k$. We see the potential pieces cancel (this is the origin of the sign flip in the in front of the potential piece in the expression for $Q_{n}(m)$)
\begin{align}
     \int dk dm e^{+\frac{1}{g}\left( V(k)-km \right)} P_{n}(k) Q_{j}(m)  = & (2\pi g) \int dz e^{+\frac{1}{g}\left( V(z) \right)} [z^n] [(g \partial_{z})^{j} \delta(z)] [e^{-\frac{1}{g}\left( V(z))\right)}] \\
     =  & (2\pi g) \int dz  [z^n] [(+g \partial_{z})^{j} \delta(z)]\\
     = & (2 \pi g) \int dz [(g \partial_{z})^{j} z^{n}] \delta(z) \\
     = & (2 \pi g) g^{n} n! \delta_{n j}.
\end{align}

\subsection{Duality of the polynomials}

In the $K,M$ matrix model calculations, we encounter the following integral
\begin{equation} 
  \int dk dm e^{+\frac{1}{g}\left( V(k)-k(m-y) \right)} Q_{n}(m).
\end{equation}
In the 1-Matrix case with source, this is the equivalent of Fourier-transforming the orthogonal polynomial weighted by the potential. 
By almost identical manipulations as the proof of bi-orthogonality, one shows that 
 \begin{equation} 
P_{n}(y) = \frac{1}{(2\pi g ) }\int dk dm e^{-\frac{1}{g}\left( V(k)-k(m-y) \right)} Q_{n}(m).
\end{equation}
The integral with the opposite sign for the potential in the dual matrix model acts as an "inverse kernel" taking $P_{n} \rightarrow (2\pi g)Q_{n}$.
\begin{align}
     \int da db e^{-\frac{1}{g}\left(V(a) +a(b-x)\right)} P_{n}(b) = & \int da  e^{-\frac{1}{g}\left(V(a) -a x)\right)}\int db e^{-\frac{1}{g}ab}b^n \\
     = & \int da  e^{-\frac{1}{g}\left(V(a) -a x)\right)} (-g \partial_{a})^{n}\int db e^{\frac{1}{g}a b} \\
     = & \int da \left[(+g \partial_{a})^{n}e^{+\frac{1}{g}\left(V(a) -a x)\right)} \right] (2 \pi g) \delta(a)\\
     =& ( 2\pi g) Q_{n}(x).
\end{align}
Deep down, this is the mechanics underlying the duality from the point of view of orthogonal polynomials. 

\subsection{Compact branes \& their wavefunctions} \label{sec:FZZTcalcs}

Consider the following matrix integral which we may interpret as the wavefunction for $Q$ compact branes:
\begin{equation} \label{eq:FZZTstart}
   \braket{\prod_{c=1}^{Q} \det(y_c -K)} = \frac{1}{Z_{N}}\int dK dM_{N \times N} e^{-\frac{1}{g} Tr \left( V(K)-KM \right)} \prod_{c=1}^{Q} \det(y_c -K).
\end{equation}
We wish to prove the following identity:
\begin{equation}
\braket{\prod_{c=1}^{Q} det(y_c -K)} = \frac{\det_{c,d}\left[P_{c+N-1}(y_{d})\right]}{\Delta(y)}. 
\end{equation}

Starting with Eq.~(\ref{eq:FZZTstart}) and using the HCIZ-integral to do the integral over the relative unitary between $K$ and $M$, we can obtain an expression purely in terms of the eigenvalues of $K$ and $M$ \footnote{The HCIZ integral reads $\int dU e^{+\frac{1}{g} \Lambda_{K} U \Lambda_{M} U^{\dagger}} = C_{N,g}\frac{\det\left[e^{k_{j}m_{l}}\right]}{\Delta(k)\Delta()}$. Given this is multiplying a the remaining Van-der-Monde which is completely anti-symmetric, we can trade the $\det\left[e^{k_{j}m_{l}}\right]$ for $e^{\sum_{i}k_{j}m_{j}}$ under the integral sign, with an additional factor of $N!$, see \cite{Morozov:1995pb}. We follow the conventions of \cite{Zubernotes}, so that the constant $C_{N,g}$ reads $C= g^{N(N-1)/2}\prod_{j=1}^{N-1}j!$.}.
\begin{align}
       \braket{\prod_{c=1}^{Q} \det(y_c -K)} = & \frac{1}{Z_{g,N}} \int \prod_{i=1}^{N} dk_{i} dm_{i} e^{+\frac{1}{g}\left( V(k_{i})- k_{i}m_{i} \right)} \Delta(k)^2 \Delta(m)^2 \frac{C_{N,-g} N!}{\Delta(k) \Delta(m)} \prod_{c=1}^{Q} \prod_{j=1}^{N} (y_{c}-k_j). 
\end{align}
Note we are left with a single factor of the Van-der-Monde determinant for both $K$ and $M$. 
The idea is now to combine the $N$ $k_{i}$ and $Q$ $y_{c}$ eigenvalues into some joint object, an $N+Q$ dimensional vector $z_{\alpha}$.
\begin{equation}
    z_{\alpha} = k_{i} \, \thinspace \text{for} \thinspace \alpha \in 1,..,N;  \quad   z_{\alpha} = y_{c} \thinspace  \text{for} \thinspace  \alpha \in N+1,..,N+Q. 
\end{equation}
Then note that\footnote{For Q=1, i.e. the case just one brane insertion, we have simply $ \Delta(b) \prod_{j=1}^{N} (y-a_{j}) = \prod_{\alpha < \beta}(z_{\beta}-z_{\alpha})$.} 
\begin{equation} \label{eq:jointvdmy}
    \Delta(b) \prod_{c=1}^{Q} \prod_{j=1}^{N} (y_{c}-k_{j}) =  \frac{\prod_{\alpha < \beta}(z_{\beta}-z_{\alpha})}{\Delta(y)}.
\end{equation}
and recall the fact that we can rewrite the Van-der-Monde in terms of the orthogonal polynomials
\begin{equation}
    \Delta(k)=\det\left[Q_{i-1}(k_{j})\right]; \quad \quad \frac{\prod_{\alpha < \beta}(z_{\beta}-z_{\alpha})}{\Delta(y)} = \frac{\det\left[P_{\alpha-1}(z_{\beta})\right]}{\Delta(y)}
\end{equation}
so that our integral becomes 
\begin{align}
     \braket{\prod_{c=1}^{Q} det(y_a -K)} = & \frac{ C_{N,-g} N!}{\Delta(y) Z_{N}} \int \prod_{i=1}^{N} dk_{i} dm_{i} e^{+\frac{1}{g}\left( V(k_{i})- k_{i}m_{i} \right)}  \det(Q_{i-1}(m_{j})) \det(P_{\alpha-1}(z_{\beta})) \\
     = & \frac{ C_{N,-g} N!}{\Delta(x) Z_{N}} \int \prod_{i=1}^{N} dk_{i} dm_{i} e^{+\frac{1}{g}\left( V(k_{i})- k_{i}m_{i} \right)} \epsilon^{j_{1}...j_{N}} Q_{j_{1}-1}(m_{1})...Q_{j_{N}-1}(m_{N}) \\
     & \times  \epsilon^{\alpha_{1}...\alpha_{N+Q}} P_{\alpha_{1}-1}(k_1)...P_{\alpha_{N}-1}(k_N)P_{\alpha_{N+1}-1}(y_1)...P_{\alpha_{N+Q}-1}(y_Q).
\end{align}

We can pull out all the polynomials $P$ which depend on $y$ outside of the integral and then use the bi-orthogonality condition in each factor, so that 
\begin{align}
    \braket{\prod_{c=1}^{Q} det(y_c -K)} = & \frac{ C_{N,-g} N!}{\Delta(y) Z_{N}} (\prod_{i=0}^{N-1}h_{i})\epsilon^{j_{1}..j_{N}} \epsilon^{j_{1}...j_{N}\alpha_{N+1}..\alpha_{N+Q}} P_{\alpha_{N+1}-1}(y_1)...P_{\alpha_{N+Q}-1}(y_Q) \\ 
    = & \left(1=\frac{N!^{2} C_{N,-g} \prod_{i=0}^{N-1}h_{i} }{Z_{N}} \right)  \frac{\det_{c,d \in 1,..,Q}\left[P_{c+N-1}(y_{d})\right]}{\Delta(y)},
\end{align}
where we used that $ \epsilon^{j_{1}..j_{N}} \epsilon^{j_{1}...j_{N} \alpha_{N+1}..\alpha_{N+Q}} = N! \epsilon^{(c_{1}+N)...(c_{Q}+N)}$ with $c_{a+N}\in 1,...,Q$ (since all the first $N$ indices range over $1,..,N$, the $\alpha_{N+1...\alpha_{N+Q}}$ must all range in $N+1,...,N+Q$ - and we can make that shift manifest by shifting the index). Finally we used the fact that the partition function can be written as $Z_{N,g}=N!^2 C_{N,-g} \prod_{i=0}^{N-1}h_{i}$.

Physically, we can think of these determinant insertions as adding $Q$ extra branes on top of the Fermi-sea created by the $N$ eigenvalues of the $K,M$ matrices. Following \cite{exact}, the expectation value of these determinant operators can be interpreted as a wavefunction in $Y$ space for the $Q$ compact branes. Indeed, it takes the form of a Slater determinant for the wavefunctions in the energy band $N+1,...,N+Q$. 
\begin{equation}
    \boxed{ \braket{\prod_{a=1}^{Q} det(y_a -K)} = \frac{\det_{c,d}\left[P_{c+N-1}(y_{d})\right]}{\Delta(y)}. }
\end{equation}

For the case of $Q=1$, from Eq.~(\ref{eq:jointvdmy}) we can see there will be no divergent factor of the $\Delta(y)$ and the computation reduces to 
\begin{equation}
    \boxed{\braket{\det(y-K)}_{N} = P_{N}(y).}
\end{equation}
We can in fact further simplify the multiple compact brane wavefunction by exploiting the fact that the orthogonal polynomials $P_{n}(y)=y^{n}$, i.e. are simple monomials:
\begin{eqnarray}
  \frac{\det_{c,d}\left[P_{c+N-1}(y_{d})\right]}{\Delta(y)} = & \frac{\epsilon^{c_{1}...c_{Q}}y_{1}^{c_{1}+N-1}...y_{Q}^{c_{Q}+N-1}}{\det\left[ P_{a-1}(y_{b}) \right]} \\
  = & \frac{\epsilon^{c_{1}...c_{Q}}y_{1}^{c_{1}-1}...y_{Q}^{c_{Q}-1} y_{1}^{N}..y_{Q}^N}{\det(P_{a-1}(y_{b})} \\
  = & \frac{\det\left[ P_{a-1}(y_{b}) \right]}{\det\left[ P_{a-1}(y_{b}) \right]} \times \prod_{a=1}^{Q} y_{a}^{N} \\
  = & \prod_{a=1}^{Q} \braket{\det(y_{a}-K)}_{N}.
\end{eqnarray}

One thing this exact result teaches us is that the compact branes do no interact amongst themselves (if no other branes are present) since their $Q$-point function factorizes into $Q$ $1$-point functions.
An alternative viewpoint on this result comes from looking directly at the matrix integral and noting that without any non-compact brane insertions (determinants involving the matrix $M$), the $M$-integral simply enforces a delta-function constraint forcing $A$ to vanish.

\subsection{The partition function solely with sources ($X=0,Y \neq 0$)}

One of the most important things to understand is the meaning of the source $Y$. We will see that ultimately the partition function solely with the source $Y$ turned on does not actually depend on $Y$, but the way in which this happens is illuminating. 

We now wish to show
\begin{equation}
     \frac{1}{Z_{N}}\int dK dM_{N \times N} e^{- \frac{1}{g} \left( Tr V(K) - K(M-Y) \right)} = \frac{\det \left[  P_{i-1}(y_{j})\right]}{\Delta(y)}= 1.
\end{equation}
We can again reduce this matrix integral to eigenvalues of $K$,$M$ and $Y$, but will now need an additional HCIZ integral to do the integral over the relative unitary between $K$ and $Y$.
\begin{equation}
     \frac{1}{Z_{N}} \int \prod_{i=1}^{N} da_{i} db_{i} \Delta(a)^2 \Delta(b)^2 e^{-\frac{1}{g}V(a_{i})} \frac{C_{N,g}\det\left[ e^{+\frac{1}{g} a_{k}b_{l}} \right]}{\Delta(a)\Delta(b)} \frac{C_{N,-g}\det\left[ e^{-\frac{1}{g} a_{k}y_{l}} \right]}{\Delta(a)\Delta(y)} 
\end{equation}
which upon using the anti-symmetry imposed by the remaining Van-der-Monde becomes \cite{Morozov:1995pb},
\begin{equation}
    \frac{C_{N,-g}C_{N,g}N!^2}{\Delta(y)Z(Y=0)} \int \prod_{i=1}^{N} da_{i} db_{i} \Delta(b) e^{-\frac{1}{g}\left( V(a_{i}) - a_{i}(b_{i}-y_{i}) \right)}. 
\end{equation}

Now we rewrite the single Van-der-Monde for $b$ as $\Delta(b)=\det\left[ Q_{i-1}(b_{j})\right] = \epsilon^{k_{1}...k_{N}} Q_{k_{1}-1}(b_1)...  Q_{k_{N}-1}(b_N)$ and obtain
\begin{equation}
     \frac{C_{N,g}C_{N,-g}N!^2}{\Delta(y)Z_{N}} \epsilon^{j_{1}...j_{N}} \prod_{i=1}^{N} \int dk_{i} dm_{i} e^{+\frac{1}{g}\left( V(k_{i}) - k_{i}(_{i}-y_{i}) \right)} Q_{j_{i}-1}(m_{i}).
\end{equation}
At this point, we rely on the fact that the integral (with source) acts as a kernel taking $Q_{j}(m)$ into $P_{j}(y)$:
\begin{equation}
\int dk_{i} dm_{i} e^{+\frac{1}{g}\left( V(k_{i}) - k_{i}(_{i}-y_{i}) \right)} Q_{j_{i}-1}(m_{i}) = (2\pi g) P_{j_{i}-1}(y_{i}),
\end{equation}
so that we obtain 
\begin{equation}
     \frac{C_{N,g}C_{N,-g}N!^2 (2\pi g)^N}{\Delta(y)Z_{N}} \epsilon^{j_{1}...j_{N}} P_{j_{1}-1}(y_{1})...P_{j_{N}-1}(y_{N}).  
\end{equation}
which simplifies to
\begin{equation}
 \boxed{\frac{1}{Z_{N}}\int dK dM_{N \times N} e^{+ \frac{1}{g} \left( Tr V(K) - K(M-Y) \right)} = \frac{\det \left[  P_{i-1}(y_{j})\right]}{\Delta(y)} = 1,}
 \end{equation}
 where, in the last equality, we used the fact that the Van-der-Monde determinant could equally be written as a determinant over the $P$-orthogonal polynomials:  $\Delta(y)=\det\left[P_{i-1}(y_{j})]\right]$. 

This looks exactly like the wavefunction for compact branes (recall Eq.~(\ref{eq:multFZZT})), except without the shift by $N$ in the degree of the polynomials. Usually, this shift is interpreted in the fermion picture of eigenvalues. The first $N$ levels are already filled by the branes "making up the matrix model", and the determinant insertions create additional fermions occupying higher levels. Using that reasoning, Eq.~(\ref{eq:justsource}) looks as if one had taken the eigenvalues and turned them into compact branes located at the $y_i$. Since we are using the eigenvalues making up the matrix model, we just populate the lowest $N$ levels, therefore explaining the lack of shift.

\subsection{Non-compact brane wavefunctions ($X \neq 0, Y=0$)} \label{sec:dualfzztcalcs}

We now wish to establish the identity
\begin{equation}
    \braket{\prod_{c=1}^{Q} \det(x_c -M)} = \frac{\det\left[ Q_{a+N-1}(x_{b}) \right]}{\Delta(x)},
\end{equation}
by computing the right hand side of the following equation 
\begin{equation}
   \braket{\prod_{c=1}^{Q} \det(x_c -M)} = \frac{1}{Z_{N}}\int dK dM_{N \times N} e^{+\frac{1}{g} Tr \left( V(K)-KM \right)} \prod_{c=1}^{Q} \det(x_c -M),
\end{equation}
in terms of the bi-orthogonal polynomials.

The calculation follows along very similar lines to those in Sec.\ref{sec:FZZTcalcs}. Using the HCIK-integral to do the integral over the relative unitary between $K$ and $M$, we can again obtain an expression purely in terms of the eigenvalues of $K$ and $M$:
\begin{align}
       \braket{\prod_{c=1}^{Q} \det(x_c -M)} = & \frac{1}{Z_{N}} \int \prod_{i=1}^{N} dk_{i} dm_{i} e^{+\frac{1}{g}\left( V(k_{i})- k_{i}m_{i} \right)} \Delta(k)^2 \Delta(m)^2 \frac{C_{N,-g} N!}{\Delta(k) \Delta(m)} \prod_{c=1}^{Q} \prod_{j=1}^{N} (x_{c}-m_j).
\end{align}
The idea this time is to combine the $N$ $m_{i}$ and $Q$ $x_{c}$ eigenvalues into an $N+Q$ dimensional vector $z_{\alpha}$
\begin{equation}
    z_{\alpha} = m_{i} \, \thinspace \text{for} \thinspace \alpha \in 1,..,N;  \quad   z_{\alpha} = x_{c} \, \thinspace  \text{for} \thinspace  \alpha \in N+1,..,N+Q. 
\end{equation}
Then note that 
\begin{equation}
    \Delta(b) \prod_{c=1}^{Q} \prod_{j=1}^{N} (x_{c}-m_{j}) = \frac{\prod_{\alpha < \beta}(z_{\beta}-z_{\alpha})}{\Delta(x)},
\end{equation}
and we can rewrite the Van-der-Monde in terms of orthogonal polynomials
\begin{equation}
    \Delta(k)=\det\left[P_{i-1}(k_{j})\right] \quad \quad \frac{\prod_{\alpha < \beta}(z_{\beta}-z_{\alpha})}{\Delta(x)} = \frac{\det\left[Q_{\alpha-1}(z_{\beta})\right]}{\Delta(x)}.
\end{equation}
so that our integral becomes 
\begin{align}
     \braket{\prod_{c=1}^{Q} det(x_c -M)} = & \frac{C_{N,-g} N!}{\Delta(x) Z_{N}} \int \prod_{i=1}^{N} dk_{i} dm_{i} e^{+\frac{1}{g}\left( V(k_{i})- k_{i}m_{i} \right)}  \det(P_{i-1}(k_{j})) \det(Q_{\alpha-1}(z_{\beta})) \\
     = & \frac{C_{N,-g} N! }{\Delta(x) Z_{N}} \int \prod_{i=1}^{N} dk_{i} dm_{i} e^{+\frac{1}{g}\left( V(k_{i})- k_{i}m_{i} \right)} \epsilon^{j_{1}...j_{N}} P_{j_{1}-1}(k_{1})...P_{j_{N}-1}(k_{N}) \\
     & \times  \epsilon^{\alpha_{1}...\alpha_{N+Q}} Q_{\alpha_{1}-1}(m_1)...Q_{\alpha_{N}-1}(m_N)Q_{\alpha_{N+1}-1}(x_1)...Q_{\alpha_{N+Q}-1}(x_Q).
\end{align}

Of course, we can pull out all the factors of $Q$ which depend on $x$ outside of the integral and then use the bi-orthogonality condition in each factor
\begin{equation}
    \int dk dm e^{+\frac{1}{g}\left(V(k)- k m \right)} P_{j-1}(k) Q_{\alpha-1}(m)= \delta_{j, \alpha} h_{j-1},
\end{equation}
so that 
\begin{align}
    \braket{\prod_{a=1}^{Q} \det(x_a -M)} = & \frac{C_{N,-g} N!}{\Delta(x) Z_{N}} (\prod_{i=0}^{N-1}h_{i})\epsilon^{i_{1}..i_{N}} \epsilon^{i_{1}...i_{N}\alpha_{N+1}..\alpha_{N+Q}} Q_{\alpha_{N+1}-1}(x_1)...Q_{\alpha_{N+Q}-1}(x_Q) \\ 
    = & \left(\frac{C_{N,-g} N!^2 \prod_{i=0}^{N-1}h_{i} }{Z_{N}} =1 \right) \frac{\det_{c,d\in 1,..,Q}\left[Q_{c+N-1}(x_{d})\right]}{\Delta(x)},
\end{align}
where we used that $ \epsilon^{i_{1}..i_{N}} \epsilon^{i_{1}...i_{N} \alpha_{N+1}..\alpha_{N+Q}} = N! \epsilon^{(c_{1}+N)...(c_{Q}+N)}$ with $c_{a+N}\in 1,...,Q$ (since all the first $N$ indices range over $1,..,N$, the $\alpha_{N+1...\alpha_{N+Q}}$ must all range in $N+1,...,N+Q$ - and we can make that shift manifest by shifting the index). Finally we used the fact that the partition function can be written as $Z_{N,g}=N!^2 C_{N,g} \prod_{i=0}^{N-1}h_{i}  $. 

We thus have obtained
\begin{equation}
\boxed{  \braket{\prod_{a=1}^{Q} \det(x_a -M)} = \frac{\det\left[ Q_{c+N-1}(x_{d})\right]}{\Delta(x)} }.
\end{equation}

%
%
%
%
%

\subsection{Proof of the general duality $X \neq 0, Y \neq 0 $ via orthogonal polynomials}

We wish to show that
\begin{eqnarray} 
   & \frac{1}{Z_{N}} \int dK dM_{N \times N} e^{+ \frac{1}{g} \Tr \left(  V(K) - K(M-Y) \right)} \prod_{a=1}^{Q} \det(x_{a}-M) & \\
    = & \frac{1}{\Delta(x) \Delta(y)} \det \left[ \begin{array}{c|c} 
    P_{i-1}(y_{j}) & Q_{i-1}(x_{b}) \\ \hline
    P_{a+N-1}(y_{j}) & Q_{a+N-1}(x_{b}) 
    \end{array}\right] \\
     & \frac{(-1)^{NQ}}{Z_{Q}}\int dA dB_{Q \times Q} e^{- \frac{1}{g} \Tr \left( V(A) - A(B+X) \right)} \prod_{i=1}^{N} \det(y_{i}-B), &
\end{eqnarray}
by computing the top and bottom lines explicitly in terms of orthogonal polynomials. Many of the steps are familiar from Secs. \ref{sec:FZZTcalcs} and \ref{sec:dualfzztcalcs}, so we will be brief. 

\subsubsection{Starting from the $K,M$ matrix model}
 
We first start with 
\begin{equation}
   \frac{1}{Z_{N}}\int dK dM_{N \times N } e^{+\frac{1}{g} \Tr \left( V(K)-K(M-Y) \right) } \prod_{c=1}^{Q} det(x_{c}-M),
\end{equation}
which becomes after two applications of the HCIZ integral:
\begin{align}
    \frac{1}{Z_{N}}\int \prod_{i=1}^{N} dk_{i} dm_{i} \Delta(k)^2 \Delta(m)^2 e^{+\frac{1}{g}V(k_{i})} \frac{C_{N,-g}\det\left[ e^{-\frac{1}{g} k_{n}m_{l}} \right]}{\Delta(k)\Delta(m)} \frac{C_{N,+g}\det\left[ e^{+\frac{1}{g} k_{n}y_{l}} \right]}{\Delta(k)\Delta(y)} \prod_{c=1}^Q \prod_{j=1}^{N} (x_{c}-m_{j}).
\end{align}

By combining the $m_{i}$ and $x_{c}$ into $z_{\alpha}$ as in Sec. \ref{sec:dualfzztcalcs}, writing $\Delta(z)= \det\left[ Q_{\alpha -1}(z_{\beta})\right]$ and using the anti-symmetry of the remaining Van-der-Monde determinants, we arrive at  
\begin{align}
   & \frac{C_{N,-g}C_{N,g} N!^2}{Z_{N} \Delta(x) \Delta(y)}\int \prod_{i=1}^{N} dk_{i} dm_{i}  e^{+\frac{1}{g}\left( V(k_{i})-k_{i}(m_{i}-y_{i}) \right)} \det\left[ Q_{\alpha -1}(z_{\beta})\right] \\
      = & \frac{C_{N,-g}C_{N,g} N!^2}{Z_{N}\Delta(x) \Delta(y)} \epsilon^{\alpha_{1}...\alpha_{N+Q}} Q_{\alpha_{N+1}}(x_1)...Q_{\alpha_{N+Q}}(x_{Q}) \\
     &  \times \prod_{i=1}^{N} \int dk_{i} dm_{i}  e^{+\frac{1}{g}\left( V(k_{i})-k_{i}(m_{i}-y_{i}) \right)} Q_{\alpha_{i}-1}(m_{i}).
\end{align}
Now we use the fact the integral will map $Q_{\alpha_{i}-1}(m_{i}) \rightarrow (2\pi g)P_{\alpha_{i}-1}(y_{i})$ to obtain
\begin{align}
   & \frac{C_{N,-g}C_{N,g} N!^2 (2\pi g)^{N}}{Z_{N}\Delta(x) \Delta(y)} \epsilon^{\alpha_{1}...\alpha_{N+Q}} P_{\alpha_{1}-1}(y_{1})...P_{\alpha_{N}-1}(y_{N}) Q_{\alpha_{N+1}}(x_1)...Q_{\alpha_{N+Q}}(x_{Q})\\
   & = \boxed{\frac{1}{\Delta(x) \Delta(y)} \det \left[ \begin{array}{c|c} 
    P_{i-1}(y_{j}) & Q_{i-1}(x_{b}) \\ \hline
    P_{a+N-1}(y_{j}) & Q_{a+N-1}(x_{b}) 
    \end{array}\right]}.
\end{align}

\subsubsection{Starting from the $A,B$ matrix model}

We now start from 
\begin{equation} \label{eq:dualmmstart}
    \frac{(-1)^{NQ}}{Z_{Q}}\int dA dB_{Q \times Q} e^{- \frac{1}{g} \left( \Tr V(A) - A(B+X) \right)} \prod_{i=1}^{N} \det(y_{i}-B).
\end{equation}
If one were not careful, one would \textit{incorrectly} guess that one should obtain
\begin{equation}
    \frac{1}{\Delta(x) \Delta(y)} \det \left[ \begin{array}{c|c} 
    Q_{i-1}(y_{j}) & P_{i-1}(x_{b}) \\ \hline
    Q_{a+N-1}(y_{j}) & P_{a+N-1}(x_{b}) 
    \end{array}\right].
\end{equation}
We will see the change in sign in front of the potential actually saves us and instead gives us the correct answer. After two HCIZ integrals becomes, Eq.~(\ref{eq:dualmmstart}) becomes
\begin{align}
    \frac{(-1)^{NQ}}{Z_{Q}}\int \prod_{i=1}^{Q} da_{i} db_{i} \Delta(a)^2 \Delta(b)^2 e^{-\frac{1}{g}V(a_{i})} \frac{C_{Q,+g}\det\left[ e^{+\frac{1}{g} a_{k}b_{l}} \right]}{\Delta(a)\Delta(b)} \frac{C_{Q,-g}\det\left[ e^{-\frac{1}{g} a_{k}x_{l}} \right]}{\Delta(a)\Delta(x)} \prod_{i=1}^{N} \prod_{c=1}^{Q} (y_{i}-b_{c}).
\end{align}

We now combine the $a_{c}$ and $y_{i}$ into the joint object $z_{\alpha}$ (with $z_{\alpha \in 1,..,Q}= a_{c}$ and $z_{\alpha \in Q+1,..,N+Q}= y_{i}$), and write $\Delta(s)\prod_{i=1}^{N} \prod_{=1}^{Q} (y_{i}-b_{c}) =  \frac{\prod_{\alpha <\beta} (z_{\beta}-z_{\alpha})}{\Delta(y)}$. 
The "trick" will be to write $\Delta(z)=\det\left[z_{\beta}^{\alpha-1}\right] = \det\left[ P_{\alpha-1}(z_{\beta})\right]$ (and not for example $\det \left[ Q_{\alpha-1}(z_{\beta}) \right]$),
so that we obtain 
\begin{align}
    \frac{C_{Q,+g}C_{Q,-g} (-1)^{NQ} N!^2}{Z_{-g,Q} \Delta(x) \Delta(y)}\int \prod_{i=1}^{Q} da_{i} d_{i} e^{-\frac{1}{g}\left(V(a_{i}) +a_{i}(b_{i}-x_{i}))\right)} & & \nonumber \\
    \times \epsilon^{\alpha_{1}...\alpha_{N+Q}} P_{\alpha_{1}-1}(s_{1})..P_{\alpha_{Q}-1}(s_{Q})P_{\alpha_{Q+1}-1}(y_1)...P_{\alpha_{Q+N}-1}(y_N).& &
\end{align}

Obviously, we can simply pull out the $P(y)$'s from the integral. As for the remaining $s$-dependent terms, we note that there is now a "reverse" duality between the orthogonal polynomials. More precisely, the integral with the other sign for the potential now acts as a kernel taking $P_{k}(s) \rightarrow (2\pi g)Q_{k}(x)$
We therefore arrive at 
\begin{align}
    & \frac{C_{Q,+g}C_{Q,-g} (-1)^{NQ} N!^2 (2 \pi g)^Q}{Z_{Q} \Delta(x) \Delta(y)} \epsilon^{\alpha_{1}...\alpha_{N+Q}} Q_{\alpha_{1}-1}(x_{1})..Q_{\alpha_{Q}-1}(x_{Q})P_{\alpha_{Q+1}-1}(y_1)...P_{\alpha_{N+Q}-1}(y_N) \\
    & =  \boxed{\frac{1}{\Delta(x) \Delta(y)} \det \left[ \begin{array}{c|c} 
    P_{i-1}(y_{j}) & Q_{i-1}(x_{b}) \\ \hline
    P_{a+N-1}(y_{j}) & Q_{a+N-1}(x_{b}) 
    \end{array}\right]}.
\end{align}
which proves the result.

\bibliographystyle{utphys}
\bibliography{refs}

\end{document}